\documentclass[12pt]{iopart}

\usepackage[colorlinks,allcolors=blue]{hyperref}
\usepackage{graphicx}
\usepackage{braket}
\usepackage[numbers,sort&compress]{natbib}
\usepackage{amssymb}
\usepackage{booktabs}

\expandafter\let\csname equation*\endcsname\relax
\expandafter\let\csname endequation*\endcsname\relax

\usepackage{amsmath}

\begin{document}

\title{Efficient light harvesting and photon sensing via engineered cooperative effects}

\author{Francesco Mattiotti$^{1,2,3}$, Mohan Sarovar$^4$, Giulio G.~Giusteri$^5$, Fausto Borgonovi$^{2,6}$ and G.~Luca Celardo$^{7,8}$}

\address{$^1$ ISIS (UMR 7006) and icFRC, University of Strasbourg and CNRS, 67000 Strasbourg, France}
\address{$^2$ Dipartimento di Matematica e Fisica and Interdisciplinary Laboratories for Advanced Materials Physics, Universit\`a Cattolica del Sacro Cuore, via della Garzetta~48, Brescia I-25123, Italy}
\address{$^3$ Department of Physics, University of Notre Dame, Notre Dame, IN 46556, USA}
\address{$^4$ Extreme-scale Data Science and Analytics, Sandia National Laboratories, Livermore, California 94550, USA}
\address{$^5$ Dipartimento di Matematica and Padua Quantum Technologies Research Center, Universit\`a degli Studi di Padova, Via Trieste 63, 35121, Padova, Italy}
\address{$^6$ Istituto Nazionale di Fisica Nucleare, Sezione di Milano, via Celoria 16, Milano I-20133, Italy}
\address{$^7$ Benem\'erita Universidad Aut\'onoma de Puebla, Apartado Postal J-48, Instituto de F\'isica,  72570, Mexico}
\address{$^8$ Dipartimento di Fisica e Astronomia, Universit\`a di Firenze, Via Sansone 1, 50019 Sesto Fiorentino, Firenze, Italy}
\ead{mattiotti@unistra.fr}

\begin{abstract}
Efficient devices for light harvesting and photon sensing are fundamental building blocks of basic energy science and many essential technologies.
Recent efforts have turned to biomimicry to design the next generation of light-capturing devices, partially fueled by an appreciation of the fantastic efficiency of the initial stages of natural photosynthetic systems at capturing photons. In such systems extended excitonic states are thought to play a fundamental functional role, inducing cooperative coherent effects, such as superabsorption of light and supertransfer of photoexcitations.
Inspired by this observation, we design an artificial light-harvesting and photodetection device that maximally harnesses cooperative effects to enhance efficiency. The design relies on separating absorption and transfer processes (energetically and spatially) in order to overcome the fundamental obstacle to exploiting cooperative effects to enhance light capture: the enhanced emission processes that accompany superabsorption. 
This engineered separation of processes greatly improves the efficiency and the scalability of the system.
\end{abstract}


\maketitle

Since the discovery of coherent features 
in natural light-harvesting
complexes~\cite{photo,photoT,Collini2010,photo2,photo3,Rebentrost2009,Caruso2009,Ishizaki2010} and 
subsequent studies of the functional role of these features,  
there has been great
interest in engineering biomimetic devices for photon sensing or light harvesting,
able to exploit
coherent quantum features~\cite{creatore, sarovarbio, superabsorb}
even in ambient conditions. 
Natural light-harvesting complexes are composed of organic chromophores, each characterized by a dipole moment that determines its coupling to the electromagnetic field (EMF) and its interaction with neighboring systems. Once light is
absorbed, the induced photoexcitation is transmitted to another
molecular aggregate, called the reaction center (RC), where charge
separation occurs, which drives subsequent steps in the photosynthesis chain.
The main quantum coherent effects that are thought to contribute to the high efficiency of natural photosynthetic complexes are induced by the delocalization of the excitation over many molecules~\cite{schulten,kaplan1,kaplan2,srlloyd,sr2,vangrondelle,srfmo,srrc, caoJPCA2009,caoNJP2013}. 
Such delocalized excitonic states can have an enhanced dipole strength that strongly couples them to the EMF. Thus, these states are able to absorb light at a rate much larger than the single-molecule absorption rate. Indeed, the absorption rate of a single delocalized excitonic state can increase linearly with the number of molecules over which the excitation is
delocalized. On the other hand, the states that absorb light efficiently also exhibit enhanced emission rates (termed superradiance) since the process is time reversible. The time-reversal character of absorption/emission processes is one of the main reasons for the Shockley-Queisser efficiency limit for photocell devices~\cite{SQ}.  

Several recent proposals have aimed to suppress re-emission in systems composed of few molecules.  
Specifically, it has been discussed how coherent effects can suppress re-emission 
leaving absorption intact for a molecular dimer~\cite{scullyPRL,scully1,scully2}, and
 a mechanism for suppressing re-emission, exploiting bright and dark states and fast thermal relaxation, has been devised for 
  two~\cite{erik2017,creatore} and three coupled molecules~\cite{zhang}.   
Moreover, in Ref.~\cite{superabsorb} it is shown how to maximize cooperative absorption and engineer super-absorbing many-atom structures that avoid superradiance by controlling the structure of a collection of atoms/molecules and engineering their vibrational environment to achieve delicately tuned thermal transition rates.
The above results suggest that one can in principle exploit cooperative effects to enhance light capture and transfer by designing engineered structures that avoid detrimental effects such as superradiance.

In this work we propose a molecular architecture that is able to suppress re-emission, while leaving absorption intact. The general idea of our device is based on engineering a super-absorbing state at high energy. Once the excitation is absorbed, it is transferred to the low-energy states by thermal relaxation, which, being much faster then re-emission, prevents radiative losses. The low-energy states transfer the excitation to the central core absorber with reduced radiative losses due to the fact that their dipole strength is smaller  than the high-energy absorbing state.  

Specifically, our light-harvesting device is composed of a ring of $N$  molecules surrounding a central core absorber, similar to photosynthetic complexes found in purple bacteria. 
The molecular arrangement is engineered so that we have three bright excitonic states with orthogonal dipole moments, one at high energy and two at low energy. Only the low-energy states are coupled to the central core absorber. 
By changing the orientation of the molecular dipoles, we can control the brightness of these states. If we make the high-energy state the brightest, absorption mainly occurs through it. That state is not directly coupled to the central core absorber, but fast thermal relaxation funnels the excitation towards the low-energy states. The brightness of these cannot be zero if we want to exploit the radiative coupling with the central core absorber to transfer and trap the excitation. Nevertheless, to minimize re-emission, it is convenient to keep the low-energy brightness rather small. Balancing these two requirements, one can find optimal parameter ranges in which transfer is maximized.

The efficiency of our proposed architecture is analyzed as a function of the ring radius, which determines the number of molecules needed to keep a given density, and the  orientation of the molecules in the ring.
Related work providing theoretical insight into
the role of fold symmetry in
promoting efficient energy transfer in LH2 can be found in \cite{caoPNAS2013}.
Since our proposed architecture suppresses radiative recombination, the advantages of our design are present only if radiative recombination is the main cause of efficiency losses. For instance, for very large trapping rates at the RC, radiative losses become negligible. 
Note that the mechanism proposed here extends the design of Ref.~\cite{creatore} for two molecules to a device composed of an arbitrary number of molecules. Moreover, while the transfer mechanism in Ref.~\cite{creatore} was not radiative, here we consider that the coupling to the central core absorber is radiative in nature, as it happens in natural photosynthetic systems.

Our model device can be tuned to mimic natural light-harvesting systems, and we compare the performance of the optimized device to that of a model of a natural system, under weak laser excitation and a realistic model for natural sunlight.
The efficiency of our device, in the optimal size range $N\approx 50$, is found to be more than two orders of magnitude larger than that of a single absorber and enhanced by a factor larger than $N$.
In the same regime, as we will show below, the efficiency of models mimicking natural systems
is enhanced only by a factor equal to the number of absorbers.
We also show that, under natural sunlight excitation, the decoupling mechanism of our model leads to an improvement in the efficiency of about five times with respect to a non-decoupled configuration.

\section{The structural model}

\begin{figure}
  \centering
    \includegraphics[width=0.6\textwidth]{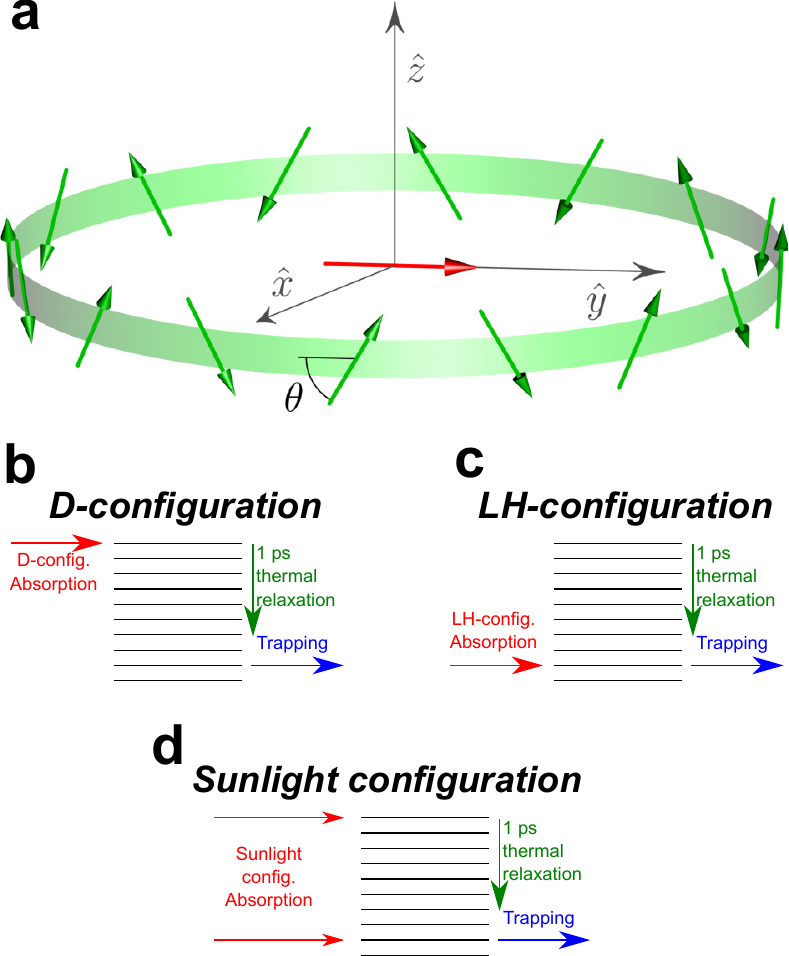}
  \caption{{\it Dipole arrangement of the device considered in this paper and level schemes}.
    (a) The dipoles are tangential to the ring with a component along the $z$ direction with alternating signs, see details in the text.
    Dipole disposition of the model is shown for $N=16$ and $\theta=\pi/3$, see
    Eq.~\eqref{eq:dip}. The central dipole lies along the $y$ axis and is
    parallel to the dipole of the first excited state of the ring.
(b-d) In the lower part a schematic representation of the excitonic
eigenstates of the model is shown, for the three kinds of absorption considered. (b) In the \emph{D-configuration} the polarization of the incident light is such that the main absorbing
state is separated from the trapping state, and transfer of energy
between the two states is driven by fast thermal relaxation processes. (c) In the \emph{LH-configuration} the polarization of the incident light is such that the main absorbing
state coincides with the trapping state. (d) In the \emph{Sunlight configuration}, absorption happens at both edges of the spectrum and fast thermalization drives the whole absorbed excitation to the low-energy trapping state.
}
  \label{fig:dip}
\end{figure}

We consider a molecular complex composed of $N$ molecules (or point absorbers) placed on a ring of radius $R$ with constant density $d$, sketched in Fig.~\ref{fig:dip}a.
The excitation energy of each molecule, $\hbar \omega_0$, is assumed to be the same while their dipole moments $\vec{\mu}_n$ have the same modulus, $\mu$, but possibly different orientations, indicated by the unit vector $\hat{p}_n$. The properties of the system will be studied for different system sizes, keeping the density $d$ fixed, while varying the radius.  The role of the ring structure is to absorb the electromagnetic radiation and transfer the excitation to a central core absorber, such as the reaction center of natural photosynthetic complexes, that we mimic with an additional central site, coupled to an external environment where the excitation can be irreversibly trapped. The whole system is also coupled to a thermal bath at fixed temperature $T$. 

The molecules lie on a ring in the $xy$ plane with their dipole moments tangential to the circumference and tilted
by an angle $\pm\theta$ with respect to the $xy$ plane, see Fig.~\ref{fig:dip}a. Moreover, the vertical components are
alternated upwards and downwards, so that the normalized dipole moment orientations are
\begin{eqnarray}
  \label{eq:dip}
  \hat{p}_n = \cos \theta \, \hat{\phi} + (-1)^n \sin \theta \, \hat{z} \, ,
\end{eqnarray}
where $\hat{\phi}$ and $\hat{z}$ are the unit vectors corresponding,
respectively, to the azimuthal and to the vertical direction of
a cylindrical coordinate system.  
Due to the discrete rotational invariance of the system around the $z$ axis, the eigenstates and their dipoles, in the single excitation manifold (that is appropriate to describe the weak-field limit) take the form
$\ket{E_\alpha} = \sum_n
c_n(E_\alpha) \ket{n}$, with $c_n(E_\alpha) =
\frac{1}{\sqrt{N}}\exp({i2\pi\alpha n / N})$, and the corresponding dipoles are 
$\vec{p}_\alpha = \sum_n c_n(E_\alpha) \hat{p}_n$,
with a square modulus ranging from zero to $N$. 
Here, $\ket{n}$ is the state where the $n$-th molecule is excited and all the other ones are in their ground state.
For any $\theta$, the dipole strength $|\vec{p}_\alpha|^2$ is non-vanishing only for three excitonic states: $\ket{E_2}, \ket{E_3}$ and $\ket{E_N}$, where we have ordered the excitonic states by increasing energy.
Actually, $\ket{E_2}$ and $\ket{E_3}$ span a degenerate subspace and, without loss of generality, we choose them to be two orthogonal states in this subspace such that their dipole moments are
\begin{subequations}
\label{dips}
  \begin{eqnarray}
    \vec{p}_{E_2} &= \sqrt{\frac{N}{2}} \cos \theta \, \hat{y}\,, \\
    \vec{p}_{E_3} &= \sqrt{\frac{N}{2}} \cos \theta \, \hat{x} \, ,
  \end{eqnarray}
  with $\hat{x}$ and $\hat{y}$ being the unit vectors of the planar
  axes. The
  third bright eigenstate is the highest-energy exciton, whose dipole moment is
  \begin{eqnarray}
    \vec{p}_{E_N} = \sqrt{N} \sin \theta \, \hat{z} \, ,
  \end{eqnarray}
  perpendicular to the ring plane.
\end{subequations}
Since the emission rate from a
state is proportional to $|\vec{p}_\alpha|^2$, the states $\ket{E_{2,3,N}}$ are also
superradiant, with an emission rate proportional to $N$. All the other
states are subradiant with zero emission rate (dark states). 
Moreover, the enhanced dipole of the high-energy state $\ket{E_N}$ is orthogonal to the $xy$ plane, and to the dipoles of the other bright states: this guarantees the  separation of excitation and transfer processes.

We add a RC to this model by an additional site $\ket{rc}$ placed at the center of the ring, with excitation energy
$\hbar \omega_{\rm rc}$, and
coupled to an external environment (sink) where the excitation can be
trapped at rate $\kappa$. The RC is dipole coupled to the choromophores on the ring, and we choose the dipole moment of the RC along the $y$ axis, $\hat{p}_{rc} = \hat{y}$. We also set the energy of the RC site to be resonant with the first excited state
of the ring. As a consequence, only the state $|E_2\rangle$, with 
dipole moment along the $y$ direction,
has a non-vanishing coupling strength to the reaction center (see~\ref{app:Oc}),
\begin{eqnarray}
  \Omega_C = \frac{\mu^2}{\epsilon_r R^3} \sqrt{\frac{N}{2}} \cos \theta \, .
\end{eqnarray}
Since the density of dipoles on the ring $d = N/(2\pi R)$ is kept constant, then $\Omega_C$ scales with $N$ as
\begin{eqnarray}
  \Omega_C = \frac{\mu^2 (2 \pi d)^3 \cos \theta}{\sqrt{2} \epsilon_r}
  N^{-5/2} \, .
\label{OC}
\end{eqnarray}
This coupling determines the transfer between the ring and the RC.

Since  the sum of the dipole strengths of all
the eigenstates must be constant, $\sum_\alpha |\vec{p}_\alpha|^2=N$,
increasing $\theta$ from 0 to $\pi/2$, the dipole strength of the high-energy state ($\ket{E_N}$) is also increased, so that the smallest dipole strength 
of the low-energy states is decreased [see Eqs.~\eqref{dips}]. This configuration limits radiation losses together with ring--RC transfer. Nevertheless, as we discuss in the manuscript, the transfer is much faster than the radiative losses below a critical ring size, thus preserving the trapping efficiency.

In order to describe this ring+RC system interacting with an electromagnetic field and with a thermal phonon reservoir, we use a master equation, see section~\ref{sec:hme}.  The reservoir coupling is assumed to be Markovian, and each molecule is assumed to couple to an independent Ohmic bath at the same temperature. The parameters of the bath, see~\ref{sec:mas}, have been chosen so that 
thermal relaxation among exciton states occurs in about a picosecond at room
temperature for $N=32$. This is comparable with
estimates for natural photosynthetic systems reported in
literature~\cite{schulten1a,schulten1b}, and it is much faster than the
emission timescales, of the order of nanoseconds. 

As a measure of efficiency of our device we use the 
stationary current
transmitted from the central site to the sink, while the system is driven by the
EMF, defined as
\begin{eqnarray}
  \label{eq:cur}
  I = \lim_{t \to \infty} \kappa \rho_{rc}(t) \, ,
\end{eqnarray}
where $\rho_{rc}(t)$ is the population of the RC at time
$t$. 
The current is  further divided by the maximal stationary current, $I_s$, of the RC alone (in absence of the ring) under the same illumination conditions.
In this way the normalized current $I/I_s$ 
measures the increased efficiency of our network of dipoles with respect to a
single site. Since the excitation can only be trapped
in the central site, a normalized current larger than unity indicates increased
effectiveness of the network of sites in absorbing and transferring the
excitation.

\section{Hamiltonian and Master Equation}
\label{sec:hme}
In our model we choose the values of parameters to be close to realistic values found in natural complexes such as LHI-LH2 in purple bacteria~\cite{schulten1a,schulten1b}:
squared dipole moment $\mu^2 = 519310$~\AA$^3$~cm$^{-1}$ (corresponding to $\mu \approx 10$~D);
excitation energy of the ring sites $\hbar \omega_0 = 12911$~cm$^{-1}$ (corresponding to a single-site transition wavelength $\lambda_0 \approx
775$~nm); site density $d=32/(2 \pi
R_0)$ with $R_0=5$~nm. 

The dynamics of our model is described by the following master equation, written in a rotating frame with respect to the driving field mode frequency $\omega$~\cite{kaiser1,kaiser2,mukamelspano,lloydfmo}:
\begin{eqnarray}
  \frac{d\rho}{dt} = &-\frac{i}{\hbar} \left[ H_{S}, \rho \right] + \mathcal{L}_{fl}[\rho] + \mathcal{L}_{rc}[\rho] + \mathcal{R}_T [\rho] \, .
    \label{eq:mas}
\end{eqnarray}
Here, the Hamiltonian $H_S = H_0 + \Delta + H_{EM}$ captures the evolution of the system in the weak-field limit, where no more than one excitation is induced in the system. Specifically, $ H_0 = \sum_{n=1}^N \hbar (\omega_0-\omega) \ket{n}\bra{n} + \hbar (\omega_{\rm rc}-\omega) \ket{rc}\bra{rc}$ represents the site energies of the ring chromophores and RC, and $\Delta$ represents the Coulomb coupling between chromophores. Here we assume that the ring chromophores are distant from each other and from the RC, so that each molecule can be approximated as a point dipole.
Explicitly, the matrix elements of the coupling $\Delta$ are
 \begin{eqnarray}
 \label{dipdip}
    \bra{n}\Delta\ket{m} \approx& \frac{\mu^2}{\epsilon_r r_{nm}^3} \big[ \hat{p}_n \cdot
\hat{p}_m - 3 \left( \hat{p}_n \cdot \hat{r}_{nm} \right) \left( \hat{p}_m \cdot
      \hat{r}_{nm} \right) \big] (1- \delta_{n,m})\, , \nonumber \\
  \bra{rc} \Delta \ket{n} \approx& \frac{\mu^2}{\epsilon_r R^3} \hat{p}_{rc} \cdot \hat{p}_n \, .
\end{eqnarray}
Here $\hat{r}_{nm} := \vec{r}_{nm}
/ r_{nm}$ is the unit vector joining the $n$-th and the $m$-th
sites, and $\hat{p}_n :=  \vec{\mu}_n  /
\mu$ is the normalized dipole moment of the $n$-th site. The dielectric constant is $\epsilon_r=1$, which is a good approximation for molecules surrounded by air.
In principle, the nearest-neighbor coupling in the ring should be computed without using the point dipole-point dipole approximation, because the distance between the chromophores is comparable to the molecular size. Nevertheless, this is a detail which does not qualitatively change our results. For instance, the nearest-neighbor dipolar couplings used in this manuscript range between $\approx500-1200$~cm$^{-1}$, which is comparable to the $\approx 400-800$~cm$^{-1}$ couplings estimated from detailed electronic calculations in Ref.~\cite{schulten1a}. On the other hand, the coupling with the central core absorber can be safely assumed to be a point dipole-point dipole coupling (as it has been done also in Refs.~\cite{schulten1a,schulten1b}) since the molecules in the ring are far apart from the central core absorber. 
The expressions in Eq.~\eqref{dipdip} are valid in the small volume limit, where the wavelength of the optical transition is larger than the system size ($\lambda_0 \gg R$), which is the regime where natural light-harvesting complexes operate. The full expressions, without this approximation can be found in~\ref{sec:Muk}. 

The term $H_{EM}$ in the Hamiltonian describes the coupling between molecules and the continuous-wave (CW) driving laser and it is given by
\begin{eqnarray}
  \label{hem}
  H_{\it EM} = \frac{\hbar \Omega_R}{2} \sum_q (\hat{p}_q \cdot \hat{\epsilon})
  e^{-i\vec{k} \cdot \vec{r}_q} \ket{0} \bra{q} 
  + \textrm{h.c.} \, ,
\end{eqnarray}
where 
$\Omega_R = \mu E_0 / \hbar$ is the Rabi frequency, $E_0$ is the amplitude
of the electric field, $\hat{\epsilon}$ is a unit vector which specifies
the laser polarization, $\vec{k}$ is the wave vector of the laser field, $\ket{q}$ represents the system sites (either the ring sites $\ket{n}$ or the reaction center $\ket{rc}$) and
$\vec{r}_q$ is the position of the $q$-th site. $\ket{0}$ is the ground state of all molecules in the system. In our calculations we always choose $|\vec{k}| \approx 2\pi/\lambda_0$ and since we are in the small volume limit, we can approximate the matrix elements of Eq.~\eqref{hem} as
\begin{eqnarray}
      \bra{0} H_{\it EM} \ket{E_\alpha} \approx \frac{\hbar \Omega_R}{2}
      (\vec{p}_\alpha \cdot \vec{\epsilon}).
\label{Hem2}
\end{eqnarray}
We have confirmed the validity of this approximation for our ring system, as long as $R/\lambda_0
<0.1$, see~\ref{sec:Rlambda}. 

Now we return to the other terms in the master equation, Eq.~\eqref{eq:mas}: ${\cal L}_{fl}$ and ${\cal L}_{rc}$ are Lindblad dissipators derived under the Born-Markov and secular approximations~\cite{petruccione} and they describe, respectively, fluorescence emission of the molecules and transfer to the RC, while ${\cal R}_T$ is a non-secular Redfield dissipator~\cite{petruccione} modelling thermal relaxation and decoherence in the presence of a thermal bath. The dissipators read explicitly
\begin{eqnarray}
  \mathcal{L}_{fl}[\rho] = & \sum_{m,n} \Gamma_{mn} \left[ a_n
  \rho a_m^\dag  - \frac{1}{2} \left\{ a_m^\dag a_n , \rho \right\} \right] \label{Lfl} \\
\mathcal{L}_{rc}[\rho] = & \kappa \left[ a_{rc}
  \rho a_{rc}^\dag  - \frac{1}{2} \left\{ a_{rc}^\dag a_{rc} , \rho \right\} \right] \label{Lrc} \\
  {\cal R}_T[\rho] =  &\sum_{\omega,\omega'} \sum_{n} \frac{\gamma^{(p)}(\omega)}{2} \left[ A_n(\omega) \rho A_n^\dag(\omega')+ A_n(\omega') \rho A_n^\dag(\omega) \right.\nonumber \\
  &\left. \qquad \qquad \qquad - A_n^\dag(\omega') A_n(\omega) \rho- \rho A_n^\dag(\omega) A_n(\omega') \right]
    \label{LT}
\end{eqnarray}
where the sums over $m,n$ run over all the system sites (ring sites or RC), $a_n=\ket{0}\bra{n}$ (here, $\bra{n}$ can be a ring site or the RC), $a_{rc}~=~\ket{0}\bra{rc}$, and $\Gamma_{mn} \approx \gamma \hat{p}_n \cdot \hat{p}_m$ in the small volume limit ($R \ll \lambda_0$), with $\hbar\gamma=\frac{4}{3} \mu^2 k_0^3/\epsilon_r$. Here, $k_0:=\omega_0
n_r/c$,  where $c$ is the speed of light and $n_r$ the refractive
index. For the realistic parameters chosen here, 
the decay width of a single molecule is 
\mbox{$\hbar\gamma=3.7 \times 10^{-4}$ cm$^{-1}$}.
We also set $n_r=1$, which is a good
approximation when the system is surrounded by air~\footnote{This value of $\gamma$ corresponds to a fluorescence
time $\tau_{\it fl} = 14$ ns and differs from the excitation lifetime
$\sim~1$~ns  found in literature~\cite{schulten}, because here
$\gamma$  represents only the radiative
decay processes and non-radiative decay is neglected.
In the
case of pure water, one should set $n_r=1.33$ and
$\epsilon_r=n_r^2=1.77$, thus obtaining $\gamma=4.9 \times 10^{-4}$
cm$^{-1}$ and $\tau_{\it fl} = 11$ ns. For a proteic environment, instead, it
is usually set $\epsilon_r=2.3$~\cite{vangrondelle} which, keeping the
refractive index of water, gives the same $\gamma$ obtained in air.
}. 
Again, for a discussion about the regime beyond the small
volume limit see~\ref{sec:Rlambda}.
Finally, $\mathcal{R}_T$ describes dissipation due to the coupling of each molecule to an Ohmic bath, where
\begin{eqnarray}
    \gamma^{(p)}(\omega)=\frac{2\pi}{\hbar}[J(\omega)(1+n_{BE}(\omega))+J(-\omega)n_{BE}(-\omega)]
\end{eqnarray}
are the thermal rates, depending on the spectral density $J(\omega)$ and on the Bose distribution $n_{BE}(\omega)$ of the phonons which form the bath and
\begin{eqnarray}
    A_n(\omega)=\sum_{E_\alpha-E_\beta=\hbar\omega} c_n^*(E_\alpha) c_n(E_\beta) \ket{E_\beta}\bra{E_\alpha}~.
\end{eqnarray}
More details about ${\cal R}_T$ can be found in~\ref{sec:mas}.
 
Note that, for the coupling to the thermal bath, we use the non-secular Redfield dissipator ${\cal R}_T$ instead of the commonly used Lindblad dissipator since we found that, in our model, the secular approximation is not valid and produces unphysical results, a well-known issue in molecular excitonic transfer~\cite{fleming2009}. Specifically, when the coupling $\Omega_C$ is very small, see Eq.~\eqref{OC}, the secular approximation incorrectly predicts that the transfer rate between the ring and the RC becomes independent of $\Omega_C$, while the Redfield dissipator ${\cal R}_T$ correctly predicts that the transfer rate tends to zero with $\Omega_C$.
Although the Redfield master equation is known to produce negative populations in the intermediate-to-strong system-bath coupling (see Ref.~\cite{creatore} and references therein), we checked that all the steady-state populations are positive within the parameter range that we analyzed.
On the other hand, the secular approximation is valid for modeling fluorescence decay and decay to the reaction center, and thus we can keep the ${\cal L}_{fl}$ and ${\cal L}_{rc}$ dissipators in their Lindblad form, see Eqs.~\eqref{Lfl} and \eqref{Lrc}.

Our model has been derived 
under the single-excitation approximation. This is a good
approximation of a realistic situation only if the excited state population is much smaller than unity, which is true for our choice of the parameters (see~\ref{sec:site} for more details).

\section{Illumination conditions}

\begin{figure*}
  \centering
  \begin{tabular}{l|l|l}
  \textbf{\large a}~~~{\it D-configuration} &   \textbf{\large b}~~~{\it LH-configuration} &   \textbf{\large c}~~~{\it Effective three-level model} \\
 \includegraphics[width=0.29\textwidth]{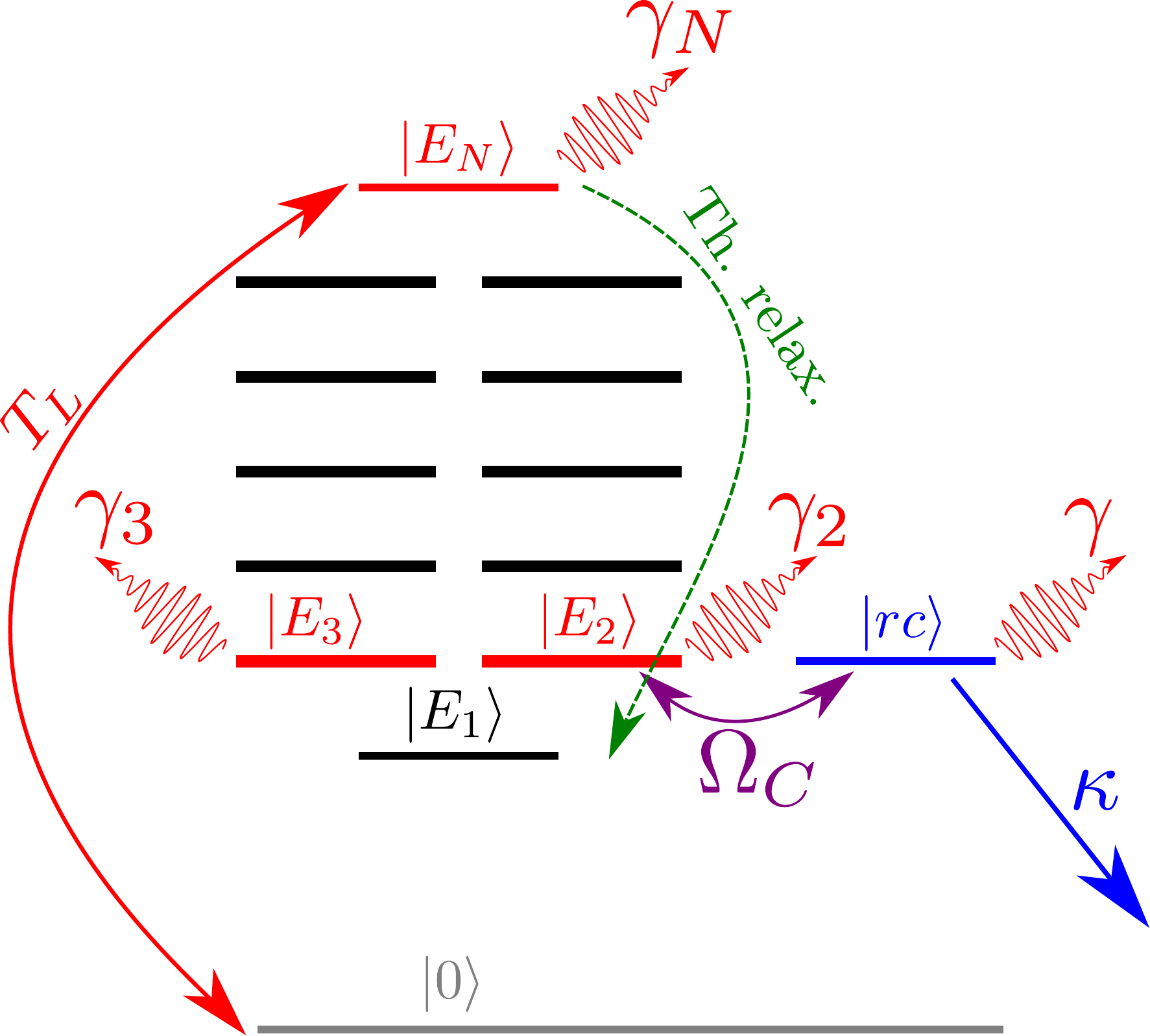} &
 \includegraphics[width=0.29\textwidth]{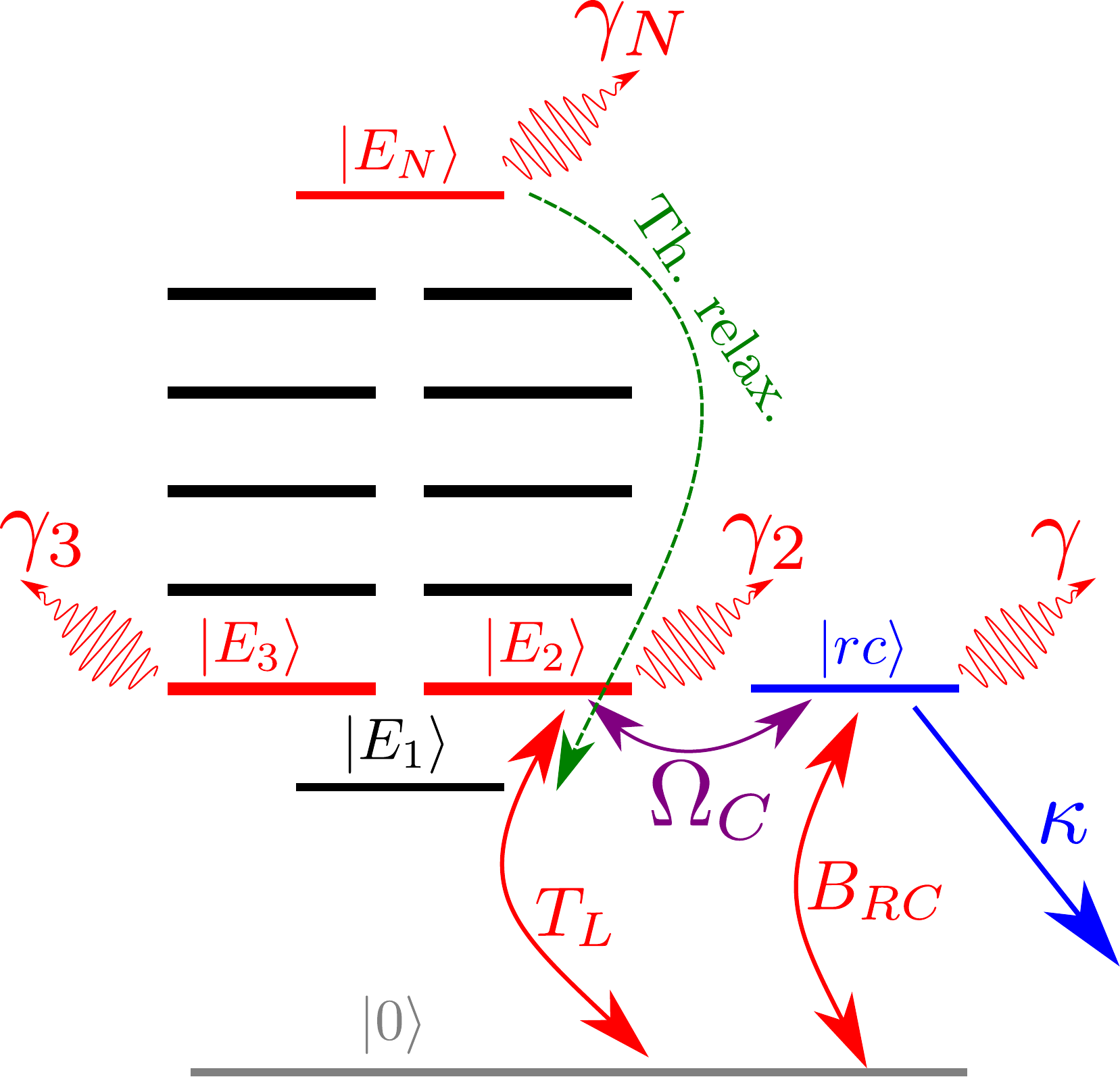} &
 \includegraphics[width=0.29\textwidth]{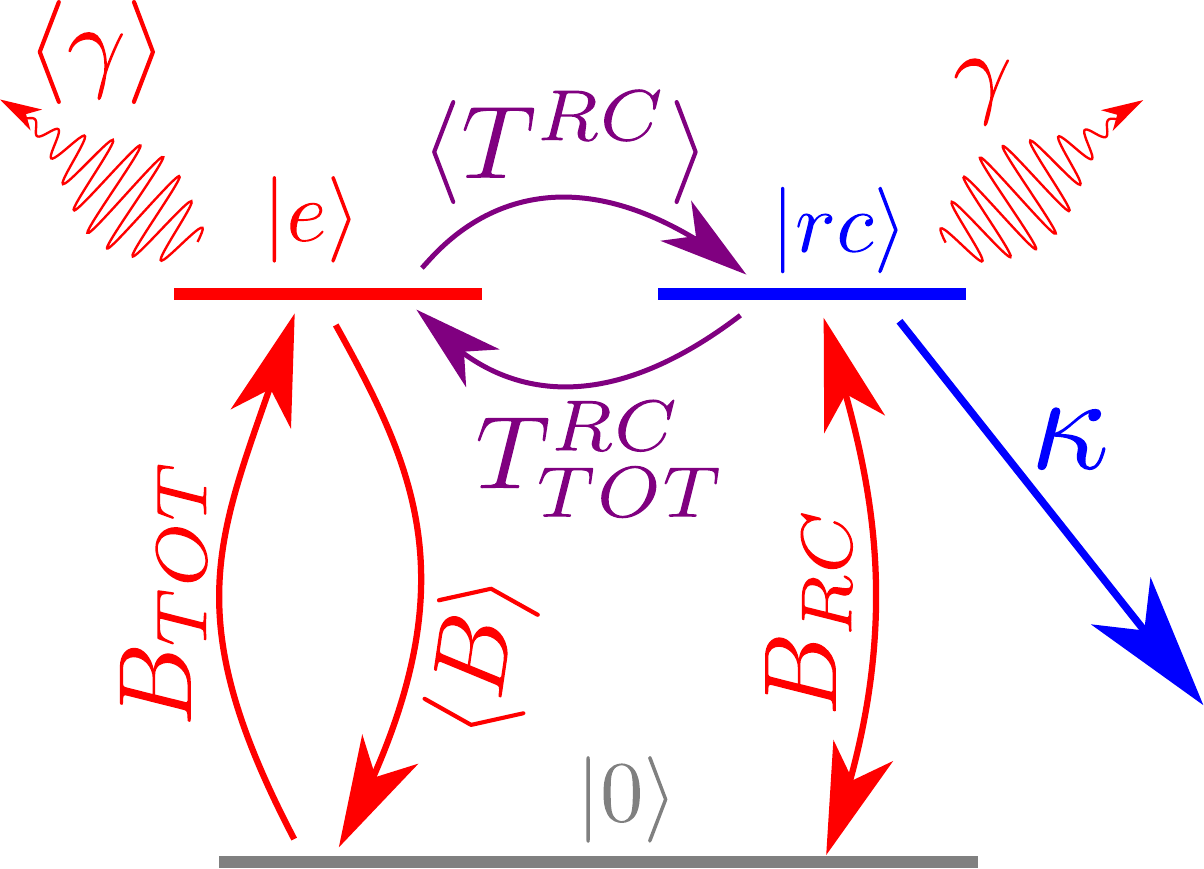}
  \end{tabular}
  \caption{{\it Main energy levels of  our device and three-level model}. 
(a,b): Schematic representation of a many-level system subjected to
    an external laser field with absorption rate $T_L$, 
    emission rates $\gamma_\alpha$, trapping rate $\kappa$, coupling $\Omega_C$ to the central site and
    thermal relaxation. Panel (a) shows illumination under the D-configuration, while panel (b) represents illumination under the LH-configuration. In both (a,b), the most important states are: the ground state, $\ket{0}$; the excitonic (dark) state having the lowest energy, $\ket{E_1}$; the lower ring eigenstate, $\ket{E_2}$, having dipole strength $\vec{p}_{E_2} = \sqrt{N/2}\cos\theta~\hat{y}$ and coupled to the RC; the RC state $\ket{rc}$, with dipole $\vec{p}_{rc}=\hat{y}$ and coupling $\Omega_C=(\mu^2/R^3) \sqrt{N/2}\cos\theta$ with $\ket{E_2}$;
the ring eigenstate $\ket{E_3}$, whose dipole strength is $\vec{p}_{E_3} = \sqrt{N/2}\cos\theta~\hat{x}$, and is decoupled from the RC; and 
the highest-energy ring eigenstate, $\ket{E_N}$, whose dipole strength
is $\vec{p}_{E_N} = \sqrt{N}\sin\theta~\hat{z}$. The radiative decay rates of the states depend on the dipole strength of the excitonic states and are given by
$\gamma_\alpha = \gamma |\vec{p}_\alpha|^2$. The absorption rates for the D-configuration (a) and LH-configuration (b) depend on the laser frequency, its intensity and polarization, see Eq.~\eqref{TL}. (c): {\it Effective three-level model}. 
Schematic representation of the effective three-level model which is able to capture
the main properties of the many-level system, see Eqs.~(\ref{eq:rate1a},\ref{eq:rate1b},\ref{eq:rate1c}). 
}
  \label{ml}
\end{figure*}

We consider three types of electromagnetic field states illuminating the device. Firstly, under what we call the \emph{D-configuration} (Fig.~\ref{fig:dip}b, more details in Fig.~\ref{ml}a), we consider a coherent, CW monochromatic polarized field as it was done in Ref.~\cite{scullyPRL}. The polarization axis is chosen to be aligned with $\hat{z}$, which means that it couples to the highest-energy excitonic state in the device, see Eqs.~\eqref{dips}. Second, under what we call the \emph{LH-configuration} (Fig.~\ref{fig:dip}c, more details in Fig.~\ref{ml}b), we consider a coherent, CW monochromatic field polarized in the $\hat{y}$ direction and incoming perpendicular to the ring. Such a field only excites  the low-energy ring eigenstate $\ket{E_2}$. As one can see from Fig.~\ref{ml}b, absorbing and transfer states coincide in this set-up, and such model is a good representative of some natural light-harvesting complexes (see~\ref{sec:lh}). Finally, under what we call the \emph{Sunlight configuration} (Fig.~\ref{fig:dip}d), we model illumination by natural sunlight, which is isotropic, unpolarized, incoherent and broad-band. This is modeled well as black-body radiation at 6000~K~\cite{whaley,caoJPCL2020}.
Specifically, in the sunlight configuration the Hamiltonian term $H_{EM}$ is not present, while we include two additional Lindblad dissipators for absorption and stimulated emission induced by sunlight,
\begin{eqnarray}
  \mathcal{L}_{sun}[\rho] = & \sum_{mn} f_Sn_S\Gamma_{mn} \left[ a_n^\dag
  \rho a_m  - \frac{1}{2} \left\{ a_m a_n^\dag , \rho \right\} \right] \nonumber\\
  & +\sum_{mn} f_Sn_S\Gamma_{mn} \left[ a_n
  \rho a_m^\dag  - \frac{1}{2} \left\{ a_m^\dag a_n , \rho \right\} \right]~,
\end{eqnarray}
where $n_S\approx 0.04$ is the Bose occupation of the Sun photons at the excitation energy $\hbar\omega_0$ and at the Sun temperature (6000~K) and $f_S=5.4\times 10^{-6}$ accounts for the Sun-to-Earth distance~\cite{biolaser}. Specifically, under sunlight illumination each eigenstate acquires absorption and stimulated emission rates, $B_\alpha=f_Sn_S\gamma |\vec{p}_{E_\alpha}|^2$, with $f_S$ representing the solid angle of the Sun as seen on Earth,
\begin{eqnarray}
    \label{fac1Sun}
    f_S = \frac{\pi r_S^2}{4\pi R_{ES}^2} = 5.4\times 10^{-6} \, ,
\end{eqnarray}
with $r_S$ being the radius of the Sun and $R_{ES}$ the Sun-to-Earth distance. Finally, the rates are proportional to the squared magnitude of the eigenstate dipole strength, so that only the states $\ket{E_{2,3,N}}$ have a non-vanishing sunlight absorption rate. 
The intensity of natural sunlight is $1365$~W/m$^2$, and we consider the same intensity also in the D-configuration and the LH-configuration. In those configurations the light intensity is encoded in the Rabi frequency, $\Omega_R=\mu E_0 / \hbar$.
By imposing the intensity of the CW laser to be $E_0^2/(4\pi)=1365$~W/m$^2$ (using Gaussian units), we determine the corresponding value of $E_0$ and, from that, the Rabi frequency, which is $\Omega_R=4.68\gamma$ (in units of the single-molecule radiative decay rate, $\gamma\approx 0.07$~ns$^{-1}$). We keep this value of $\Omega_R$ fixed in all the manuscript.

\section{Effective three-level model}

Here we show that the dynamics of the complex structure described above, under all the illumination conditions considered, can
be mapped to the dynamics of an effective three-level incoherent model with the relevant quantum effects encoded in few parameters. The three-level model is described by the zero-excitation state $|0\rangle$, a single-excitation state $|e\rangle$ for the whole ring, and a single-excitation state $|rc\rangle$ for the RC, see Fig.~\ref{ml}c.
The excitation pumped by the EMF into the ring is
quickly funneled to the low energy states by thermal relaxation. Therefore, we determine the rates between $\ket{0}$, $\ket{e}$ and $\ket{rc}$ assuming that the ring is always at thermal equilibrium with the phononic reservoir. Under this assumption (see~\ref{sec:3lev}), the emission rate from $\ket{e}$ to $\ket{0}$ is the thermal average of the ring eigenstate emission rates, $\braket{\gamma}=\sum_\alpha \frac{e^{-E_\alpha/(k_BT)}}{Z}\gamma_\alpha$, while the transfer rate from the ring to the RC is also a thermal average, $\braket{T^{RC}}=\sum_\alpha \frac{e^{-E_\alpha/(k_BT)}}{Z}T^{RC}_\alpha$, involving the transfer rates $T^{RC}_\alpha$ between each $\alpha$ eigenstate and the RC, that are proportional to the squared coupling between $\ket{\alpha}$ and $\ket{rc}$. 
Note that the $\braket{T^{RC}}$ rate is equivalent to the well-known multi-chromophoric F\"orster resonance energy transfer (MC-FRET)~\cite{silbeyPRL2004,caoNJP2013} or generalized F\"orster theory~\cite{scholesARPC2003} rate, as we show in detail in~\ref{sec:3lev}.
In our specific case, due to the ring symmetry, only the $\ket{E_2}$ eigenstate has a nonvanishing transfer rate, $T_2^{RC}=\tau_{RC}^{-1} (32/N)^5 \cos^2\theta$, where $\tau_{RC}=3.9$~ps is the transfer time between the ring and the RC at $\theta=0$ and $N=32$ (more details in~\ref{sec:3lev}), while $T_\alpha^{RC}=0$ for all $\alpha\neq 2$. On the other hand, the absorption rate is the sum of all the absorption rates, $B_{TOT}=\sum_\alpha B_\alpha$ (where the absorption rates, $B_\alpha \propto |\vec{p}_\alpha|^2$, have different expressions whether the excitation is induced by a CW laser or by sunlight, see~\ref{sec:3lev}), and the transfer rate from the RC to the ring is also the sum of all the transfer rates, $T^{RC}_{TOT}=\sum_\alpha T^{RC}_\alpha$. Finally, the stimulated emission rate is again a thermal average, $\braket{B}=\sum_\alpha \frac{e^{-E_\alpha/(k_BT)}}{Z}B_\alpha$.
The RC also can absorb the incoming radiation with an absorption rate $B_{RC}$, that accounts also for stimulated emission, while its emission rate is $\gamma$. 
This approach yields the following rate equations for the populations of the three levels,
\begin{numparts}
\begin{eqnarray}
  \label{eq:rate1a}
  \fl \qquad 
  \frac{dP_0(t)}{dt} = &- \left( B_{TOT} + B_{RC} \right) P_0(t) + \left\langle B \right\rangle P_e(t)
                 + \left\langle \gamma \right\rangle P_e(t) + \left( \kappa + \gamma \right) P_{rc}(t)~, \\
  \label{eq:rate1b}
  \fl \qquad 
  \frac{dP_{e}(t)}{dt} = &B_{TOT} P_0(t) - \left\langle B \right\rangle P_e(t)- \left\langle \gamma \right\rangle P_e(t)
                 +T_{TOT}^{RC} P_{rc}(t) - \left\langle T^{RC} \right\rangle P_e(t)~,  \\
  \label{eq:rate1c}
  \fl \qquad 
  \frac{dP_{rc}(t)}{dt} = &- T_{TOT}^{RC} P_{rc}(t) + \left\langle T^{RC} \right\rangle P_e(t)
                 - \left( \kappa + \gamma \right) P_{rc}(t) + B_{RC}P_0(t)~.
\end{eqnarray}
\end{numparts}
Solving for the steady state of these equations (details in~\ref{sec:3lev}), we obtain an approximation to the steady-state transmitted current,
\begin{eqnarray}
  \label{jj}
  \fl \qquad 
  I_3 = \frac{ \kappa \left( B_{TOT} + B_{RC} \right) }{B_{TOT} + \gamma + \kappa + 2B_{RC} + \left( B_{TOT} + B_{RC} + \left\langle B \right\rangle + \left\langle \gamma \right\rangle \right) \frac{ T_{TOT}^{RC} + B_{TOT}\frac{B_{RC} + \gamma + \kappa }{B_{TOT} + B_{RC}} }{\left\langle T^{RC} \right\rangle  + B_{RC} \frac{ \left\langle B \right\rangle + \left\langle \gamma \right\rangle}{B_{TOT} + B_{RC}}} }
\end{eqnarray}
The validity and effectiveness of this three-level model is discussed in the next
section, see also Ref.~\cite{biolaser}. 
Note that the effective three-level model presented in Eqs.~(\ref{eq:rate1a},\ref{eq:rate1b},\ref{eq:rate1c}) is able to describe the whole system, composed by the ring and the central core absorber, under both the pumping from a light source (laser or sunlight) and thermal relaxation. Our effective three-level model is based on the assumption of fast thermal relaxation, incoherent pumping and incoherent transfer between the ring and the core absorber. Specifically the coupling between the ring, assumed at thermal equilibrium, and the central core absorber is described with an approach equivalent to the the generalized F\"orster theory, see~\ref{sec:3lev}. Finally note that in literature three-level models describing exciton transport have been widely used~\cite{trimer,yang1,yang2,caoJPCA2009,caoJPCL2020}. In particular our approach is similar to the one used in Ref.~\cite{caoJPCL2020} where the pumping of sunlight on a dimer system has been considered.

\section{Results}

\subsection{Super-absorption in the low-fluence regime}

First we demonstrate that the molecular device developed above is capable of exploiting cooperative effects to enhance the absorption from a weak-intensity EMF.
For any $\theta > 0$ and $N$, only three ring eigenstates have a non zero dipole strength: two in the low-energy region ($\ket{E_2}$ and $\ket{E_3}$) and one with the highest energy ($\ket{E_N}$). Concerning the low-energy states, $\ket{E_2}$ has a polarization along $y$ while $\ket{E_3}$ along $x$. In contrast, the high-energy state has a polarization along the $z$-axis. Thus for an EMF polarized in the $z$ direction (D-configuration) the absorbing and the transferring states are separated: only the highest-energy state $\ket{E_N}$ is coupled to the EMF, while only the low-energy state $\ket{E_2}$ can transfer the excitation to the RC.

At high (\emph{e.g.,}~room) temperature the pumping rate for this system under CW laser excitation can be described 
semi-classically by the F\"orster rates~\cite{yang1}
\begin{eqnarray}
  T_L = \frac{(\Omega_R \vec{p}_{\rm abs}\cdot \hat{\epsilon})^2
    \Gamma_T}{2\left[ \Gamma_T^2 + (\omega-\omega_{\rm abs})^2 \right]}
\label{TL}
\end{eqnarray}
where $(\omega-\omega_{\rm abs})$ is the detuning frequency of the laser with respect to the
absorption frequency and $\vec{p}_{\rm abs}$ is
the dipole strength of the absorbing state.
 $\Gamma_T$ is the dephasing rate of the
coherences between the 
absorbing state and the ground state. We compute $\Gamma_T$ 
analytically in~\ref{sec:GT}, and show that it depends only on the density of states of the system and on the parameters of the bath. Critically, $\Gamma_T$ is independent of $N$ and very weakly dependent on $\theta$. Thus $T_L \propto |\vec{p}_{\rm abs}|^2 \propto N$. This demonstrates what we call \emph{superabsorption}: the  absorption is concentrated in a very specific system eigenstate characterized by a giant dipole, and the absorption rate grows proportionally to the system size. Note that our definition of superabsorption refers to the low-fluence regime, which is the focus of this manuscript. Under high fluence, cooperative absorption is instead characterized by a super-linear absorption rate, as it has been shown in Ref.~\cite{superabsorb}.

\subsection{Scalability and efficiency}

\begin{figure*}
  \centering
  \includegraphics[width=\linewidth]{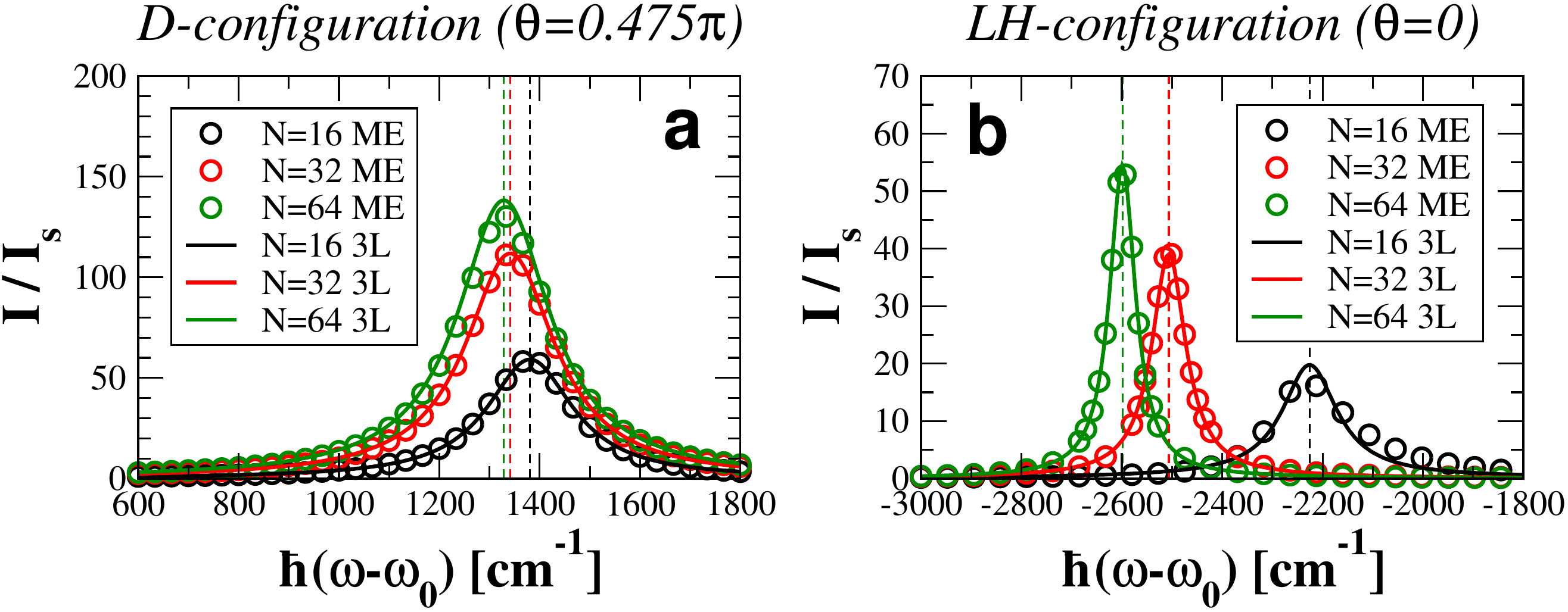}
  
  \vspace{1em}
  
  \includegraphics[width=\linewidth]{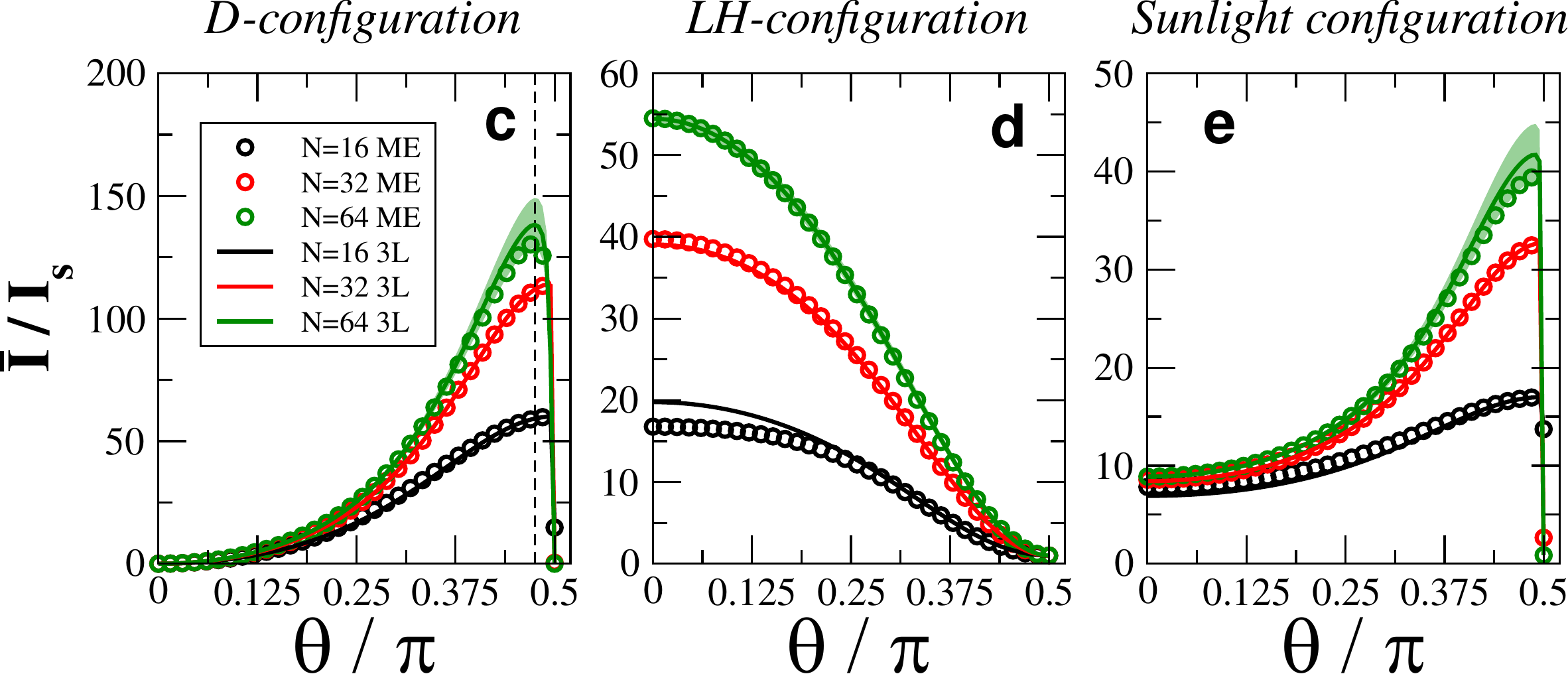}
  \caption{{\it Validity of the three-level model}.  
(a,b) Transmitted normalized current $I/I_s$ \emph{\it vs.} laser frequency under the D-configuration ($\theta=0.475\pi$) on panel (a), and LH-configuration ($\theta=0$) on panel (b), at room temperature ($T=300$~K). Different values of $N$ have
    been considered, varying the radius of the ring to keep the
    density fixed. $I_s$ is the maximal stationary current of a
    single site in the same conditions. The energies of the absorbing eigenstates are
    shown as vertical dashed lines in panels (a,b): the high-energy state $E_N$ (D-configuration) and the first-excited state $E_2$ (LH-configuration).
    Symbols represent the current $I$, see Eq.~\eqref{eq:cur}, obtained from the master equation (ME), see Eq.~\eqref{eq:mas}, while the solid curve is the three-level approximation $I_3$ (3L), see Eq.~\eqref{jj}.
    (c,d,e) Transmitted normalized current at the peak laser frequency $\bar{I}/I_s$ (corresponding to the absorbing states, see dashed lines in panels a,b) \emph{\it vs.} $\theta$ for D-configuration (c), LH-configuration (d), and Sunlight configuration (e).
    The dashed line in panel (c) represents $\theta=0.475\pi$, which is optimal for the D-configuration at $N=64$ and close to optimal for $N=16,32$, see figure. In panels (c,d) $\bar{I}_3$ (lines) is the current at the peak laser frequency obtained from the three-level solution, see Eq.~\eqref{jj}. In panel (e), there is no laser frequency in the sunlight configuration, so $\bar{I}$ coincides with $I$ and $\bar{I}_3$ coincides with $I_3$.
    Shaded areas in panels (c,d,e) represent variations in $\bar{I}_3$ produced by $\pm 20\%$ variations in $\tau_{RC}$.
    Parameters for all panels: $\kappa=10\gamma$, $\Omega_R=4.68\gamma$ 
  (laser intensity: $1365$~W/m$^2$, same as natural sunlight), 
$\tau_{RC}=3.9$~ps (for three-level model). 
    }
  \label{fig:tilt}
\end{figure*}

Now we demonstrate that superabsorption can work in concert with the engineered supertransfer from ring to RC, to result in a photocurrent that scales with the system size.

Here we analyze the efficiency of our device under a laser field
polarized along the $z$ direction and under the action of a thermal
bath at room temperature. In particular we will analyze the dependence of the
current on the laser frequency and of the peak current $\bar{I}$
(maximal current obtained at the optimal laser frequency)
as a function of the system size. Next, we
compare the efficiency of the D-configuration with an alternative illumination condition, the LH-configuration, where the
absorbing and the transferring states coincide, thus mimicking natural light-harvesting complexes more closely. Finally, we evaluate our device in the Sunlight configuration, which is a realistic model of illumination by natural sunlight. At the same time, we show that the results of the full $N$-level system can be captured by the simpler three-level system introduced in the previous section.

Fig.~\ref{fig:tilt} shows the dependence of the photocurrent on the CW laser frequency both for the D-configuration (Fig.~\ref{fig:tilt}a, where the field is assumed polarized in the $z$ direction) and for the LH-configuration (Fig.~\ref{fig:tilt}b, with the field along $y$), for different system sizes at room temperature. For the D-configuration we choose an angle $\theta=0.475\pi$ that gives the optimal current for $N=64$ and a close-to-optimal current for $N=16,32$ (see dashed line in Fig.~\ref{fig:tilt}c).
On the other hand, for the LH-configuration we choose the optimal angle $\theta=0$, where the current is maximal (see Fig.~\ref{fig:tilt}d).

The combined effect of
superabsorption (at high-energy), thermal relaxation and transfer (at low
energy) results in a peak in the transmission
spectrum at the high
energy of the absorbing state (that is higher than $\hbar\omega_0$, see vertical dashed lines in Fig.~\ref{fig:tilt}a) and not to that of the transferring states (that would be lower than $\hbar\omega_0$). Since the high-energy state is totally decoupled from the RC, the peak at its frequency can only be explained by thermal relaxation after absorption. Note also that the height of the peak increases with the system size due to cooperative absorption. 
In Fig.~\ref{fig:tilt}a we also show as continuous curves the results of our analytical three-level model Eq.~\eqref{jj}, which reproduces the current across the entire frequency range.

\begin{figure}[!ht]
  \centering
  \includegraphics[width=0.6\columnwidth]{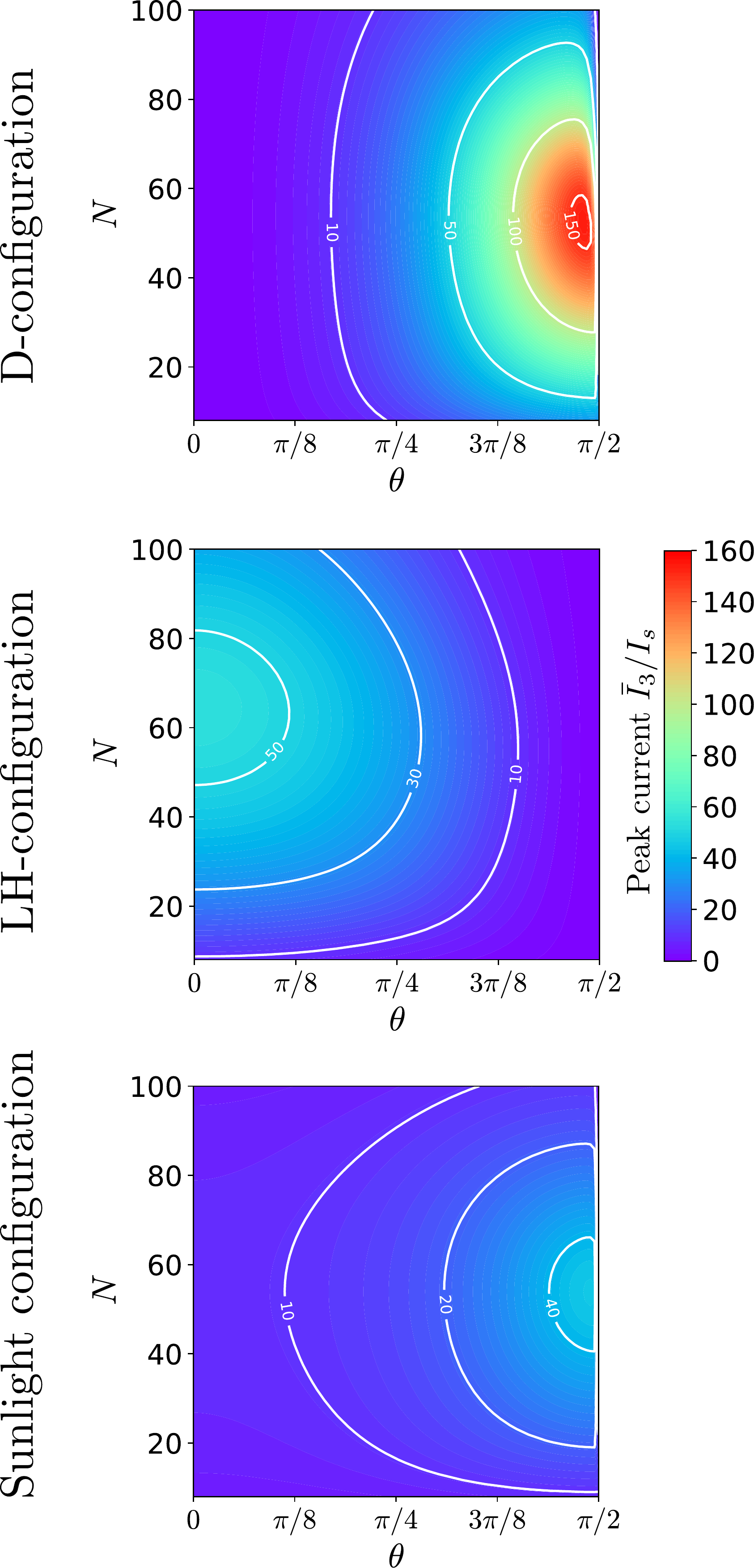}
  \caption{\textit{Peak current vs $\theta$ and $N$}. Normalized current at the peak laser frequency $\bar{I}_3/I_s$ obtained from the three-level model, see Eq.~\eqref{jj}. Different illumination conditions  
  are used (see figure). For the sunlight configuration (lowest panel), $\bar{I}_3$ coincides with $I_3$, because there is no laser frequency. Parameters: $\kappa=10\gamma$, $\Omega_R=4.68\gamma$ 
  (laser intensity: $1365$~W/m$^2$, same as natural sunlight), 
$\tau_{RC}=3.9$~ps, $T=300$~K.
}
  \label{fig:coop}
\end{figure}

In Fig.~\ref{fig:tilt}b we show the ratio between the
current $I$ and the single-site current $I_s$ [see Eq.~(\ref{eq:jsite})] for
the LH-configuration as a function of
the laser frequency at room temperature, for different system sizes.
The intensity has a peak (vertical dashed lines) when  the
laser frequency is resonant with the low-energy bright eigenstates 
of the ring, that are resonant with the RC.  
It is interesting to note that even for the LH-configuration the height of the peak increases with the system
size due to cooperative absorption.
However, the peak current obtained with the LH-configuration is about three times smaller than with the D-configuration (compare Fig.~\ref{fig:tilt}a and Fig.~\ref{fig:tilt}b).
Also for the LH-configuration, the three-level model [Eq.~\eqref{jj}] reproduces very well the results of the full system (see continuous lines in Fig.~\ref{fig:tilt}b). 

Therefore, in order to understand whether the separation of the absorbing and
transmitting states that we engineered can improve the scalability of
the system, we compute the peak transmitted current as a function of $\theta$ and of the
system size.  

In Fig.~\ref{fig:tilt}(c,d,e) we plot the peak current $\bar{I}$ \emph{vs.} $\theta$ for different values of the system size. In each panel, the results of the master equation are compared to the effective three-level model, Eq.~\eqref{jj}. Various illumination conditions are considered. For the D-configuration (Fig.~\ref{fig:tilt}c), the peak current increases with $\theta$ up to a maximum, close to $\pi/2$. This is a consequence both of the absorption rate $T_L$ increasing with $\theta$ for the D-configuration, and of the emission rate being suppressed on increasing $\theta$. On the other hand, for the LH-configuration (Fig.~\ref{fig:tilt}d) the peak current is maximal for $\theta=0$ and it decreases with $\theta$. This is a consequence of the absorption rate decreasing with $\theta$ in the LH-configuration and of the fact that absorption and emission are not decoupled. Then, for the Sunlight configuration (Fig.~\ref{fig:tilt}e)
the current is enhanced by increasing $\theta$, as a result of the suppression of emission.
In all the three cases shown, the current is enhanced on increasing $N$ from 16 to 64. Finally, as one can see from the figure, the three-level model (lines) gives a good approximation of the master equation results (symbols) in all the ranges considered.
Small deviations for large $N$ and for $\theta$ close to $\pi/2$ can be explained by a variation of the value of $\tau_{RC}$ used in the three-level model, which in the figure has been kept fixed as $N$ and $\theta$ vary. Indeed we set 
$\tau_{RC}=3.9$~ps which has been obtained by fitting the master equation for $N=32$ and $\theta=0$. Nevertheless $\tau_{RC}$ can vary by up to $20\%$ as it is shown in~\ref{sec:3lev}. If we account for those variations in $\tau_{RC}$ in our calculations of the current, we obtain a perfect agreement also for large $N$, see shaded areas in panels (c,d,e). Deviations for small $N$, see panel (d), are due to the fact that the couplings between the ring and the RC are large and the energy transfer is not fully incoherent as it is discussed in~\ref{sec:3lev}.

Since we are interested in the scalability of the device at large $N$, in the following we use the three-level system, that is much less computationally expensive than the master equation at large $N$.
In Fig.~\ref{fig:coop} the normalized maximal current $\bar{I}_3/I_s$ is shown vs.
$\theta$ and $N$ at room temperature, as obtained from the effective three level model, see Eq.~\eqref{jj} for the trapping rate $\kappa=10\gamma$ and for different pumping mechanisms (i.e. D-configuration, LH-configuration and Sunlight configuration).
In all cases, we can see that the current increases at first with $N$. Moreover, for the D-configuration the efficiency improves with increasing
$\theta$, it reaches an optimal value for $40<N<80$ and it ultimately decreases with $N$ for very large ring sizes. Such improvement with $\theta$ can also be seen for natural sunlight pumping and has been observed and commented above in Fig.~\ref{fig:tilt}(c,e).

The scaling of the current with the system size can be understood as follows. For small size, $N$, the excitation is cooperatively absorbed by the ring  and efficiently transferred to the RC where it is trapped. Indeed, a small ring radius implies a strong dipole coupling to the RC and, therefore, a fast transfer. So, the trapped current for small $N$ ultimately scales as the absorption rate, increasing with $N$. On the other hand, for large sizes $N$, a large ring radius implies a weak coupling to the RC, that decreases as $\sim |\Omega_C|^2\sim 1/N^5$, see Eq.~\eqref{OC}. Such suppression of the transfer to the RC acts as a bottleneck for large $N$, so that the trapped current decreases with $N$ for large $N$ in all cases.
Moreover for large $N$ the thermal population in the superradiant state coupled to the central absorber decreases as  $1/N$, thus quenching the current for large system sizes. 

Moreover, in Fig.~\ref{fig:coop} the normalized current $\bar{I}_3/I_s$ of the LH-configuration (y-polarized) is shown as a
function of $N$ and $\theta$ and it is compared with that of the D-configuration  for the trapping rate $\kappa=10\gamma$, that is of the same order of the emission rates ($\braket{\gamma}\approx \gamma$).

Finally, we also analyze the model in the Sunlight configuration. In this case, the pumping is incoherent, broad-band and isotropic. As a consequence, the total absorption rate of natural sunlight, $B_{TOT}=N\gamma f_S n_S$,
is proportional to $N$ and independent of $\theta$. Nevertheless, even in this case for $\kappa=10\gamma$ we see an increment of the current on increasing $\theta$, because the system benefits from the suppression of emission.

\begin{figure}
    \centering
    {\large {\bf a} {\it D-configuration}}
    
    \includegraphics[width=0.5\columnwidth]{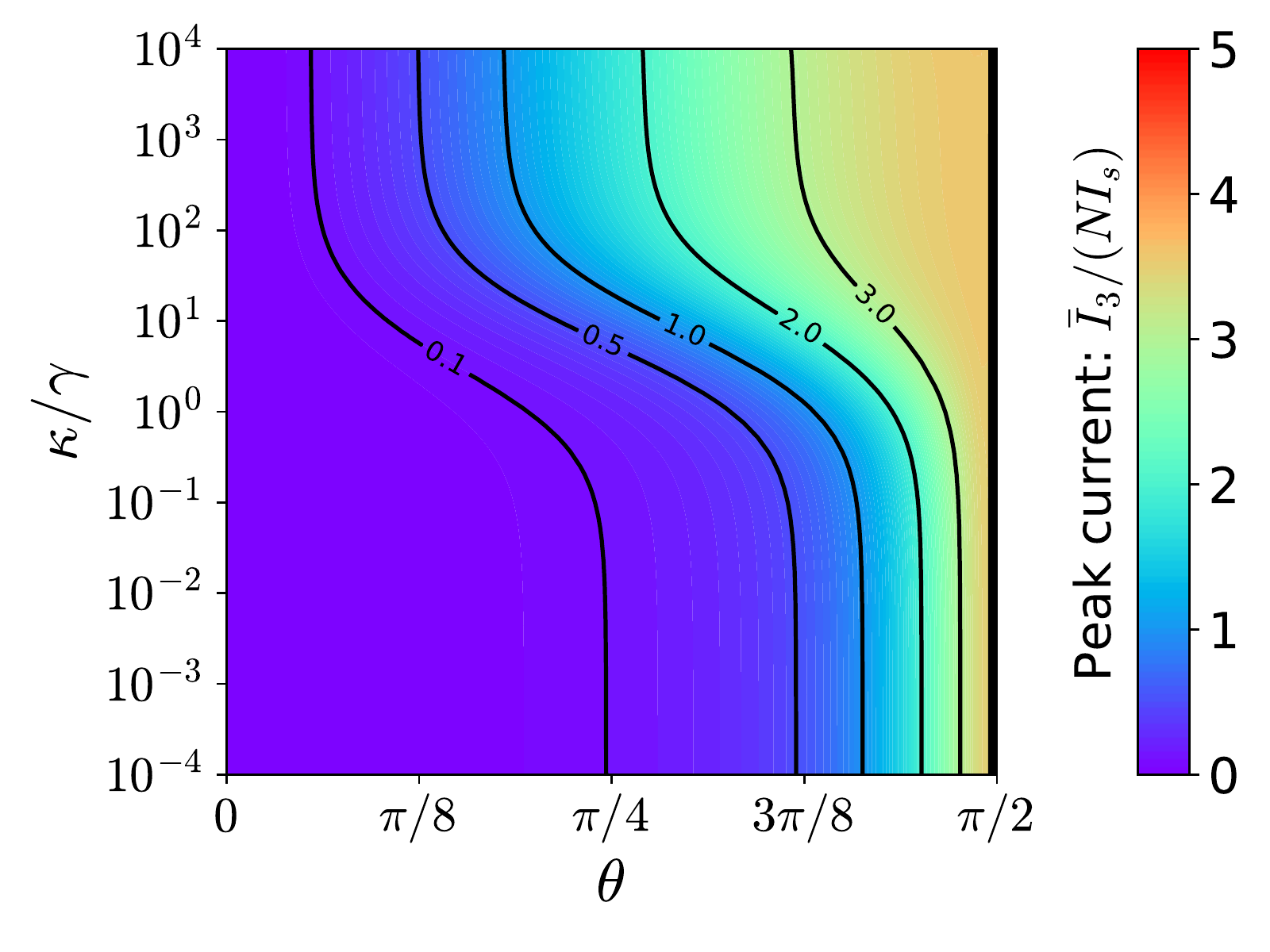}
    
    {\large {\bf b} {\it LH-configuration}}
    
    \includegraphics[width=0.5\columnwidth]{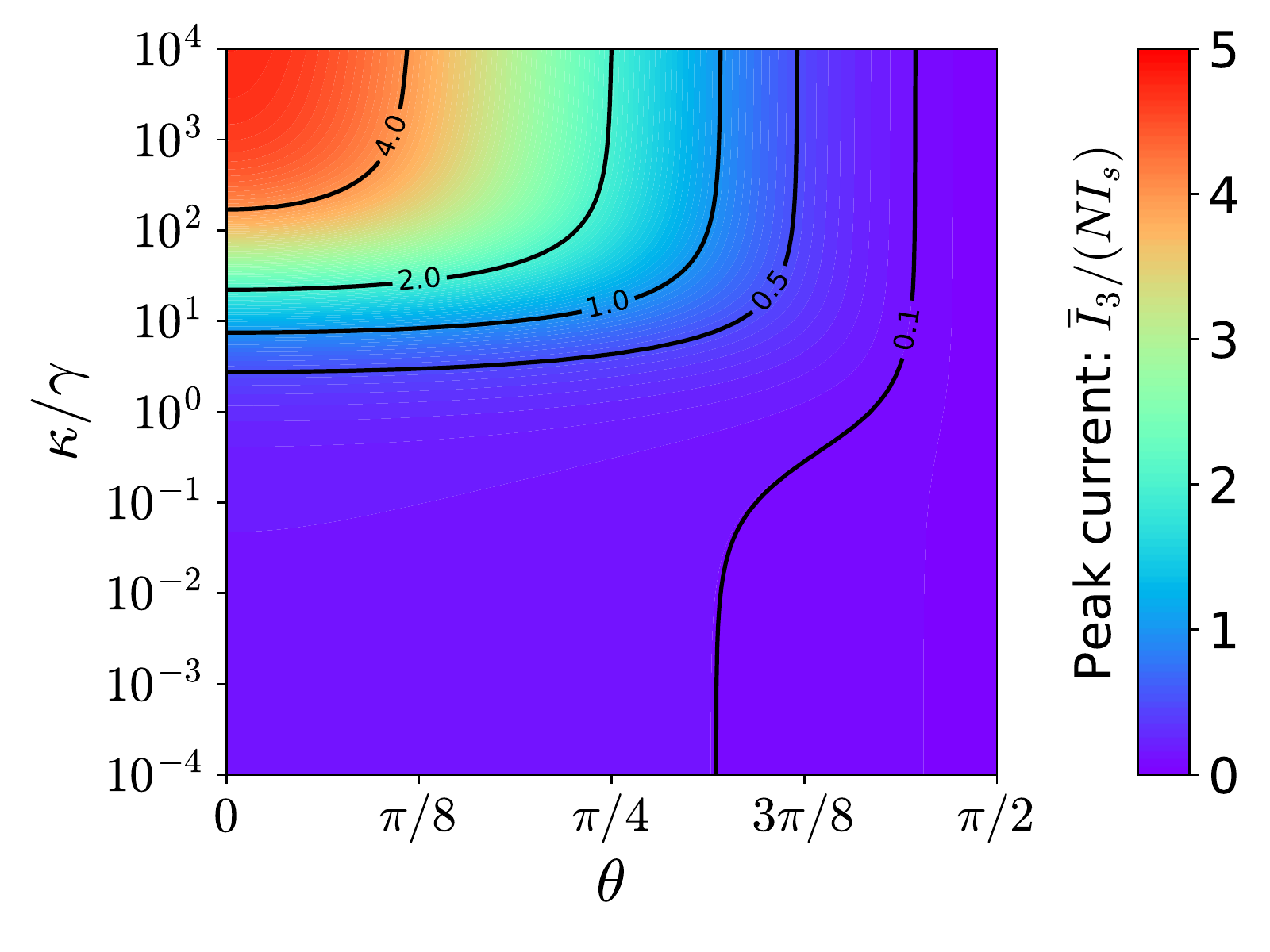}
    
    {\large {\bf c} {\it Sunlight configuration}}
    
    \includegraphics[width=0.5\columnwidth]{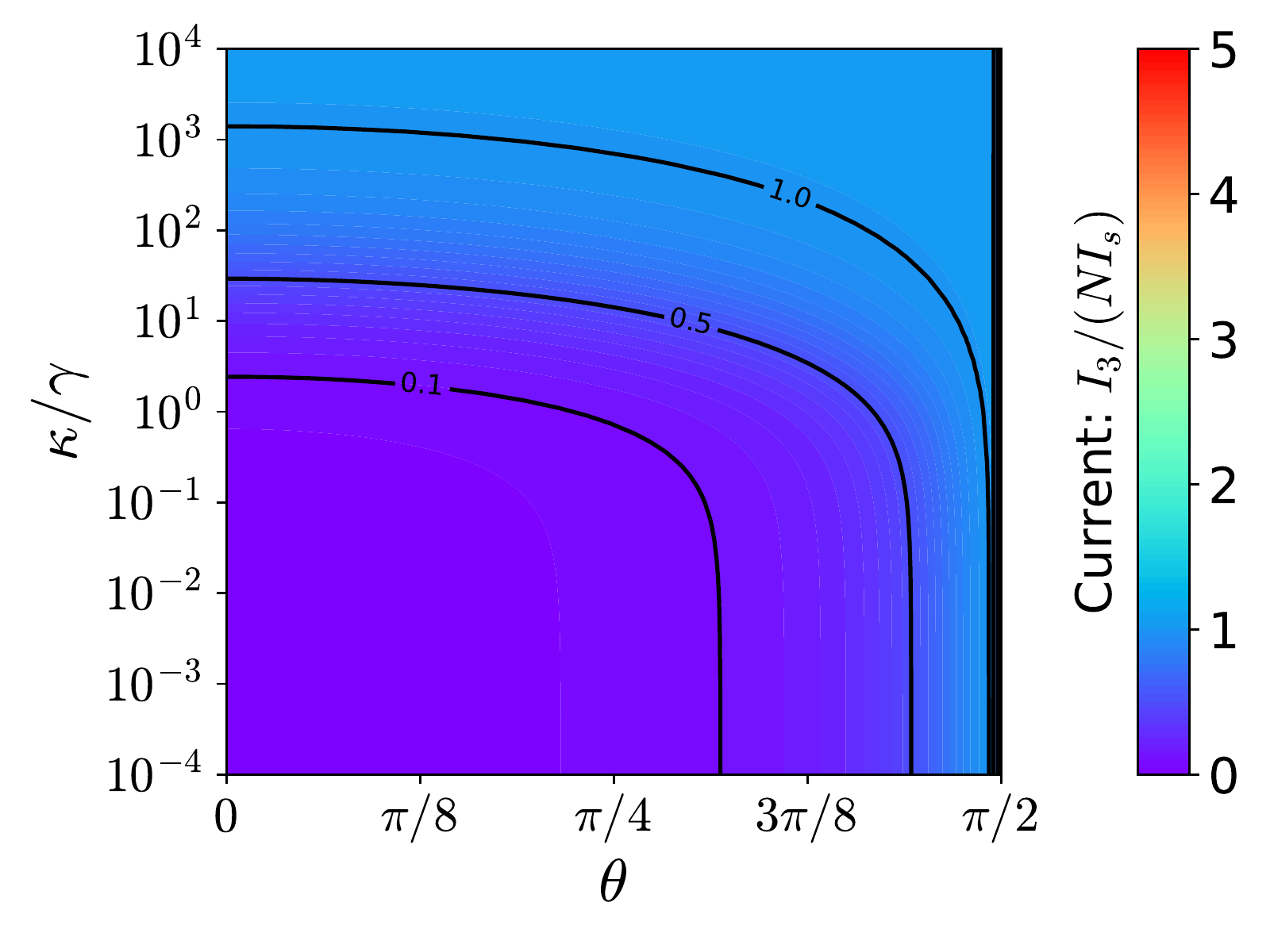}
    \caption{{\it Dependence on the trapping rate $\kappa$ and $\theta$.} Steady-state current at the peak laser frequency divided by $N$ times the single-site current, $\bar{I}_3/(NI_s)$, {\it vs.} the trapping rate $\kappa$ (divided by the monomer emission rate $\gamma$) and the angle $\theta$, computed using the three-level model, see Eq.~\eqref{jj}, for $N=32$. Each panel corresponds to a different illumination condition, as written on the top of each panel. In panel (c), $\bar{I}_3$ coincides with $I_3$, because there is no laser frequency in the sunlight configuration.  Parameters: $\Omega_R=4.68\gamma$, $\tau_{RC}=3.9$~ps, $T=300$~K.}
    \label{fig:kappa-theta}
\end{figure}

About the choice of the trapping rate we note that this is critical to the efficiency of our set-up. Indeed a very large trapping rate prevents re-emission, since the excitation is quickly trapped, and thus makes less useful the suppression of re-emission which we consider here. The trapping rate can vary a lot depending on the specific system. In photosynthetic antenna complexes a charge separated state is created very quickly once the excitation is absorbed in the reaction center (few picoseconds, corresponding to $\kappa \approx 10^4 \gamma$). On the other hand, charge transfer in the RC is much slower and the RC is not active until the charge-separated state is neutralized again, and this occurs on the order of $100~\mu$s, corresponding to $\kappa \approx 10^{-4} \gamma$. In the  figures presented here an intermediate trapping rate has been considered, $\kappa = 10 \gamma$, corresponding to a trapping time of $\approx 1$~ns. Even if this trapping rate is only slightly faster than the emission rate ($\approx 10$~ns), a considerable advantage is obtained in the D-configuration.
As one can see from Fig.~\ref{fig:coop}, for the trapping rate $\kappa=10\gamma$ the maximal current obtained from the D-configuration is about three times higher than the LH-configuration. 
Note that our proposed design is useful only if the emission plays an important role. If, instead, the trapping rate $\kappa$ is so fast to overcome any emission process,
there is no advantage in the decoupling mechanism.
 Specifically, the efficiency of the D-configuration, as opposed to the LH-configuration and to a single site, improves on decreasing the trapping rate $\kappa$, as we show in the following.

In Fig.~\ref{fig:kappa-theta} we study how the efficiency of our proposed device depends on the trapping rate $\kappa$. We analyze a broad range of reasonable values of the trapping rate: the lower bound, $\kappa=10^{-4}\gamma\approx (100~\mu$s$)^{-1}$, corresponds to the reset time in purple bacteria reaction centers~\cite{schulten}, while the upper bound, $\kappa=10^4\gamma\approx (1$~ps$)^{-1}$, is the charge separation rate in purple bacteria~\cite{schulten}.

Specifically, in Fig.~\ref{fig:kappa-theta} we plot the ratio $\bar{I}_3/(NI_s)$ between the maximal steady-state current obtained from the three-level model $\bar{I}_3$, Eq.~\eqref{jj}, divided by $N$ times the single-site current, $I_s$, for $N=32$ as a function of the trapping rate $\kappa/\gamma$ and of the angle $\theta$. Different panels represent different illumination conditions.
For the D-configuration, see Fig.~\ref{fig:kappa-theta}a, for any $\kappa$, the current is enhanced by increasing $\theta$, because the absorption is increased and at the same time the emission is suppressed. It is interesting to note that,
for any fixed $\theta$, the current increases with $\kappa$. Indeed, for these parameters, the emission rate in our device is of order $\approx N\gamma\cos^2\theta\approx \gamma$ (for $N=32$, as in Fig.~\ref{fig:kappa-theta}), and therefore a fast trapping rate $\kappa\gg\gamma$ allows to overcome the emission rate.

Similar comments can be applied to the sunlight configuration, see Fig.~\ref{fig:kappa-theta}c. Also in such case, for fixed $\theta$, the current increases with $\kappa$, because the fast trapping rate overcomes the emission. In the sunlight configuration, for $\kappa=10^4\gamma$ the current becomes basically independent of $\theta$ and equal to $N$ times the single-site current, see also Fig.~\ref{fig:theta}c and~\ref{sec:klarge} for more details.

For the LH-configuration, see Fig.~\ref{fig:kappa-theta}b, a similar pattern can be seen: the current increases with $\kappa$ for $\theta$ fixed.
This configuration does not have a decoupling mechanism between absorption and transfer, and therefore the LH-configuration shows the best performance under the trivial conditions: (a) optimal absorption rate at $\theta=0$ and (b) maximal trapping rate (in the figure, $\kappa=10^4\gamma$).

We also point out that, both for the D- and the LH-configuration, there is a broad range in the $(\kappa,\theta)$ parameter space where the current is more than $N$ times larger than the single-site current $I_s$.
Interestingly, the normalized current $\bar{I}_3/(NI_s)$ in Fig.~\ref{fig:kappa-theta}a-b can be larger than unity for the LH-configuration only for large trapping rates, while for the D-configuration the normalized current can be larger than unity even for small trapping rates.

\subsection{Robustness to disorder}

The efficiency of our proposed device will be affected by disorder, that can be due for instance to fluctuations in the positions of the sites, in the orientations of their transition dipoles, or in the site energies. Here we study how the efficiency of our device, measured by the peak steady-state current $\bar{I}/I_s$, is affected by disorder in the dipolar orientation. We introduce the angular disorder as follows: each dipole is given a random orientation inside a cone  centered on the precise orientation of Eq.~\eqref{eq:dip}. All the cones have the same solid angle, that can take values from 0 to $4\pi$. The magnitude of the solid angle represents the disorder strength: a vanishing solid angle represents no disorder, while the maximal solid angle $4\pi$ represents completely disordered dipoles. The energy of the RC is always equal to the energy of the first excited state of the ring without disorder and, similarly, the frequency of the CW laser is determined by the energies of the ring at zero disorder: the CW laser is resonant to the highest-energy ring eigenstate for the D-configuration, and it is resonant to the first-excited state of the ring for the LH-configuration.

\begin{figure}
    \centering
    \includegraphics[width=0.5\columnwidth]{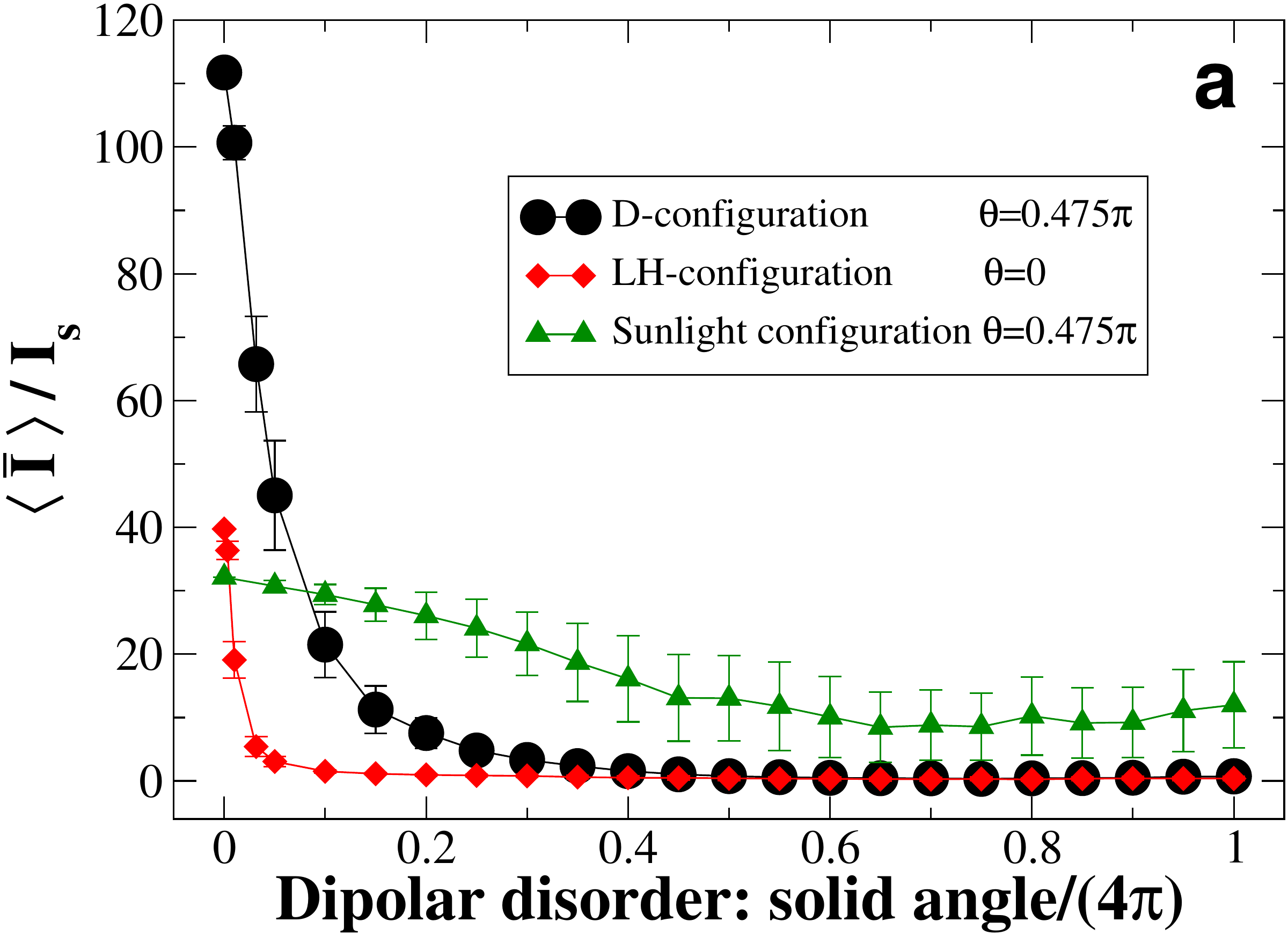}
    \includegraphics[width=0.5\columnwidth]{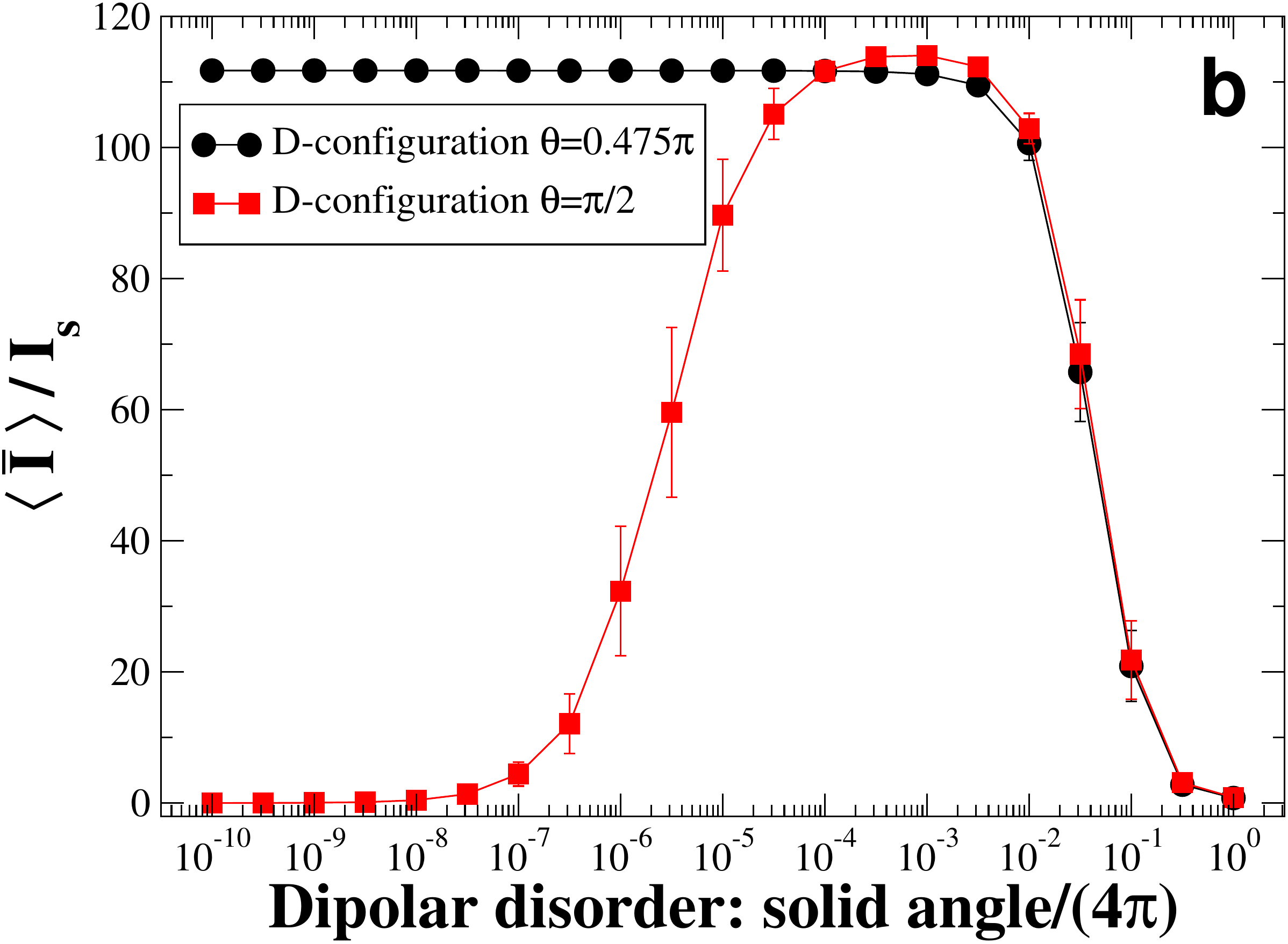}
    \caption{{\it Current vs angular disorder.} Average steady-state current divided by the single-site current, $\braket{\bar{I}}/I_s$, {\it vs.} angular disorder. Each dipole has a random orientation inside a cone centered on the precise orientation of Eq.~\eqref{eq:dip} and with a solid angle varying from 0 (no disorder) to $4\pi$ (completely disordered dipoles). Results obtained with the master equation, see Eq.~\eqref{eq:mas}. The symbols indicate the average and the error bars indicate one standard deviation among 100 disorder realizations. In the D-configuration and LH-configuration, the laser frequency corresponds to the peak at zero disorder. Note that the horizontal axis in panel (b) is in logarithmic scale. Parameters: $N=32$, $\kappa=10\gamma$, $\Omega_R=4.68\gamma$.
    }
    \label{fig:robust}
\end{figure}

In Fig.~\ref{fig:robust}a we plot the average normalized peak steady-state current, $\braket{\bar{I}}/I_s$, against the strength of angular disorder for $N=32$,  different illumination conditions and the corresponding optimal angles $\theta$: for the D- and sunlight configurations we consider the optimal angle $\theta=0.475\pi$ discussed above (see Fig.~\ref{fig:tilt}c), while for the LH-configuration we show the case $\theta=0$ (which is optimal, see Fig.~\ref{fig:tilt}d). As one can see from Fig.~\ref{fig:robust}a, in all cases the current is suppressed by disorder. Such suppression is very sharp for the D- and LH- configurations, and milder for the sunlight configuration. Specifically, the efficiency of our system is robust to very high disorder for the sunlight configuration, but it is also robust for the D-configuration. Indeed, as one can see from Fig.~\ref{fig:robust}a, the current for the D-configuration remains very high for angular disorder as large as $10\%$ of the full solid angle. On the other hand, the LH-configuration appears less robust to disorder, and the current quickly decreases for angular disorder larger than $1\%$ of the full solid angle (see Fig.~\ref{fig:robust}a).

Moreover, in Fig.~\ref{fig:robust}b we consider the D-configuration for $\theta=\pi/2$ compared to the optimal case ($\theta=0.475\pi$). In the case $\theta=\pi/2$, at zero disorder, the only ring eigenstate with non-vanishing dipole strength is the highest-energy ring eigenstate, with polarization perpendicular to the ring plane, see Eq.~\eqref{dips}. Therefore, at zero disorder the ring is decoupled from the RC and there is no current. However, one can see that a small amount of angular disorder is able to increase the current up to very high values, which are comparable to the optimal configuration $\theta=0.475\pi$ in absence of disorder, see Fig.~\ref{fig:robust}b. The reason is that a small angular disorder slightly increases the dipole strength of the lowest-energy states of the ring: in this way, the coupling between the ring and the RC is activated giving rise to  a current.
Actually, the emission from the ring remains low, because most of the dipole strength remains concentrated in the highest-energy state (for small disorder). These results, therefore, suggest an alternative way to engineer our device: instead of a finely-tuned configuration with $\theta$ very close to the optimal value $0.475\pi$, one can use an existing ring where all the dipoles are perpendicular to the plane ($\theta=\pi/2$) and a bit of angular disorder is present. In such configuration, the emission is suppressed while it is possible to transfer the excitation to the RC via the low-energy ring states.

\begin{figure*}
    \centering
    \begin{tabular}{cc}
        {{\bf a} \it D-configuration ($\theta=0.475\pi$)} & {{\bf b} \it LH-configuration ($\theta=0$)} \\
        \includegraphics[width=0.47\columnwidth]{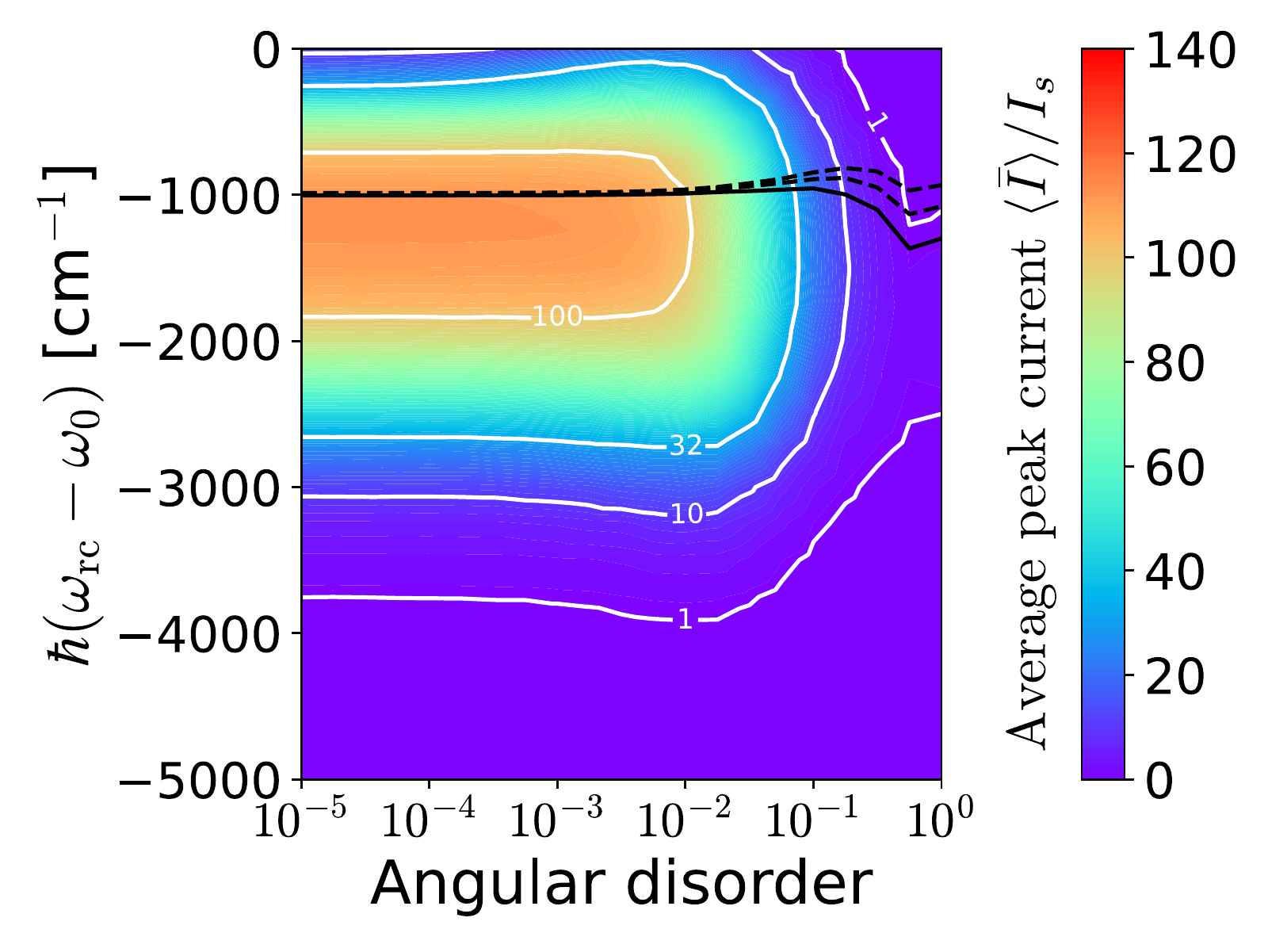} & \includegraphics[width=0.47\columnwidth]{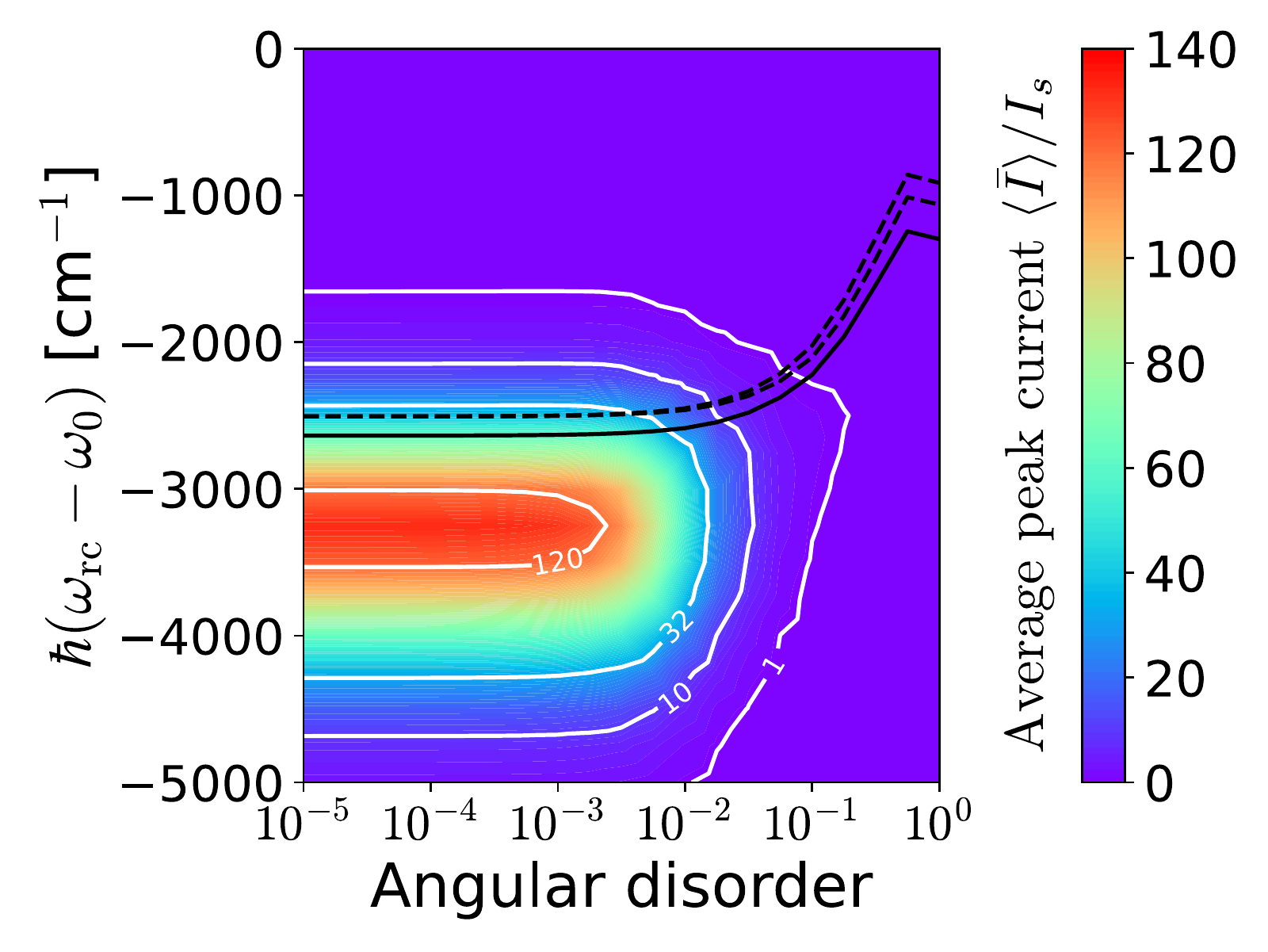} \\
        {{\bf c} \it Sunlight configuration ($\theta=0.475\pi$)} & {{\bf d} \it Sunlight configuration ($\theta=0$)} \\
        \includegraphics[width=0.47\columnwidth]{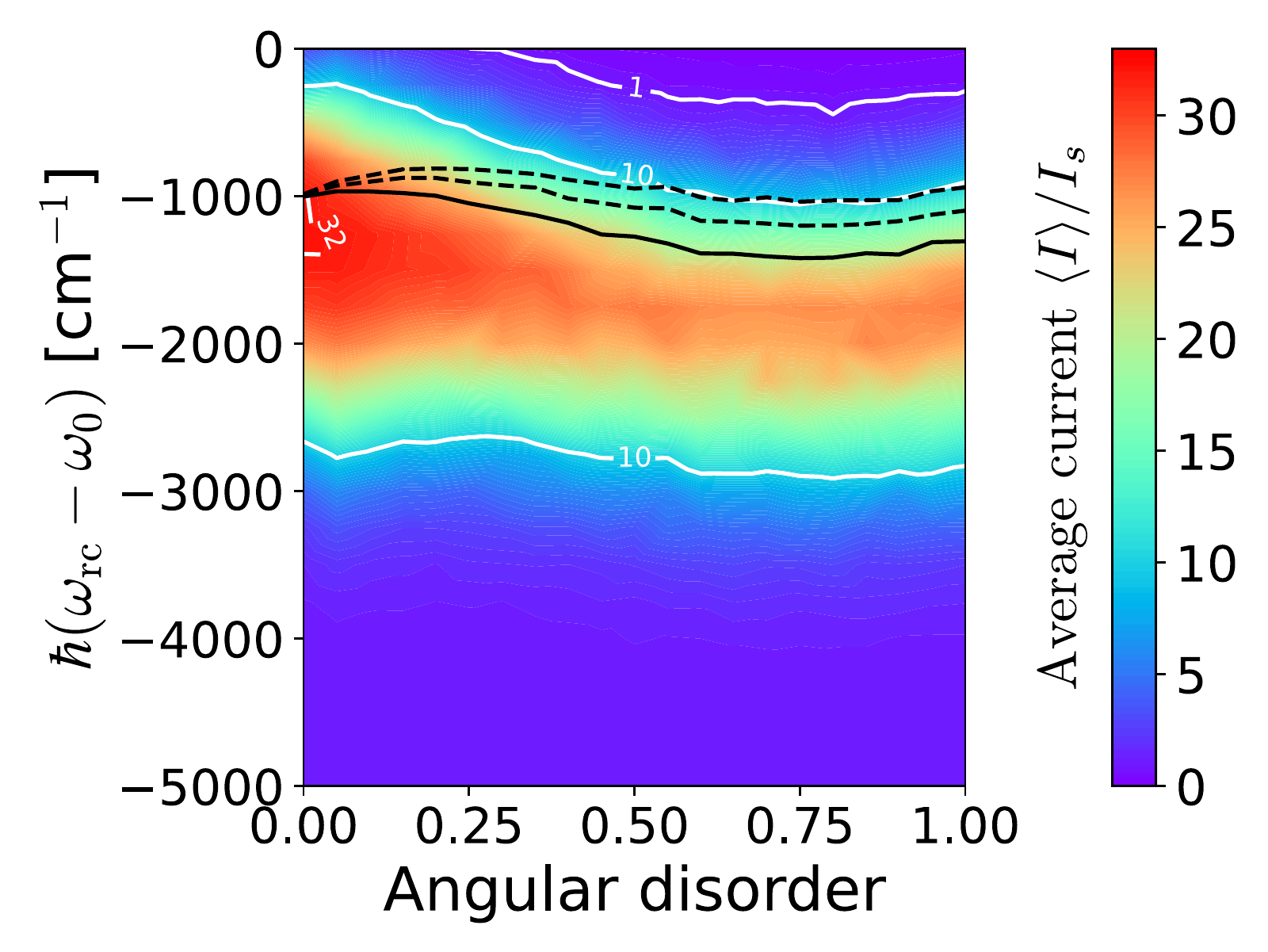} & \includegraphics[width=0.47\columnwidth]{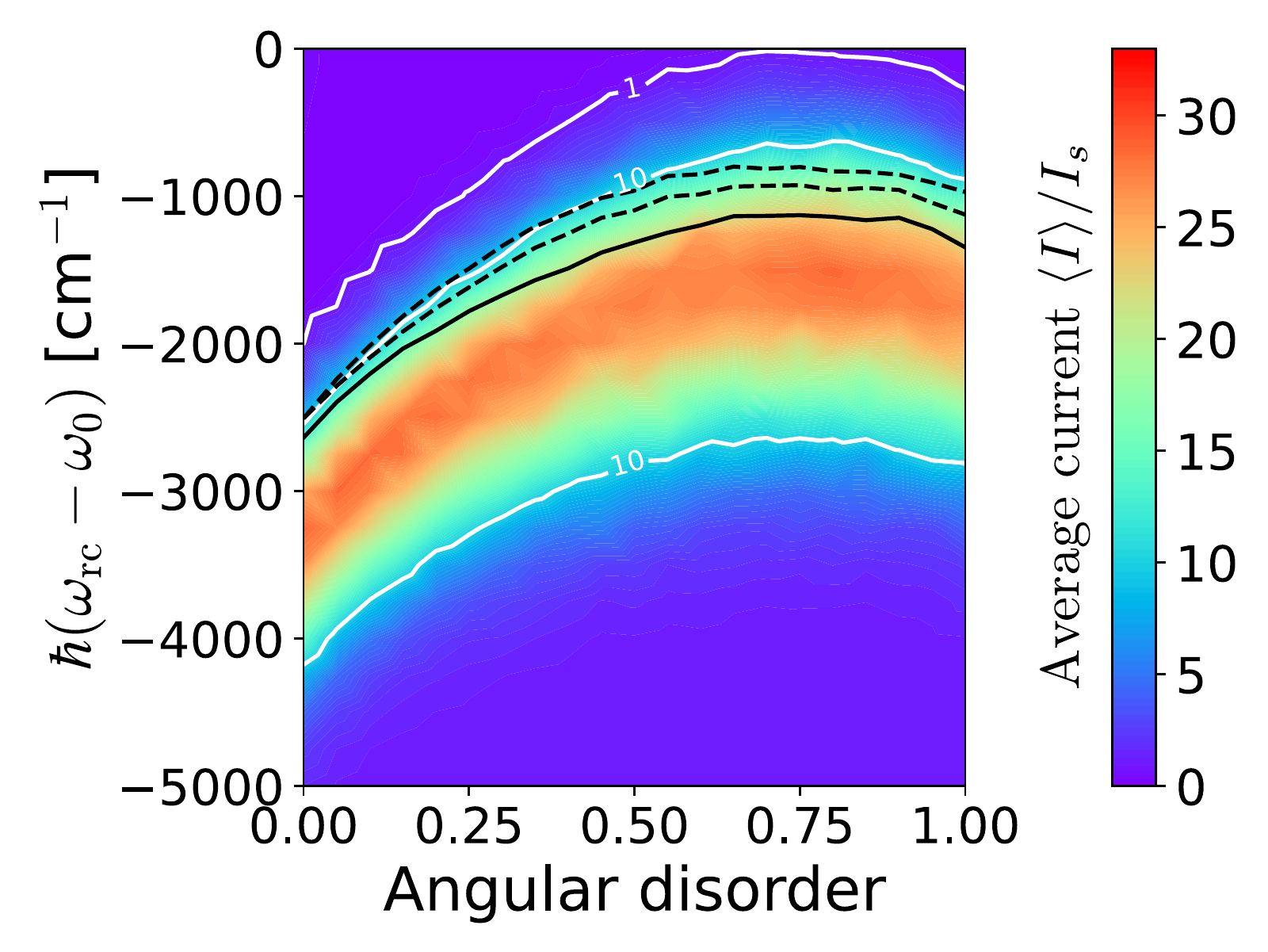} 
    \end{tabular}
    \caption{Average peak steady-state current, $\braket{\bar{I}}/I_s$, {\it vs.} the angular disorder and the energy of the RC $\hbar\omega_{\rm rc}$ (measured with respect to the site energy $\hbar\omega_0$). In each panel, the solid black line represents the average energy of the lowest ring eigenstate {\it vs.} disorder, while the dashed lines are the first and second excited states of the ring. The current is averaged over 100 disorder realizations, in each panel. Results obtained from Eq.~\eqref{eq:cur} solving the master equation, see Eq.~\eqref{eq:mas}. In panels (a,b), the laser frequency corresponds to the peak at zero disorder. Parameters: $N=32$, $\kappa=10\gamma$, $\Omega_R=4.68\gamma$, $T=300$~K. 
    }
    \label{fig:robust2}
\end{figure*}

So far, we kept the energy of the RC fixed and at resonance with the first and second excited states of the ring, in the absence of disorder. However, it is known that the spectral width is affected by disorder, therefore one may ask whether putting the RC at resonance with the zero-disorder energy levels is the best choice in the presence of disorder. Therefore, in Fig.~\ref{fig:robust2} we show how the peak current changes as a function of the angular disorder and of the energy of the RC. To guide the eye, in each panel we plot the average energies of the three lowest eigenstates of the ring as a function of disorder: the lowest excitonic state $\ket{E_1}$ (black continuous line), and the first and second excited states $\ket{E_2}$ and $\ket{E_3}$ (black dashed lines).   Analyzing Fig.~\ref{fig:robust2}, one can see that, for the LH- and sunlight configurations, the optimal efficiency is always obtained with the RC energy red-shifted with respect to the ring states by some thousands of cm$^{-1}$. Indeed if the RC is at resonance there is a large backward transfer to the ring states and excitation can be lost by re-emission. If the detuning is too large, then transfer to the RC is suppressed and again re-emission lowers the efficiency. The optimal position of the energy of the RC follows the ring ground state energy (black continuous lines in each panel) as it is modified by disorder. As a final note we stress that in absence of disorder the LH-configuration with optimal detuning can even be more efficient than our proposed device configuration, compare panels a) and b) in Fig.~\ref{fig:robust2}.  This shows that several paths are available to suppress re-emission, and red-shifting the central core absorber energy can also be very effective in suppressing re-emission. This mechanism has also been discussed and exploited in Ref.~\cite{biolaser} by some of the authors of this manuscript.  As a final remark, let us note that exploiting thermal relaxation to suppress re-emission requires detailed knowledge of the system--bath coupling and bath structure, while the mechanism proposed by us in this manuscript is more direct and does not depend on the details of the system--bath  coupling to be effective.

\section{Conclusions and perspectives}

In many natural light-harvesting complexes, most of the dipole
strength is concentrated in few states that absorb
light and, at the same time,  transmit the excitation to an external
trapping environment. The large dipole strength of such states favour
the absorption of light  but it also induces
losses by re-emission of the excitation, thus  limiting the
efficiency of the energy transfer.

Here we propose a light-harvesting device  in which the absorbing and the
transferring states are engineered to be different by structural arrangement of chromophores. 
We proved that our engineered device  is able to improve the
efficiency of light-harvesting complexes by several orders of magnitude both when the
interaction with a polarized monochromatic field is considered and under natural sunlight. 
Since the solar spectrum is broad,
to use the proposed device as solar light-harvesting complex,  
an ensemble of devices absorbing at different
frequencies should be considered. 
The proposed device can also be extended to unpolarized light by arranging the rings
on a spherical shell, in a similar way to the arrangement found in the chromophores of purple
bacteria~\cite{schulten}. 

Our approach allowed us also to study the scalability of our device as the number of light-harvesting chromophores is increased by increasing the ring radius and keeping a fixed chromophore density. We have shown the existence of  an optimal size. The reason for that is that  even if increasing the ring radius improves absorption, it also suppresses transfer towards the central core absorber. As a future development we plan to consider different architectures to make the system scalable to larger sizes without losing efficiency. In particular, instead of increasing the ring radius, a network of smaller rings where the excitation is efficiently transferred between them and finally concentrated in the central core absorber should be able to improve the efficiency of our device at larger sizes. Indeed this is precisely the architecture of several photosynthetic natural systems~\cite{schulten,schulten1a,schulten1b}. A similar idea has been successfully employed by some of the authors of this paper in Ref.~\cite{biolaser} where a bio-inspired sunlight-pumped laser has been proposed.

In order to realize the proposed complex molecular structure a precise control over molecular orientation and structure is required. In this context,  modern molecular synthesis and modification techniques can easily  meet these needs. Several nanostructures, including nanotubes~\cite{Zhou2009}, DNA proteins~\cite{Boulais2017}, and viruses~\cite{Witus2011,Park2015} can be precisely functionalized with organic molecules using an impressive variety of bioconjugation tools. Engineering of synthetic molecular aggregates in linear, circular, and other geometric configurations, with nanometer separation between molecules, is commonplace. The major challenge facing these techniques is controlling energetic and structural disorder after functionalization. However, recent experiments demonstrating the use of functionalized DNA proteins for light harvesting~\cite{Boulais2017} constitute a proof-of-principle confirmation of the promise of such synthetic molecular engineering techniques for light capture technologies. We also note that the effect of structural disorder could also be exploited at our advantage. As shown above, in a perfect H-aggregate, \emph{i.e.} a molecular aggregate characterized by a single bright state above the energy of the monomer absorption peak (Fig.~\ref{fig:robust}b, $\theta=\pi/2$), all the dipole strength is concentrated in the highest excitonic state, and structural disorder is able to add some dipole strength  to other states  allowing  the lowest excitonic state to also transfer energy to the central core absorber. In this way  we achieve the same separation of absorption from transfer shown to be so effective in improving light-harvesting efficiency. 
Finally, a lesson can be drawn   about natural photosynthetic antenna complexes from our analysis: natural LH2 systems can be very efficient at absorbing and transporting light excitations if emission is efficiently suppressed either by a large trapping rate, see Fig.~\ref{fig:kappa-theta}c,  or by properly detuning the RC energy, see Fig.~\ref{fig:robust2}d. In this sense it is likely that natural systems also exploit suppression of superradiant emission while using super-absorption to enhance their efficiency.

\ack 
This research was supported, in part, by the Center for Research Computing of the University of Notre Dame through access to key computational resources.
We acknowledge financial support from Fondazione E.U.L.O. 
in the frame of the project 
``Trasporto quantistico in sistemi nanostrutturati con applicazioni ai biosistemi''.
FB and FM acknowledge support by the Iniziativa Specifica INFN-DynSysMath.   This publication has been financially supported by the
Catholic University of Sacred Heart within the program of promotion and diffusion of scientific research.
Research has been financially supported by Ministero dell’Istruzione, dell’Università e della Ricerca within the project PRIN 20172H2SC4. 
GLC~acknowledges the funding of ConaCyt Ciencia Basica project A1-S-22706.
MS~ acknowledges support from the U.S. Department of Energy, Office of Science,  Office of Basic Energy Sciences under the Materials and Chemical Sciences Research for Quantum Information Research program, and from the DARPA DETECT program.

Sandia National Laboratories is a multimission laboratory managed and operated by National Technology \& Engineering Solutions of Sandia, LLC, a wholly owned subsidiary of Honeywell International Inc., for the U.S. Department of Energy's National Nuclear Security Administration under contract DE-NA0003525. This paper describes objective technical results and analysis. Any subjective views or opinions that might be expressed in the paper do not necessarily represent the views of the U.S. Department of Energy or the United States Government.

\appendix

\section{Effective Hamiltonian }
\label{sec:Muk}

The effective Hamiltonian of an aggregate interacting with an electromagnetic field can be written as~\cite{mukameldeph,mukamelspano,kaiser1,kaiser2}
\begin{subequations}
\label{heff}
\begin{eqnarray}
  H_{\rm eff} &= H_0+\Delta
-\frac{i}{2}\Gamma \\
&=\sum_n \hbar \omega_0 \ket{n}\bra{n} + \sum_{m,n} \hbar\left(\Delta_{nm} - \frac{i}{2}\Gamma_{nm} \right) \ket{n}\bra{m} \, . 
\end{eqnarray}
\end{subequations}
where the latter expression refers explicitly to the site basis $|n\rangle$.
In Eq.~\eqref{heff} $\hbar \omega_0$ is the energy of each site.
In the limit where only one excitation is present in the
system~\cite{mukameldeph,mukamelspano,kaiser1,kaiser2}
the diagonal and off-diagonal  matrix elements are given respectively by,
\begin{subequations}
\label{Hmuk}
\begin{eqnarray}
  \Delta_{nn} &= 0 \, , \\
  \Gamma_{nn} &= \gamma \, , 
\end{eqnarray}
and
\begin{eqnarray}
\label{dmuk}
  \Delta_{nm} &= \frac{3\gamma}{4} \left[ \left( -\frac{\cos x_{nm}}{x_{nm}} +
    \frac{\sin x_{nm}}{x_{nm}^2} + \frac{\cos x_{nm}}{x_{nm}^3} \right)
    \hat{p}_n \cdot \hat{p}_m \right. \nonumber \\
    &\left. - \left( -\frac{\cos x_{nm}}{x_{nm}} + 3\frac{\sin x_{nm}}{x_{nm}^2} +
    3\frac{\cos x_{nm}}{x_{nm}^3}\right) \left( \hat{p}_n \cdot \hat{r}_{nm}
    \right) \left( \hat{p}_m \cdot \hat{r}_{nm} \right) \right]\\
  \Gamma_{nm} &= \frac{3\gamma}{2} \left[ \left( \frac{\sin x_{nm}}{x_{nm}} +
    \frac{\cos x_{nm}}{x_{nm}^2} - \frac{\sin x_{nm}}{x_{nm}^3} \right)
    \hat{p}_n \cdot \hat{p}_m \right. \nonumber \\
    &\left.- \left( \frac{\sin x_{nm}}{x_{nm}} + 3\frac{\cos x_{nm}}{x_{nm}^2} -
    3\frac{\sin x_{nm}}{x_{nm}^3}\right) \left( \hat{p}_n \cdot \hat{r}_{nm}
    \right) \left( \hat{p}_m \cdot \hat{r}_{nm} \right) \right] \label{gmuk} 
\end{eqnarray}
\end{subequations}
where $\hbar\gamma = \frac{4}{3} \mu^2 k_0^3/\epsilon_r$,
$\mu=|\vec{\mu}|$ is the transition dipole matrix element, $\epsilon_r$
the relative dielectric permittivity, $\hat{p}_n :=  \vec{\mu}_n  /
\mu$  the normalized dipole moment of the $n$-th site, $x_{nm} = k_0 r_{nm}$, $\hat{r}_{nm} := \vec{r}_{nm}
/ r_{nm}$  the unit vector joining the $n$-th and the $m$-th
sites, and $k_0=\omega_0/c$.
A derivation of the expressions in Eqs.~\eqref{dmuk} and \eqref{gmuk} is presented in~\ref{sec:sun}.

In the small volume limit $x_{nm} \ll 1$, that is when the system size is much smaller than the wavelength $\lambda_0 = 2\pi/k_0$ connected with the optical transition,
the matrix elements can be approximated as
\begin{subequations}
  \begin{eqnarray}
    \label{Hm2}
    \hbar\Delta_{nm} &\approx \frac{\mu^2}{\epsilon_r} \frac{\hat{p}_n \cdot \hat{p}_m - 3(\hat{p}_n \cdot \hat{r}_{nm})(\hat{p}_m \cdot \hat{r}_{nm})}{r_{nm}^3} (1-\delta_{nm}) \\
    \Gamma_{nm} &\approx \gamma (\hat{p}_n \cdot \hat{p}_m). \label{eq:gamma}
  \end{eqnarray}
\end{subequations}

\section{Emission rates and coupling to the laser beyond the small volume limit}
\label{sec:Rlambda}

Eigenvalues and eigenstates can be obtained by diagonalizing the effective Hamiltonian, Eq.~\eqref{heff}. From them it is possible to define the emission rate $\gamma_\alpha$ associated with each eigenstate with eigenvalue $E_\alpha-i\hbar \gamma_\alpha/2$. In the small volume limit one can approximate the emission rate as
\begin{eqnarray}
\label{galpha}
\gamma_\alpha \approx \gamma |\vec{p}_\alpha|^2 \, .
\end{eqnarray}

For any value of the angle $\theta$ between the dipoles and the ring plane, the dipole strength of the ring eigenstates is non-vanishing for just
three bright eigenstates. Two of them correspond to the degenerate subspace of the first
and second excited states. Without loss of generality, we choose
two combinations of those two states so that their dipole moments are
\begin{subequations}
\label{dips2}
  \begin{eqnarray}
    \vec{p}_{E_2} &= \sqrt{\frac{N}{2}} \cos \theta \, \hat{y} \quad \text{and} \\
    \vec{p}_{E_3} &= \sqrt{\frac{N}{2}} \cos \theta \, \hat{x} \, ,
  \end{eqnarray}
  with $\hat{x}$ and $\hat{y}$ being the unit vectors of the planar
  axes (note that $E_1$ is the excitonic ground state energy). The
  third bright eigenstate is the highest-energy one, whose dipole moment is
  \begin{eqnarray}
  \label{dips2z}
    \vec{p}_{E_N} = \sqrt{N} \sin \theta \, \hat{z} \, ,
  \end{eqnarray}
  perpendicular to the ring plane. 
\end{subequations}
For the above  three states, the dipole strength increases
with the system size due to cooperative effects induced by the
symmetric arrangement of the dipoles in the ring.

\begin{figure}
  \centering
  \includegraphics[width=0.6\columnwidth]{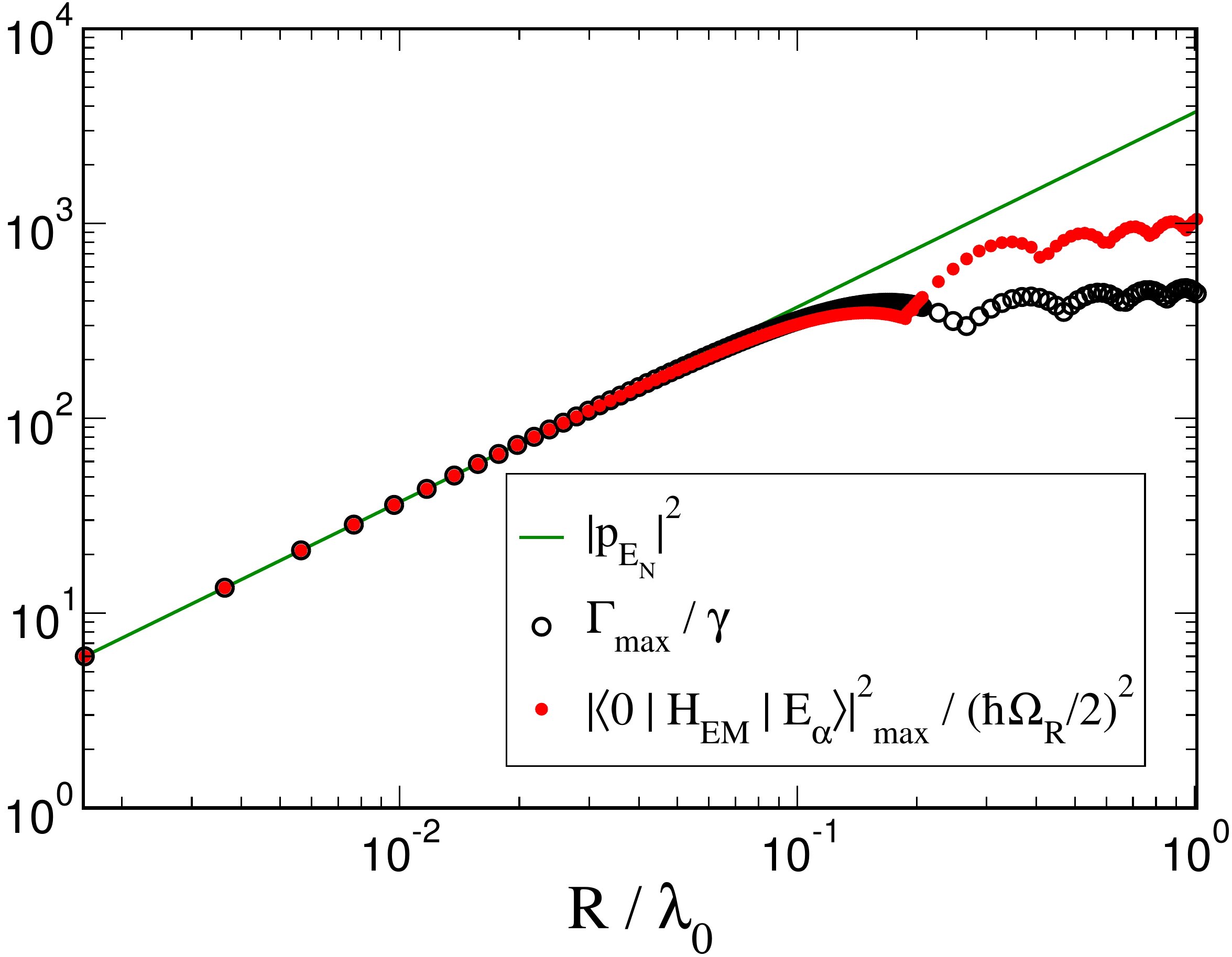}
  \caption{Maximal decay rate $\Gamma_{max}$ of the effective Hamiltonian, see Eqs.~\eqref{heff} and \eqref{Hmuk}, normalized to the single molecule
    decay rate $\gamma$ (open black circles)   {\it vs.} the ratio of the radius of the
    ring over the wavelength, $R/\lambda_0$. Here the density is kept
    fixed and equal to $d=32/(10 \pi)$~nm$^{-1}$. The green continuous
    line corresponds to the dipole strength of the state having the highest energy, see Eq.~\eqref{dips2z}, which is also the maximal dipole strength of the system for the  chosen parameter $\theta=\pi/3$. The red dots are the maximal squared coupling induced by a laser polarized along $z$, see Eq.~\eqref{cem}, normalized by the single-site squared coupling to the laser $(\hbar\Omega_R/2)^2$.}
  \label{coopR}
\end{figure}

In order to study the range of validity of the small volume approximation, in
Fig.~\ref{coopR} we compare the normalized maximal decay width $\Gamma_{max}/\gamma$ obtained from the complex eigenvalues
of $H_{\rm eff}$ (open circles) with that obtained 
in the small volume limit using Eq.~\eqref{galpha} (continuous line) for different values of $R/\lambda_0$ where $R$ is the radius of the system ring. Note that we increase $R$ keeping a fixed density, meaning that the number of sites $N$ increases proportionally to $R$.
 As one can see significant deviations appear already for $R/\lambda_0 > 0.1$. 

In the previous paragraph we analyzed the emission rate of the system eigenstates. Now we focus on the absorption rate. In particular, let us assume that the ring system is coupled to a laser, described as a monochromatic electromagnetic wave. The interaction between the ring sites and the laser is modeled by an additional term
\begin{eqnarray}
  \label{hamlas}
  H_{EM} = \frac{\hbar\Omega_R}{2} \sum_n (\hat{p}_n \cdot \hat{\epsilon})
      e^{-i\vec{k} \cdot \vec{r}_n} \ket{0}\bra{n} + \text{h.c.}
\end{eqnarray}
to the effective Hamiltonian~\eqref{heff}. Here $\Omega_R=\mu E_0 / \hbar$ is the Rabi frequency ($E_0$ being the amplitude of the electric field), $\hat{\epsilon}$ the laser polarization and $\vec{k}$ the wave vector of the incident field. $H_{\it EM}$ describes the
photon absorption and stimulated emission, and it induces coherent oscillations between
the ground state $\ket{0}$ (i.e. the state without any excitation) and the single excitation states $\ket{n}$.

The absorption rate of each eigenstate $\ket{E_\alpha}$ is proportional to the square of the following matrix element:
    \begin{eqnarray}
      \bra{0} H_{\it EM} \ket{E_\alpha} = \frac{\hbar \Omega_R}{2}
      \sum_n c_n(E_\alpha)(\hat{p}_n \cdot \hat{\epsilon})
      e^{-i\vec{k} \cdot \vec{r}_n}  \, ,
\label{cem}
\end{eqnarray}
  where we used the decomposition
  \begin{eqnarray}
  \label{eigr}
  \ket{E_\alpha} = \sum_n c_n(E_\alpha) \ket{n}  \quad \text{with} \quad c_n(E_\alpha) =
\frac{e^{i2\pi\alpha n / N}}{\sqrt{N}}
\end{eqnarray}
of the ring eigenstates in the site basis.
     When the size of the system is smaller than
    the wavelength of the laser ($\vec{k} \cdot \vec{r}_n
    \ll 1$), the coupling can be expressed using the dipole strength of an eigenstate
\begin{eqnarray}
  \vec{p}_\alpha = \sum_n c_n(E_\alpha) \hat{p}_n \, ,
\end{eqnarray}
so that
    \begin{eqnarray}
    \label{hlas}
      \left|\bra{0} H_{\it EM} \ket{E_\alpha}\right|^2 \approx \left(\frac{\hbar \Omega_R}{2}\right)^2 \left|\vec{p}_\alpha \cdot \hat{\epsilon}\right|^2~.
    \end{eqnarray}
Since $|\vec{k}_0|\approx |\vec{k}|$, for small ring sizes ($|\vec{k}_0|R\ll 1$) we are in small volume limit approximation, therefore both Eq.~\eqref{hlas} and Eq.~\eqref{galpha} are satisfied. This means that for each eigenstate, the absorption  and the emission rate are both proportional to the squared dipole strength.
    In Fig.~\ref{coopR} we plot the 
    squared dipole strength of the eigenstate with highest energy (as a continuous line) {\it vs.} the
    ratio $R/\lambda_0$. In the same figure we also plot the maximal normalized emission $\Gamma_{max}/\gamma$ (open black circles) and the maximal normalized absorption $\left|\bra{0} H_{\it EM} \ket{E_\alpha}\right|^2_{max} /\left(\frac{\hbar \Omega_R}{2}\right)^2$ (full red dots). We can observe that, for $R/\lambda_0 < 0.1$ both the maximal emission $\Gamma_{max}/\gamma$ and the maximal absorption are well approximated by the squared dipole strength.
Since the maximal dipole strength is
proportional to $N$ (see Eqs.~\eqref{dips2}) we have a   cooperative coupling to the laser. On the other hand, when $\vec{k} \cdot \vec{r}_n
\approx \vec{k}_0 \cdot \vec{r}_n \gtrsim 1$, the approximations~\eqref{galpha} and~\eqref{hlas} are not valid and both the maximal absorption and
the maximal emission rate grow slower than linearly with  $N$.

\section{Analytical coupling to the RC}
\label{app:Oc}

Here we discuss the coupling between the ring
eigenstates and the reaction center (RC) in the small volume limit.
The reaction center is modeled by a site with excitation energy $\hbar\omega_{\rm rc}$ and the same transition dipole moment $\mu$ of the ring sites.
The dipole-dipole coupling between two sites is given by Eq.~\eqref{Hm2} and the
ring eigenfunctions by Eq.~\eqref{eigr}. The coupling between a ring eigenstate and the RC is
\begin{eqnarray}
  \label{eq:gen}
  \bra{rc} \Delta \ket{E_\alpha} =  \frac{\mu^2}{\epsilon_r} \sum_n c_n(E_\alpha) \frac{\hat{p}_{rc} \cdot \hat{p}_n -3 \left( \hat{p}_{rc} \cdot \hat{r}_n \right) \left( \hat{p}_n \cdot \hat{r}_n \right)}{R^3} \, .
\end{eqnarray}
 In our model, the radial component of the dipoles is vanishing, so that
\begin{eqnarray}
  \hat{p}_n \cdot \hat{r}_n = 0 \qquad \forall n=1,\dots,N \, .
\end{eqnarray}
This leads to a simplification of the expression~\eqref{eq:gen}, which becomes
\begin{eqnarray}
  \bra{rc} \Delta \ket{E_\alpha} &= \frac{\mu^2}{\epsilon_r R^3} \hat{p}_{rc} \cdot \left( \sum_n c_n(E_\alpha) \hat{p}_n \right) = \frac{\mu^2}{\epsilon_r R^3} \hat{p}_{rc} \cdot \vec{p}_\alpha \, .
\end{eqnarray}

We have the reaction center dipole oriented along the $y$ axis, i.e.
\begin{eqnarray}
  \hat{p}_{rc} = \hat{y} \, ,
\end{eqnarray}
so that only one of the $N$ eigenstates (which belongs to the doubly degenerate subspace of the first and second excited states and which we call $\ket{E_2}$) is coupled to the reaction center, with a
coupling 
\begin{eqnarray}
  \Omega_C = \bra{rc} \Delta \ket{E_2} = \frac{\mu^2}{\epsilon_r R^3} \sqrt{\frac{N}{2}} \cos \theta \, .
\end{eqnarray}
In this manuscript, the density $d = N/(2\pi R)$ is kept constant, so that $\Omega_C$ scales with $N$ as
\begin{eqnarray}
  \Omega_C = \frac{\mu^2 (2 \pi d)^3 \cos \theta}{\sqrt{2} \epsilon_r}
  N^{-5/2} \, .
\label{OC2}
\end{eqnarray}

\section{The LH-configuration as a representative of natural light-harvesting complexes}
\label{sec:lh}

\begin{figure}
  \centering
  \begin{tabular}{cc}
    {\bf LHI complex} & {\bf LH-configuration ($\theta=0$)} \\
    \includegraphics[width=0.35\columnwidth]{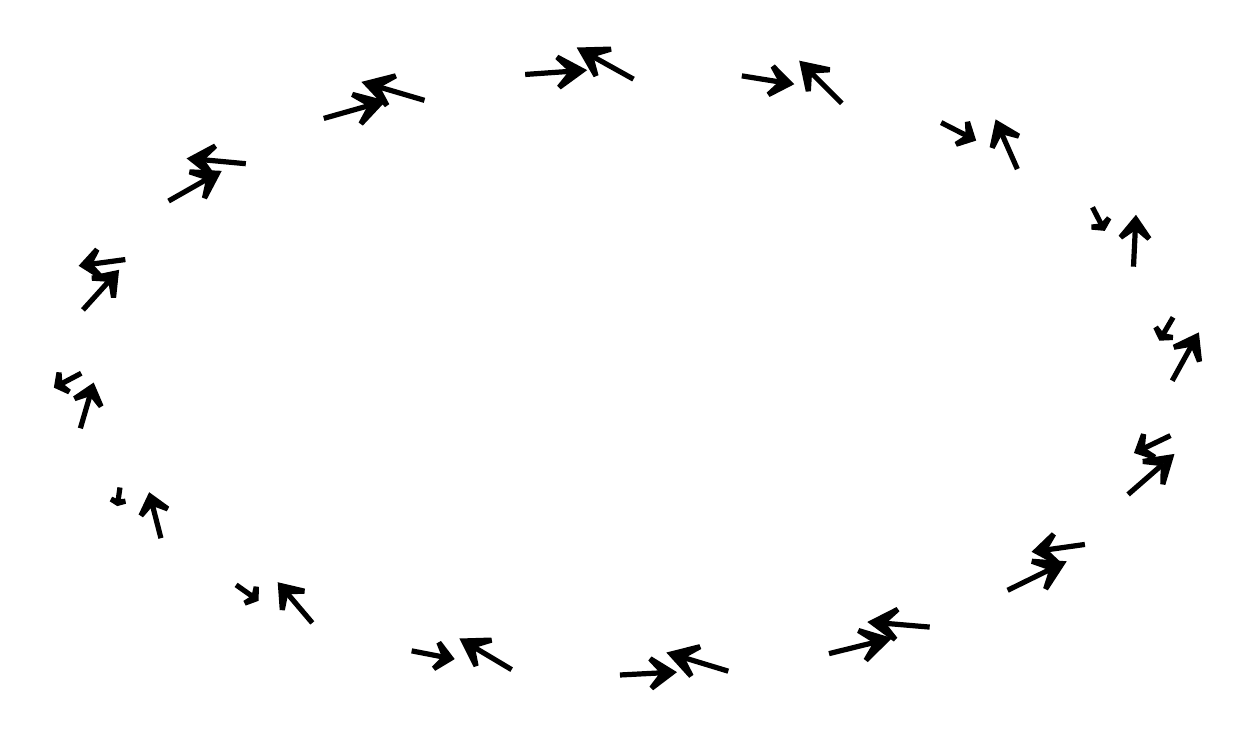} & \includegraphics[width=0.35\columnwidth]{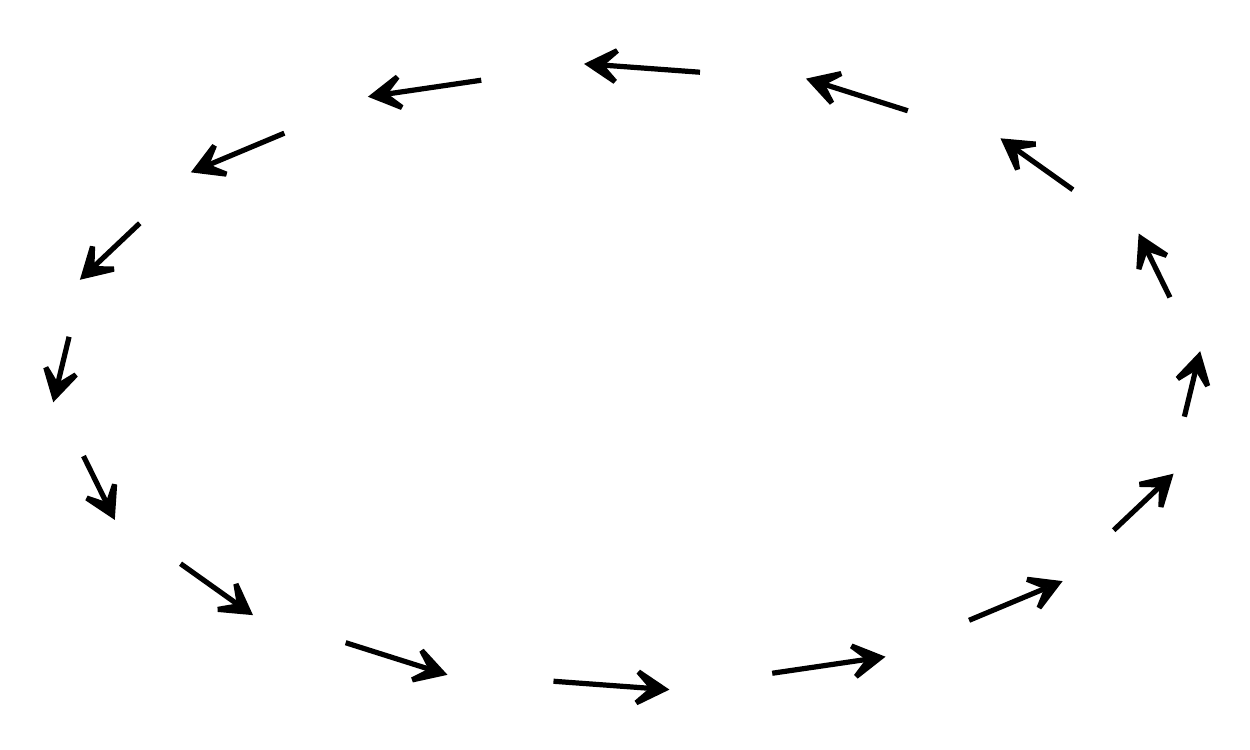}\\
    \includegraphics[width=0.45\columnwidth]{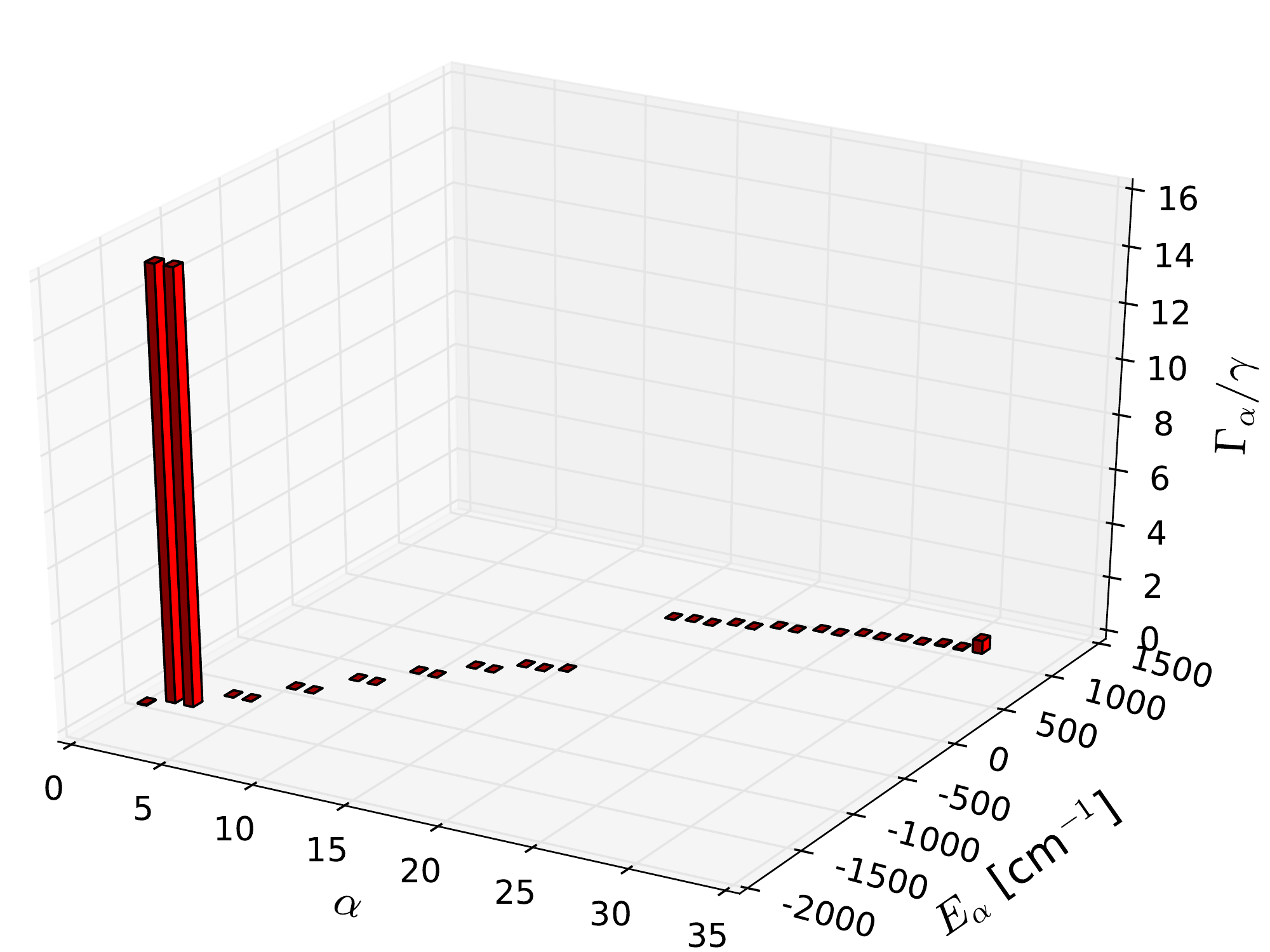} & \includegraphics[width=0.45\columnwidth]{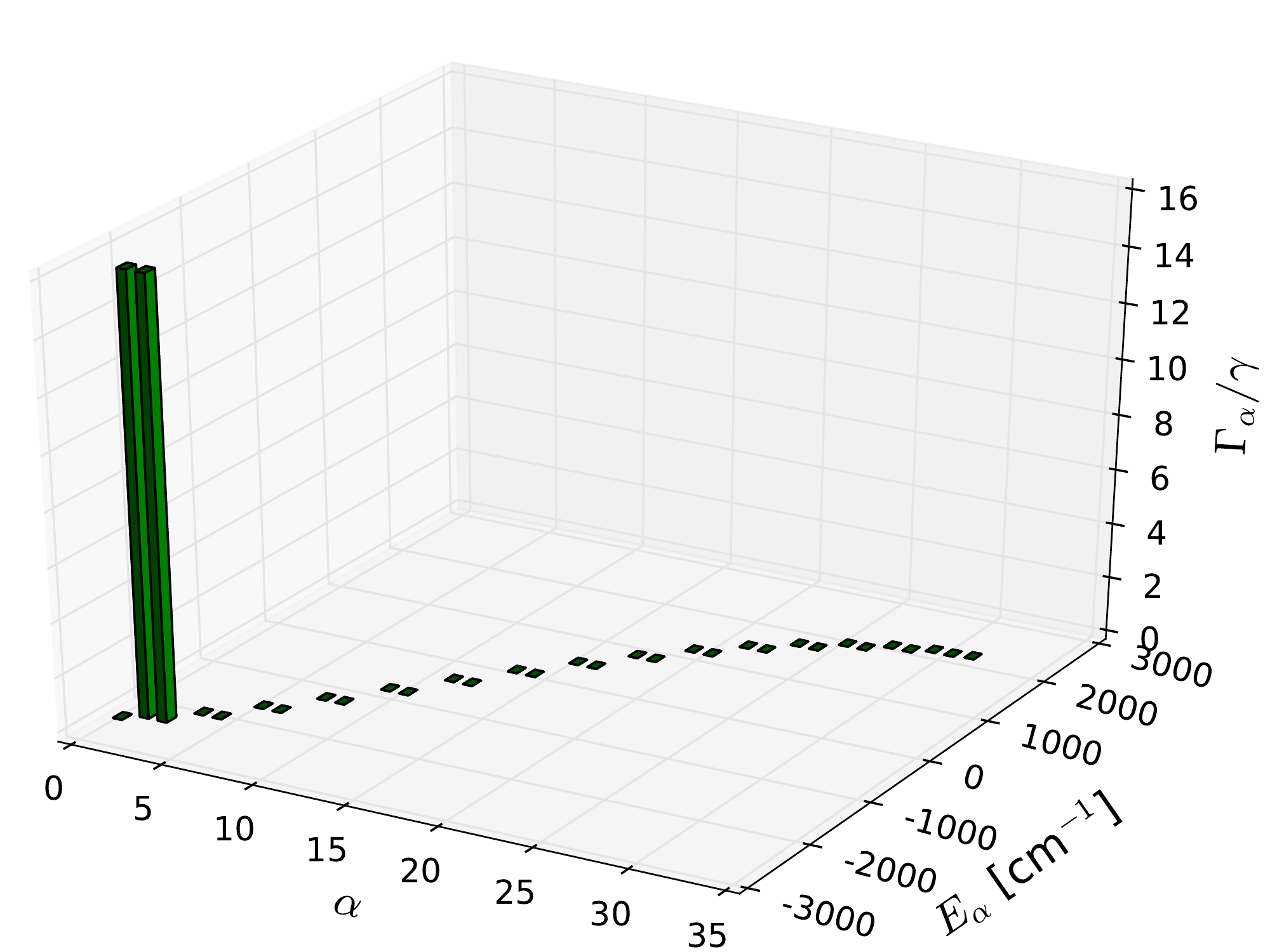}
  \end{tabular}
  \caption{\emph{Top panels:} schematic representation of the
    directions of the transition dipoles for the purple bacteria LHI complex taken from Ref.~\cite{schulten2} (left) and for the LH-configuration
    (right). \emph{Lower panels:} normalized decay rates $\Gamma_\alpha$ of the eigenstates {\it vs.} their energy $E_\alpha$ and {\it vs.} the index $\alpha = 1, \dots, N$. The Hamiltonian for the LHI complex is taken from Ref.~\cite{schulten1a}. Here the system size is $N=32$ for both the LHI complex and the LH-configuration. }
\label{lh1}
\end{figure}

The molecular structure introduced in the main text in the LH-configuration is a good representative of some natural light-harvesting
complexes. In Fig.~\ref{lh1} on the left panels we show the LHI
complex of purple bacteria~\cite{schulten1a,schulten2}, while on right panel we show the
LH-configuration with $\theta=0$. On the top panels we show a schematic
representation of the directions of the transition dipoles in the
models, while in the lower graphs we show the normalized decay rates of the
eigenstates of the systems {\it vs.} their energy. The positions and dipole orientations of the LHI complex have been taken from Ref.~\cite{schulten2} and the Hamiltonian parameters from Ref.~\cite{schulten1a}. In
both configurations the decay rates are concentrated in the first and second
excited states, which are degenerate and have the same decay rate $\approx N\gamma/2$.

\section{Master equation}
\label{sec:mas}

The interaction with the laser field is described by Eq.~\eqref{hamlas}. Note that we neglected the counter-rotating terms, according to the rotating wave approximation (RWA,~\cite{kaiser1,kaiser2}). Moreover, by a unitary transformation, the time dependence of the laser term has been removed, leading to a shift of the diagonal terms in the Hamiltonian (i.e. the site energies), which become
\begin{eqnarray}
  H_0 \rightarrow H_0 = \sum_{n=1}^N \hbar (\omega_0-\omega) \ket{n}\bra{n}
  + \hbar (\omega_{\rm rc}-\omega) \ket{rc}\bra{rc} \, .
\end{eqnarray}

In order to describe the dynamics of our model coupled to a thermal
bath, we consider the following Hamiltonian: 
\begin{eqnarray}
    H_S=H_0+\Delta+H_{\it EM}
\end{eqnarray}
where $\Delta$ is defined in Eq.~(\ref{Hm2}) and we use
the following master equation~\cite{kaiser1,kaiser2,mukamelspano,lloydfmo},
\begin{eqnarray}
  \label{eq:mas2}
  \frac{d\rho}{dt} = &-\frac{i}{\hbar} \left[ H_{S}, \rho \right] + \mathcal{L}_{fl}[\rho] + \mathcal{L}_{rc}[\rho] + \mathcal{R}_T [\rho]~,
\end{eqnarray}
where the last three terms describe, respectively, the fluorescence, the trapping in the RC and the thermal dissipation in presence of a thermal bath. They are given by
\begin{eqnarray}
  \mathcal{L}_{fl}[\rho] = & \sum_{m,n} \Gamma_{mn} \left[ a_n
  \rho a_m^\dag  - \frac{1}{2} \left\{ a_m^\dag a_n , \rho \right\} \right] \\
\mathcal{L}_{rc}[\rho] = & \kappa \left[ a_{rc}
  \rho a_{rc}^\dag  - \frac{1}{2} \left\{ a_{rc}^\dag a_{rc} , \rho \right\} \right] \\
  {\cal R}_T[\rho] =  &\sum_{\omega,\omega'} \sum_{n} \frac{\gamma^{(p)}(\omega)}{2} \left[ A_n(\omega) \rho A_n^\dag(\omega')+ A_n(\omega') \rho A_n^\dag(\omega) \right.\nonumber \\
  &\left. \qquad \qquad \qquad - A_n^\dag(\omega') A_n(\omega) \rho - \rho A_n^\dag(\omega) A_n(\omega') \right]
    \label{LT2}
\end{eqnarray}
where $\Gamma_{mn}$ are given by Eq.~\eqref{gmuk}, $a_n=\ket{0}\bra{n}$, $a_{rc}~=~\ket{0}\bra{rc}$ and $\mathcal{R}_T [\rho]$
is the thermal dissipator. In this scheme, each site is assumed to be coupled to an independent  Ohmic bath with linear coupling. Nevertheless they all follow the same dynamics, having the same temperature, spectral density and coupling strength~\cite{flemprall,flemyang}. Specifically, the Hamiltonian of the system coupled to the independent thermal baths reads
\begin{eqnarray}
    H_{SB} = H_S + \sum_{k,n} \omega_k b_{k,n}^\dag b_{k,n} + \sum_{k,n} g_k \ket{n}\bra{n} X_{k,n}
\end{eqnarray}
where $g_k$ is the linear coupling strength of a site with the harmonic oscillator with frequency $\omega_k$. Here, $X_{k,n}=\sqrt{\hbar/(2M\omega_k)}\left( b^\dag_k + b_k \right)$ is the position of the harmonic oscillator ($M$ here is the mass of the oscillator). The Redfield dissipator, Eq.~\eqref{LT2}, is derived under the Born-Markov approximations~\cite{petruccione}, assuming weak system--bath coupling and fast bath relaxation time, but without applying the secular approximation, as discussed in the main text.
Taking the continuum limit for the sum over the bath frequencies $\omega_k$ and defining the spectral density $J(\omega)$ through
\begin{eqnarray}
\label{Jomega}
    \sum_k g_k^2 \frac{\hbar}{2 M\omega_k} \rightarrow \int_0^\infty d\omega~J(\omega)
\end{eqnarray}
we obtain Eq.~\eqref{LT2}~\cite{lloydfmo,flemcho}, with
\begin{eqnarray}
  \gamma^{(p)}(\omega) =
  \frac{2\pi}{\hbar}[J(\omega)(1+n_{BE}(\omega))+J(-\omega)n_{BE}(-\omega)]~.
\label{gp}
\end{eqnarray}
In Eq.~(\ref{gp})
$n_{BE}$ is  the standard Bose-Einstein distribution of the phonons  
\begin{eqnarray}
  n_{BE}(\omega) = \frac{1}{e^{\hbar\omega/k_B T}-1}
\end{eqnarray}
and the spectral density is chosen as~\cite{lloydfmo}
\begin{eqnarray}
  \label{jbath}
  J(\omega)=
  \begin{cases}
    E_R \frac{\omega}{\omega_c} e^{-\omega/\omega_c} & \text{if }
    \omega>0 \\
    0 & \text{if } \omega<0
  \end{cases}
  \, .
\end{eqnarray}
The linear dependence of the spectral density at small frequencies, $J(\omega)\sim \omega$ for $\omega \ll \omega_c$, results from the $\sim 1/\omega$ scaling of the squared system-bath coupling (see Eq.~\eqref{Jomega}, left-hand side), multiplied by the density of modes of the oscillators in 3D, $\sim\omega^2d\omega$. We use an exponential cut-off at large frequencies, $J(\omega)\sim e^{-\omega/\omega_c}$ for $\omega \gg \omega_c$, because it has been used to reproduce spectroscopic results in similar molecular aggregates~\cite{lloydfmo,flemcho}.
As regards the bath parameters, we set the reorganization energy to $E_R=200$~cm$^{-1}$ and the cut-off frequency
to $\omega_c=333$~cm$^{-1}$. With this choice of the parameters
the thermal relaxation among exciton states occurs in about 1~ps at room
temperature for $N=32$, which is comparable with the
estimates for natural photosynthetic systems reported in literature~\cite{schulten1a,schulten1b}, and which is much faster than the times obtained by the radiative emission rates $\gamma_\alpha\sim 1$~ns$^{-1}$.
The operators in Eq.~\eqref{LT2} can be expressed as
\begin{eqnarray}
  A_n(\omega) = \sum_{\tilde{E}_\beta-\tilde{E}_\alpha=\hbar\omega} \tilde{c}_n(\tilde{E}_\alpha) \tilde{c}^*_n(\tilde{E}_\beta)
  \ket{\tilde{E}_\alpha}\bra{\tilde{E}_\beta} \, .
\end{eqnarray}
Here $\ket{\tilde{E}_\alpha}$ are the eigenstates of $(H_0+\Delta)$, according to
\begin{eqnarray}
  (H_0 + \Delta) \ket{\tilde{E}_\alpha} = \tilde{E}_\alpha \ket{\tilde{E}_\alpha}
\end{eqnarray}
and $\tilde{c}_n(\tilde{E}_\alpha)=\braket{n \vert \tilde{E}_\alpha}$.
Note that here the system includes both the ring and the RC, thus the expression of the coefficients $\tilde{c}_n(\tilde{E}_\alpha)$ is different from the one given in Eq.~\eqref{eigr}, where just the ring is considered.

\section{Thermal dephasing rate}
\label{sec:GT}

Using the expressions in~\ref{sec:mas}, we have that the dynamics of the coherences between the ground state $\ket{0}$ and the ring eigenstates in the small volume limit follow
\begin{eqnarray}
      \frac{d\rho_{0\alpha}}{dt} = -\frac{i}{\hbar} \bra{0} \left[ H, \rho \right] \ket{E_\alpha} -\frac{\gamma|\vec{p}_\alpha|^2}{2}\rho_{0\alpha} -\frac{1}{2} \sum_{\beta,\delta} \gamma^{(p)}(E_\beta-E_\delta) \Lambda_{\alpha \beta}^{\delta \delta} \rho_{0 \beta}
      \label{eq:ddeph}
\end{eqnarray}
with the overlap coefficients
\begin{eqnarray}
  \label{lambR}
  \Lambda_{\alpha \beta}^{\delta \delta} = \sum_n c_n(E_\alpha)^* c_n(E_\beta) |c_n(E_\delta)|^2~.
\end{eqnarray}
Eq.~\eqref{eq:ddeph} can be simplified in the present case, under polarized CW laser excitation. Coherences between the ground state and the ring eigenstates are created by the laser hamiltonian term $H_{EM}$. In our analysis, such coherent coupling involves just one ring eigenstate: either $\ket{E_N}$ for the D-configuration, or $\ket{E_2}$ for the LH-configuration. No coherent coupling with the ground state is present in the Sunlight configuration.
Therefore, in Eq.~\eqref{eq:ddeph} the only non-zero term is where $\alpha=\beta$, coinciding with the absorbing state, that we label ``abs'' in the following. Thus, after re-labelling the summed index $\delta\to\beta$, Eq.~\eqref{eq:ddeph} simplifies to
\begin{eqnarray}
\fl \quad
      \frac{d\rho_{0,\rm abs}}{dt} = -\frac{i}{\hbar} \bra{0} \left[ H, \rho \right] \ket{E_{\rm abs}} -\frac{\gamma|\vec{p}_{\rm abs}|^2}{2}\rho_{0,\rm abs} -\frac{1}{2} \sum_{\beta} \gamma^{(p)}(E_{\rm abs}-E_\beta) \Lambda_{{\rm abs}, \beta} \rho_{0, \rm abs}
      \label{eq:ddeph2}
\end{eqnarray}
with the simplified overlap coefficients
\begin{eqnarray}
  \label{lambL}
  \Lambda_{\rm abs,\beta}= \sum_n |c_n(E_{\rm abs})|^2 |c_n(E_\beta)|^2~.
\end{eqnarray}
In a more compact expression, we can write
\begin{eqnarray}
      \frac{d\rho_{0,\rm abs}}{dt} = -\frac{i}{\hbar} \bra{0} \left[ H, \rho \right] \ket{E_{\rm abs}} - \Gamma_T \rho_{0, \rm abs}~,
      \label{eq:ddeph3}
\end{eqnarray}
where we define the dephasing rate
\begin{eqnarray}
    \Gamma_T=\frac{\gamma|\vec{p}_{\rm abs}|^2}{2}+\frac{1}{2} \sum_{\beta} \gamma^{(p)}(E_{\rm abs}-E_\beta) \Lambda_{{\rm abs}, \beta}~.
    \label{eq:gdeph}
\end{eqnarray}
Using the spectral density, Eq.~\eqref{jbath}, we obtain:
\begin{eqnarray}
    \gamma^{(p)}(E_{\rm abs}-E_\beta) = \frac{\pi E_R}{\hbar^2 \omega_c} \frac{(E_{\rm abs}-E_\beta) e^{-|E_{\rm abs}-E_\beta|/(\hbar \omega_c)}}{1 - e^{-(E_{\rm abs}-E_\beta)/(k_BT)}}
\, ,
\end{eqnarray}
where $E_R$ is the reorganization energy, $\omega_c$ is the cut-off
frequency, $T$ is the bath temperature, and the coefficients $\Lambda_{\rm abs,\beta}$ are given in Eq.~\eqref{lambL}.

In the case of the device described in the main text (D-configuration), when the
absorbing state is the highest-energy state, we can make some
approximations. The coefficients~\eqref{lambL} have the value $\Lambda_{\rm abs,\beta}=1/N$ for any $\beta$ so, using Eq.~\eqref{dips2} for $\vec{p}_{\rm abs}$ and integrating over the spectrum we have
\begin{eqnarray}
  \label{GTintapp}
  \Gamma_T = \frac{N\gamma\sin^2\theta}{2}  + \frac{\pi E_R}{N\hbar^2 \omega_c} \int_{E_{1}}^{E_{N}} dE \, \varrho(E) \frac{(E_N-E) e^{-|E_N - E|/(\hbar \omega_c)}}{1 - e^{-(E_N - E)/(k_BT)}} \, ,
\end{eqnarray}
where $\varrho(E)$ is the density of states. Since in our case the spectral extension $E_{N}-E_{1}$ is independent of $N$, then the average density of states is
\begin{eqnarray}
  \overline{\varrho(E)} = \frac{N}{E_{N} - E_{1}} \propto N \, .
\end{eqnarray}
Moreover, in realistic situations ($T=300$~K and the parameters $E_R$ and $\omega_c$ from the literature) the first term in Eq.~\eqref{GTintapp} can be neglected for $N \ll 10^6$. Thus, under these approximations, we can claim that $\Gamma_T$ is independent of $N$.

For the LH-configuration we need to use the full expression given in
Eq.~(\ref{eq:gdeph}) which can be obtained by the parameters of the
system and by the diagonalization of the system Hamiltonian. Also in
this case we have found (numerically) that $\Gamma_T$ is independent of $N$, as shown in the
following.

\begin{figure}
  \centering
  \includegraphics[width=0.6\columnwidth]{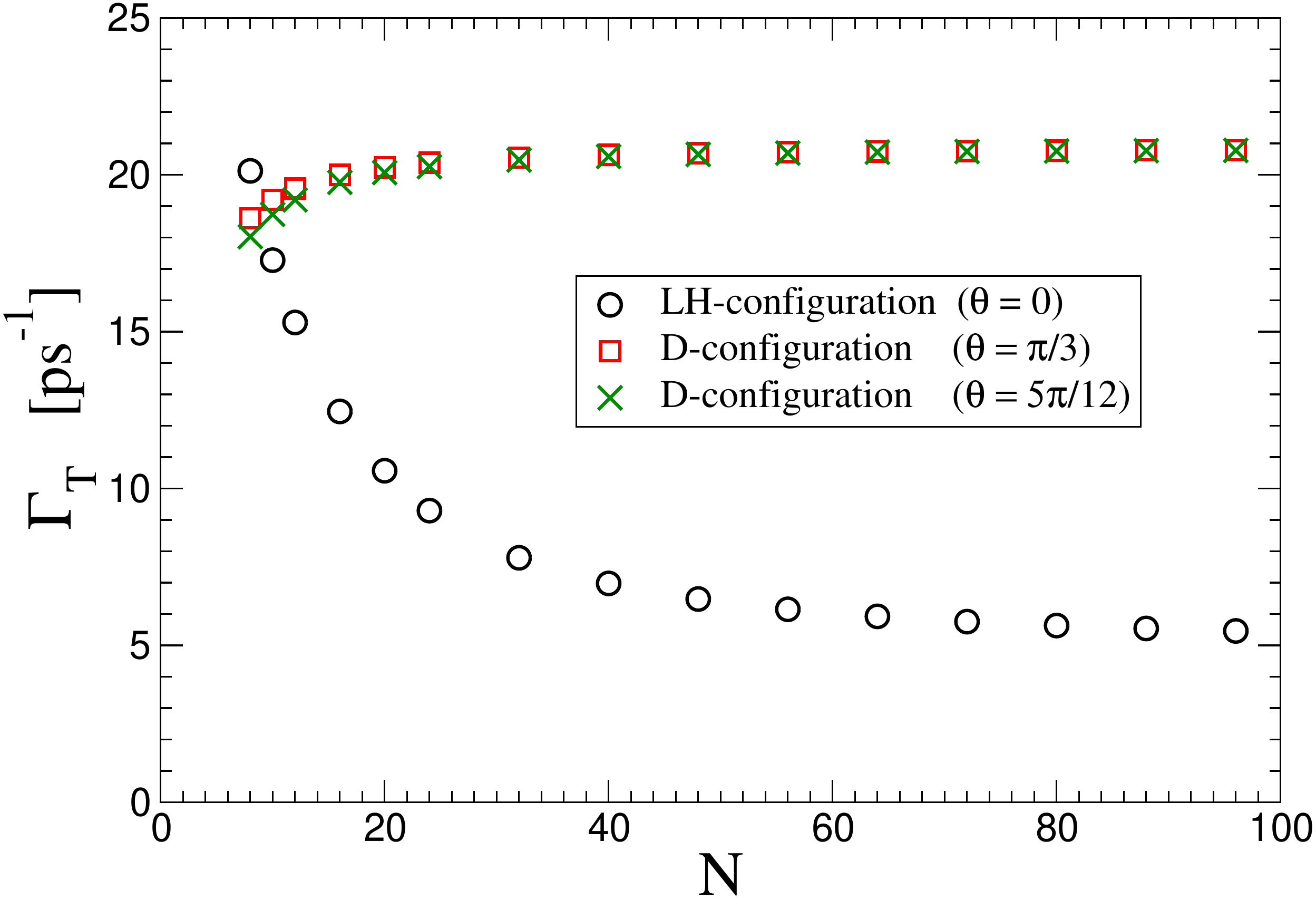}
  \caption{Dephasing rate (see Eq.~\eqref{eq:gdeph}) {\it vs.} the system size $N$ for LH-configuration ($\theta=0$) and D-configuration ($\theta=\pi/3,5\pi/12$). Parameters: $E_R = 200$~cm$^{-1}$, $\omega_c=333$~cm$^{-1}$, $T=300$~K.}
  \label{fig:g0nN}
\end{figure}

In Fig.~\ref{fig:g0nN} we plot 
$\Gamma_T$ {\it vs.} $N$ for the LH-configuration ($\theta=0$, open circles) and the D-configuration
(squares for $\theta=\pi/3$ and crosses for $\theta=5\pi/12$). 
As one can see, for $N \gtrsim 20$ the thermal dephasing rate is weakly dependent of $N$ in all cases.
 Moreover, the two values of $\theta$ considered for the D-configuration
 exhibit the same dephasing rate for any $N \gtrsim 20$. 

\begin{figure}
  \centering
  \includegraphics[width=0.6\columnwidth]{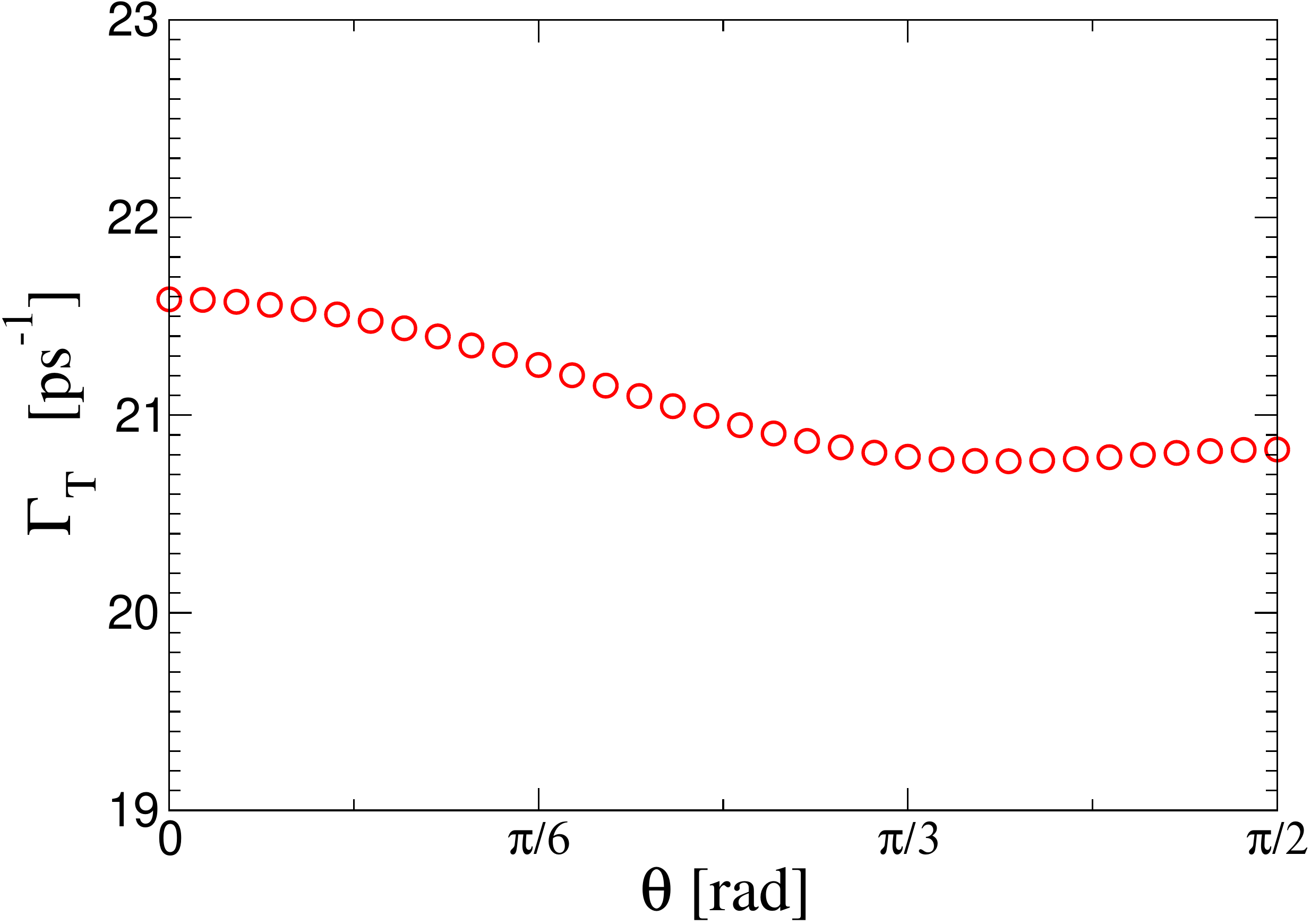}
  \caption{Dephasing rate (see Eq.~\eqref{eq:gdeph}) {\it vs.}~$\theta$
    for D-configuration.  Parameters: $N=100$, $E_R = 200$~cm$^{-1}$, $\omega_c=333$~cm$^{-1}$, $T=300$~K.}
  \label{fig:g0nth}
\end{figure}

We also study the dependence of $\Gamma_T$ on the angle $\theta$ for the
D-configuration, see Fig.~\ref{fig:g0nth}, for $N=100$. We can see that $\Gamma_T$ has a very weak
dependence on $\theta$ (less than 3\% variation), too.

So, our analytical approximation (supported by numerical simulations)
show that the dephasing rate depends very weakly on both $N$ and
$\theta$.
For instance at $T=300$~K and for the parameters considered in the main
text, we have that 
\begin{subequations}
  \begin{eqnarray}
    &\Gamma_T \approx 6 \text{ ps}^{-1} &\text{~for LH-configuration} \\
    &\Gamma_T \approx 20 \text{ ps}^{-1} &\text{~for D-configuration} \, .    
  \end{eqnarray}
\end{subequations}

\section{Analytical results for a single site}
\label{sec:site}
Here we analyze the explicit expression of the Master Equation~\eqref{eq:mas} in the case of a single site which plays the role of an absorber, emitter and trapping state. We call respectively $\ket{0}$ and $\ket{1}$ the ground state and the excited state of the site, so that Eq.~\eqref{eq:mas} reads
\begin{subequations}
  \label{eq:site}
  \begin{eqnarray}
    \frac{d\rho_{00}}{dt} =& i\frac{\Omega_R}{2} \left( \rho_{01} - \rho_{01}^* \right) + \gamma \rho_{11} + \kappa \rho_{11} \\
    \frac{d\rho_{11}}{dt} =& i\frac{\Omega_R}{2} \left( \rho_{01}^* - \rho_{01} \right) - \gamma \rho_{11} - \kappa \rho_{11} \\
    \frac{d\rho_{01}}{dt} =& i\frac{\Omega_R}{2} \left( \rho_{00} - \rho_{11} \right) - \frac{\gamma+\kappa+\gamma^{(p)}(0)}{2} \rho_{01} +i\Delta_0 \rho_{01} \, , \label{rho01}
  \end{eqnarray}
\end{subequations}
where $\Delta_0=\omega_0-\omega$ is the detuning between the laser frequency and the transition frequency of the site. As one can see from Eq.~\eqref{rho01}, the coherence term has a dephasing rate
\begin{eqnarray}
  \label{g01}
  \overline{\Gamma}_{01}^{(p)} = \frac{\gamma + \kappa}{2} + \frac{\gamma^{(p)}(0)}{2} 
\end{eqnarray}
which can be determined by neglecting the terms proportional to $i\Omega_R$ and $i\Delta_0$, which are related to oscillations. Note that this dephasing rate is exactly the same given by Eq.~\eqref{eq:gdeph}, in the trivial case $\beta=1$. The dephasing rate of the coherence term $\rho_{01}$ can also be interpreted as the dephasing rate between the ground state and the (unique) absorbing state, which we call $\Gamma_T$.

In the following we will show that it is possible to obtain a stationary solution for Eq.~\eqref{eq:site}. Using~\eqref{gp}, Eq.~\eqref{g01} can be rewritten as
\begin{eqnarray}
\label{GTsite}
  \Gamma_T = \frac{\gamma + \kappa}{2} + \pi \frac{E_R}{\hbar} \frac{k_B T}{\hbar \omega_c} \, .
\end{eqnarray}
The stationary current $I_s$ of a single molecule
at fixed temperature $T$ is defined as:
\begin{eqnarray}
  I_s = \kappa \rho_{11}^\infty \, ,
\end{eqnarray}
where $\rho_{11}^\infty$ is the steady-state value of $\rho_{11}(t)$. 
The explicit expression for $I_s$ can be derived
analytically by setting the derivatives in  Eq.~\eqref{eq:site} to zero,
\begin{eqnarray}
  \label{eq:jsite}
  I_s = \frac{\kappa}{2} \frac{\Omega_R^2}{\Delta_0^2(\gamma + \kappa)/\Gamma_T + \Omega_R^2 + \Gamma_T(\gamma + \kappa)}~.
\end{eqnarray}

It is well known that, for sufficiently large dephasing ($\Gamma_T \gg \Omega_R/2$), the full quantum master equation is well approximated by a set of rate equations with suitably defined rates~\cite{yang1}. 
Specifically, following Ref.~\cite{yang1}, we derive the pumping rate $T_L$ between the states $0,1$ as
\begin{eqnarray}
  \label{TL1}
  T_L = \frac{\Omega_R^2 \Gamma_T}{2\left(\Delta_0^2 + \Gamma_T^2\right)} \, .
\end{eqnarray}
With such rate we proceed to write a probability-preserving rate equation for the probabilities $P_0$ and $P_1$ to be in ground and in excited state respectively,
\begin{subequations}
  \label{eq:rate2}
  \begin{eqnarray}
    \frac{dP_0}{dt} &= T_L P_1 - T_L P_0 + (\gamma + \kappa) P_1 \\
    \frac{dP_1}{dt} &= T_L P_0 - T_L P_1 - (\gamma + \kappa) P_1~,
  \end{eqnarray}
\end{subequations}
where we have taken into account the pumping rate $T_L$~\eqref{TL1}.
The incoherent transmitted current
\begin{eqnarray}
  I_s^{(inc)} = \kappa P_1^\infty \, ,
\end{eqnarray}
where $P_1^\infty$ is the stationary value of the probability $P_1(t)$,
can be easily obtained from~\eqref{eq:rate2} and it is given by
\begin{eqnarray}
  I_s^{(inc)} = \frac{\kappa T_L}{2T_L + \gamma +
    \kappa}~.
\label{sseq}
\end{eqnarray}
Interestingly, the result in Eq.~\eqref{sseq} coincides with the
exact quantum result given in Eq.~\eqref{eq:jsite} for any value of the parameters. 
Note that in general (when the number of sites is larger than one) a set of effective rate equations like Eqs.~\eqref{eq:rate2} gives different results from a quantum master equation like Eq.~\eqref{eq:site}.

Now, having computed the stationary current, we analyze the stationary population of the excited state $\ket{1}$. Our interest is to determine when the single-excitation assumption (under which the master equation~\eqref{eq:mas} has been derived) is valid, as a function of the parameters. Let us start by considering the resonance case, $\Delta_0=0$ so that the stationary solution of Eq.~\eqref{eq:site} is
\begin{eqnarray}
  \label{rho11site}
  \rho_{11}^\infty = \frac{\Omega_R^2}{2\Omega_R^2 + 2\Gamma_T(\gamma + \kappa)} \, .
\end{eqnarray}
Let us recall that the effective Hamiltonian~\eqref{heff} and the master equation~\eqref{eq:mas} have meaning only for low excitation ($\rho_{11} \ll 1$). In our simulations
we have $\Omega_R=4.68\gamma$,
$(\gamma+\kappa)>\gamma$
and $2\Gamma_T>\gamma^{(p)}(0)\approx 2\times 10^6\gamma$. This implies that $\rho_{11}^\infty \lesssim 10^{-5}$, so that we are always in the single-excitation regime.

\section{Natural sunlight}
\label{sec:sun}

In this section we derive the master equation for a generic molecular aggregate coupled to natural sunlight.

Let us consider an aggregate of $N$ two-level systems all having the same excitation energy $\omega_0$. In these calculations we use $\hbar=1$. The aggregate interacts with the radiation emitted by the Sun, that we model as a black body at temperature $T_S$, with a correction accounting for the Sun-to-Earth distance. In this approach, the spontaneous emission process will come out naturally from the interaction with the vacuum mode of the electromagnetic field. The full Hamiltonian is
\begin{eqnarray}
  \hat{H} = \hat{H}_0 + \hat{H}_B + \hat{H}_I \, .
\end{eqnarray}
Here the site Hamiltonian is
\begin{eqnarray}
\label{site}
  \hat{H}_0 = \frac{\omega_0}{2} \sum_{j=1}^N \hat{\sigma}_j^z
\end{eqnarray}
with $\hat{\sigma}_j^z$ being the $z$ Pauli matrix for the $j$-th site. The black body Hamiltonian is
\begin{eqnarray}
  \hat{H}_B = \sum_{\vec{k},\lambda} \omega_k \hat{b}_{\vec{k},\lambda}^\dag \hat{b}_{\vec{k},\lambda}
\end{eqnarray}
where the summation runs over the modes $\vec{k}$ and the polarizations $\lambda=1,2$ of the field, the dispersion relation is $\omega_k = ck$ ($c$ is the speed of light) and the creation/annihilation operators follow to the commutation rules $[\hat{b}_{\vec{k},\lambda}, \hat{b}_{\vec{k}',\lambda'}^\dag] = \delta_{\vec{k},\vec{k}'} \delta_{\lambda,\lambda'}$. Finally, the light-matter interaction Hamiltonian is
\begin{eqnarray}
\label{hint0}
  \hat{H}_I = -\sum_{j=1}^N \hat{\vec{D}}_j \cdot \hat{\vec{E}}(\vec{r}_j)
\end{eqnarray}
where
\begin{eqnarray}
\label{Dhint0}
  \hat{\vec{D}}_j = \vec{d}_j ( \hat{\sigma}_j^+ + \hat{\sigma}_j^- )
\end{eqnarray}
is the dipole operator on the $j$-th site, $\vec{d}_j$ is the transition dipole moment of the same site, $\hat{\sigma}_j^\pm = (\hat{\sigma}_j^x \pm i\hat{\sigma}_j^y)/2$ and
\begin{eqnarray}
  \hat{\vec{E}}(\vec{r}_j) = i\sum_{\vec{k},\lambda} \sqrt{\frac{2\pi\omega_k}{V}} \vec{e}_{\vec{k},\lambda} \left[ e^{i\vec{k}\cdot \vec{r}_j} \hat{b}_{\vec{k},\lambda} - e^{-i\vec{k}\cdot \vec{r}_j} \hat{b}_{\vec{k},\lambda}^\dag \right]
\end{eqnarray}
is the electric field in the position $\vec{r}_j$, with $\vec{e}_{\vec{k},\lambda}$ being a unit vector which specifies the polarization.

Since the coupling to the EMF degrees of freedom is weak, we perform the Born-Markov and secular approximations~\cite{petruccione}. The density matrix is therefore factorized as $\hat{\rho}(t) \approx \hat{\rho}_S(t) \otimes \hat{\rho}_B$, and we get the Lindblad master equation
\begin{eqnarray}
\fl \quad \quad
  \frac{d\hat{\rho}_S(t)}{dt} = &\sum_{\omega=\pm \omega_0} \sum_{i,j} G_{ij}(\omega) \left[ \hat{A}_j(\omega) \hat{\rho}_S(t) \hat{A}_i^\dag(\omega) - \hat{A}_i^\dag(\omega) \hat{A}_j(\omega) \hat{\rho}_S(t) \right] + \text{h.c.}~,
\end{eqnarray}
where $\hat{A}_j(\omega)$ are operators acting on the sites,
\begin{eqnarray}
  \label{Asigma}
  \hat{A}_j(\omega_0) = \hat{\sigma}_j^- \, , \qquad \hat{A}_j(-\omega_0) = \hat{A}_j^\dag(\omega_0) = \hat{\sigma}_j^+~.
\end{eqnarray}
The complex rates $G_{ij}(\omega)$ are
  \begin{eqnarray}
    \label{gije}
    \fl \quad
    G_{ij}(\omega) = &\int_0^\infty d\tau \, e^{i\omega\tau}   \sum_{\vec{k},\lambda,\vec{k}',\lambda'} \frac{2\pi\sqrt{\omega_k\omega_{k'}}}{V} \left( \vec{d}_i \cdot \vec{e}_{\vec{k},\lambda}\right) \left( \vec{d}_j \cdot \vec{e}_{\vec{k'},\lambda'}\right) \nonumber \\
    &\times \left[ - e^{i(\vec{k}\cdot \vec{r}_i + \vec{k}'\cdot \vec{r}_j-\omega_k\tau)} \left\langle \hat{b}_{\vec{k},\lambda} \hat{b}_{\vec{k}',\lambda'}\right\rangle_B 
    + e^{i(\vec{k}\cdot \vec{r}_i - \vec{k}'\cdot \vec{r}_j-\omega_k\tau)} \left\langle \hat{b}_{\vec{k},\lambda} \hat{b}_{\vec{k}',\lambda'}^\dag\right\rangle_B \right. \nonumber \\
    &\left.\quad + e^{-i(\vec{k}\cdot \vec{r}_i - \vec{k}'\cdot \vec{r}_j-\omega_k\tau)} \left\langle \hat{b}_{\vec{k},\lambda}^\dag \hat{b}_{\vec{k}',\lambda'}\right\rangle_B - e^{-i(\vec{k}\cdot \vec{r}_i + \vec{k}'\cdot \vec{r}_j-\omega_k\tau)} \left\langle \hat{b}_{\vec{k},\lambda}^\dag \hat{b}_{\vec{k}',\lambda'}^\dag\right\rangle_B \right]~,
  \end{eqnarray}
where we use the notation $\braket{\dots}_B={\rm tr}_B\left\{ \dots \hat{\rho}_B\right\}$.
Now we assume that the black body, that in our case is the Sun, is at thermal equilibrium, i.e.
\begin{eqnarray}
  \hat{\rho}_B = \frac{e^{-\hat{H}_B/(k_BT_S)}}{\text{tr}_B\left\{ e^{-\hat{H}_B/(k_BT_S)} \right\}}
\end{eqnarray}
where $T_S=6000$~K is the temperature of the Sun. In this case the expectation values of the operators in Eq.~\eqref{gije} are
\begin{subequations}
  \begin{eqnarray}
    \left\langle \hat{b}_{\vec{k},\lambda} \hat{b}_{\vec{k}',\lambda'}\right\rangle_B &= 0 \\
    \left\langle \hat{b}_{\vec{k},\lambda}^\dag \hat{b}_{\vec{k}',\lambda'}^\dag\right\rangle_B &= 0 \\
    \left\langle \hat{b}_{\vec{k},\lambda} \hat{b}_{\vec{k}',\lambda'}^\dag\right\rangle_B &= \delta_{\vec{k},\vec{k}'} \delta_{\lambda,\lambda'} \left( 1 + f_Sn_S(\omega_k)\right) \\
    \left\langle \hat{b}_{\vec{k},\lambda}^\dag \hat{b}_{\vec{k}',\lambda'}\right\rangle_B &= \delta_{\vec{k},\vec{k}'} \delta_{\lambda,\lambda'} f_Sn_S(\omega_k)
  \end{eqnarray}
\end{subequations}
where we have defined the Bose-Einstein occupation of the Sun photons
\begin{eqnarray}
  n_S(\omega_k) = \frac{1}{e^{\omega_k/(k_BT_S)}-1}
\end{eqnarray}
and we have introduced the factor
\begin{eqnarray}
    f_S = \frac{\pi r_S^2}{4\pi R_{ES}^2} = 5.4\times 10^{-6}
\end{eqnarray}
that represents the ratio of the Sun solid angle as seen from the Earth over the full solid angle ($r_S$ is the Sun radius while $R_{ES}$ is the Sun-to-Earth distance). Such factor is needed to correct the model to describe natural sunlight: in the present calculations, in fact, the molecular aggregates exchange energy with all the modes of the EMF in all directions, both absorbing and emitting photons. This is correct for the spontaneous emission process, but it needs to be corrected by the factor $f_S$ for the absorption and stimulated emission processes, which depend on the number of thermal photons $n_S(\omega_k)$. Indeed, the Sun photons hit the system from a fraction $f_S$ of the whole solid angle, so that $n_S$ is multiplied by $f_S$.

Thus, defining $\vec{r}_{ij}=\vec{r}_i-\vec{r}_j$, Eq.~\eqref{gije} can be written as
\begin{eqnarray}
  G_{ij}(\omega) = &\int_0^\infty d\tau \, e^{i\omega\tau}   \sum_{\vec{k},\lambda} \frac{2\pi\omega_k}{V} \left( \vec{d}_i \cdot \vec{e}_{\vec{k},\lambda}\right) \left( \vec{d}_j \cdot \vec{e}_{\vec{k},\lambda}\right) \nonumber \\
  &\times\left[ e^{i(\vec{k}\cdot \vec{r}_{ij}-\omega_k\tau)} \left( 1 + f_Sn_S(\omega_k) \right) + e^{-i(\vec{k}\cdot \vec{r}_{ij} -\omega_k\tau)} f_Sn_S(\omega_k)\right].
\end{eqnarray}
As regards the sum over $\vec{k}$, we take the continuum limit
\begin{eqnarray}
  \frac{1}{V} \sum_{\vec{k}} \rightarrow \frac{1}{(2\pi)^3} \int d\vec{k} = \frac{1}{(2\pi c)^3} \int d\Omega \int_0^\infty d\omega_k \, \omega_k^2 \, .
\end{eqnarray}
Now, if we assume that the dipoles have all the same magnitude $\mu$ but different orientation, namely $\vec{d}_j = \mu \, \hat{p}_j$, and defining the function
\begin{eqnarray}
  \label{fij}
    F_{ij}(x) = \frac{1}{4\pi} \sum_\lambda \int_0^{2\pi} d\phi \int_{-1}^{1}d(\cos\theta)   \left( \hat{p}_i \cdot \vec{e}_{\vec{k},\lambda}\right) \left( \hat{p}_j \cdot \vec{e}_{\vec{k},\lambda}\right) e^{ix\cos\theta}
\end{eqnarray}
choosing a frame for the integration over $\vec{k}$ where the $z$ axis has the same direction as $\vec{r}_{ij}$, we have
\begin{eqnarray}
  G_{ij}(\omega) = \int_0^\infty d\tau \, e^{i\omega\tau} \int_0^\infty d\omega_k \frac{\mu^2\omega_k^3}{\pi c^3} &\left[ e^{-i\omega_k\tau} F_{ij}(k r_{ij}) \left( 1 + f_Sn_S(\omega_k) \right) \right. \nonumber \\
  &\left. + e^{i\omega_k\tau} F_{ij}(-k r_{ij}) f_Sn_S(\omega_k) \right].
\end{eqnarray}
Now we perform the integral over $\tau$ using the relation
\begin{eqnarray}
  \int_0^\infty d\tau \, e^{i\omega\tau} = \pi \delta(\omega) + i \text{P} \frac{1}{\omega}
\end{eqnarray}
where P is the Cauchy principal value. So, we can split the rates into their real and an imaginary parts,
\begin{eqnarray}
  G_{ij}(\omega) = \frac{1}{2}\Gamma_{ij}(\omega) + i S_{ij}(\omega)
\end{eqnarray}
which are, respectively,
\begin{eqnarray}
  \label{real}
  \fl \qquad
  \Gamma_{ij}(\omega) = &\int_0^\infty d\omega_k \frac{2\mu^2\omega_k^3}{c^3} \left[ \delta(\omega-\omega_k) F_{ij}(k r_{ij}) \left( 1 + f_Sn_S(\omega_k) \right) \right. \nonumber \\
                    &\left.\qquad \qquad\qquad+ \delta(\omega+\omega_k) F_{ij}(-k r_{ij}) f_Sn_S(\omega_k) \right] \\
  \label{lamb}
  \fl \qquad
  S_{ij}(\omega) = &\text{P} \int_0^\infty d\omega_k \frac{\mu^2\omega_k^3}{\pi c^3} \left[ \frac{F_{ij}(k r_{ij}) \left( 1 + f_Sn_S(\omega_k) \right)}{\omega-\omega_k} + \frac{F_{ij}(-k r_{ij}) f_Sn_S(\omega_k)}{\omega+\omega_k} \right]
\end{eqnarray}

\subsection{Real Part of the rates: absorption and decay}

Let us start from the real part~\eqref{real}. The two integrals are easily performed, taking into account that the only possible values of $\omega$ are  $\pm \omega_0$. By defining $k_0 = \omega_0/c$ we get
\begin{eqnarray}
\fl \quad \quad
  \Gamma_{ij}(\omega) = &\frac{2\mu^2\omega_0^3}{c^3} \left[ \delta_{\omega,\omega_0} F_{ij}(k_0 r_{ij}) \left( 1 + f_Sn_S(\omega_0) \right) + \delta_{\omega,-\omega_0} F_{ij}(-k_0 r_{ij}) f_Sn_S(\omega_0) \right] \,.
\end{eqnarray}
To have the explicit dependence of $\Gamma_{ij}(\omega)$ on the parameters, we evaluate $F_{ij}(x)$, that results
\begin{eqnarray}
  \label{fijx}
    F_{ij}(x) = &\left[ \frac{\sin x}{x} + \frac{\cos x}{x^2} - \frac{\sin x}{x^3} \right] \left( \hat{p}_i \cdot \hat{p}_j \right) + \nonumber \\
    &+ \left[ -\frac{\sin x}{x} -3 \frac{\cos x}{x^2} + 3\frac{\sin x}{x^3} \right] \left( \hat{p}_i \cdot \hat{r}_{ij} \right) \left( \hat{p}_j \cdot \hat{r}_{ij} \right) \, .
\end{eqnarray}

Note that $F_{ij}(x)$ is an even function of $x$ which, in our case, gives the useful equality $F_{ij}(-k_0r_{ij})=F_{ij}(k_0r_{ij})$. Moreover, one can see that $F_{ji}(x)=F_{ij}(x)$, which implies that both the matrices $\Gamma_{ij}(\omega)$ and $S_{ij}(\omega)$ are symmetric for $i\leftrightarrow j$. As regards the diagonal terms ($i=j$) we can analytically extend the function to $x=0$ thanks to the limit
\begin{eqnarray}
  \lim_{x \to 0} F_{ii}(x) = \frac{2}{3} \, .
\end{eqnarray}

The real parts of the rates are then
\begin{eqnarray}
  \Gamma_{ij}(\omega) = \frac{3\gamma}{2} F_{ij}(k_0 r_{ij}) \left[ \delta_{\omega,\omega_0}  \left( 1 + f_Sn_S(\omega_0) \right) + \delta_{\omega,-\omega_0} f_Sn_S(\omega_0)  \right]
\end{eqnarray}
where the single-molecule spontaneous decay rates are
\begin{eqnarray}
  \gamma = \frac{4}{3}\mu^2 \frac{\omega_0^3}{c^3} \, .
\end{eqnarray}
Using the symmetry properties of $\Gamma_{ij}$, the contribution from the real part of the rates to the master equation reads
\begin{eqnarray}
\label{sunreal}
  \left[\frac{d\hat{\rho}_S(t)}{dt}\right]_{real} = &\sum_{i,j} \Gamma_{ij} \left( 1 + f_Sn_S(\omega_0) \right) \left[ \hat{\sigma}_j^- \hat{\rho}_S(t) \hat{\sigma}_i^+  -\frac{1}{2} \left\{ \hat{\sigma}_i^+ \hat{\sigma}_j^-, \hat{\rho}_S(t) \right\} \right] \nonumber \\
  &+\sum_{i,j} \Gamma_{ij} f_Sn_S(\omega_0)  \left[ \hat{\sigma}_j^+ \hat{\rho}_S(t) \hat{\sigma}_i^-  -\frac{1}{2} \left\{ \hat{\sigma}_i^- \hat{\sigma}_j^+, \hat{\rho}_S(t) \right\} \right]
\end{eqnarray}
which is in the Lindblad form and where we have defined the coefficients
\begin{eqnarray}
  \Gamma_{ij} = \frac{\Gamma_{ij}(\omega_0)}{1 + f_Sn_S(\omega_0)} = \frac{\Gamma_{ij}(-\omega_0)}{f_Sn_S(\omega_0)} = \frac{3}{2}\gamma F_{ij}(k_0r_{ij}) \, .
\end{eqnarray}
The first double sum of Eq.~\eqref{sunreal} describes the spontaneous and stimulated emission processes, while the second double sum describes the absorption process of excitation from the Sun.

\subsection{Imaginary Part of the rates: radiative coupling}

Let us now focus on the imaginary part of the master equation. Thanks to the symmetry $S_{ji}(\omega)=S_{ij}(\omega)$ we have
\begin{eqnarray}
\fl \quad
  \left[\frac{d\hat{\rho}_S(t)}{dt}\right]_{imag} &= \sum_{\omega=\pm \omega_0} \sum_{i,j} iS_{ij}(\omega) \left[ \hat{A}_j(\omega) \hat{\rho}_S(t) \hat{A}_i^\dag(\omega)  - \hat{A}_i^\dag(\omega) \hat{A}_j(\omega) \hat{\rho}_S(t) \right] + \text{h.c.} \nonumber \\
  &= -i\sum_{\omega=\pm \omega_0} \sum_{i,j} S_{ij}(\omega) \left[ \hat{A}^\dag_i(\omega) \hat{A}_j(\omega),\hat{\rho}_S(t) \right] \nonumber \\
  &= - i\left[ \hat{\Delta}, \hat{\rho}_S(t) \right] \, ,
\end{eqnarray}
where we have defined the real part of the radiative Hamiltonian using~\eqref{Asigma}
\begin{eqnarray}
  \hat{\Delta} &= \sum_{\omega=\pm \omega_0} \sum_{i,j} S_{ij}(\omega) \hat{A}^\dag_i(\omega) \hat{A}_j(\omega) \nonumber \\
               &= \sum_{i,j} \left[ S_{ij}(\omega_0) \hat{\sigma}^+_i \hat{\sigma}_j^- + S_{ij}(-\omega_0) \hat{\sigma}^-_j \hat{\sigma}_i^+\right] \, .
\end{eqnarray}

Thanks to the commutation rules $\left[ \hat{\sigma}_i^+,\hat{\sigma}_j^- \right] = \delta_{ij}\hat{\sigma}_z$ we have
\begin{eqnarray}
  \label{hlsii}
  \hat{\Delta} = &\sum_i \left[ S_{ii}(\omega_0) \hat{\sigma}^+_i \hat{\sigma}_i^- + S_{ii}(-\omega_0) \hat{\sigma}^-_i \hat{\sigma}_i^+\right] +  \sum_{\substack{i,j \\ i \ne j}} \Delta_{ij} \hat{\sigma}^+_i \hat{\sigma}_j^-
\end{eqnarray}
where we have defined the matrix elements
\begin{eqnarray}
  \Delta_{ij} = S_{ij}(\omega_0) + S_{ij}(-\omega_0) \, .
\end{eqnarray}
The diagonal terms give a divergent renormalization energy that is constant for all the molecules, so we disregard it here. Then, the off-diagonal matrix elements are independent of the Sun temperature, namely
\begin{eqnarray}
  \Delta_{ij} = \frac{\mu^2}{\pi r_{ij}^3} \text{P} \int_{-\infty}^\infty dx \frac{x^3}{x_0-x}F_{ij}(x)~,
\end{eqnarray}
where $x_0=\omega_0r_{ij}/c$. The integral can be computed using contour methods, resulting in
\begin{eqnarray}
  \label{deltaij}
    \Delta_{ij} = &\frac{3\gamma}{4} \left[ -\frac{\cos x_{ij}}{k_0r_{ij}} + \frac{\sin x_{ij}}{x_{ij}^2} + \frac{\cos x_{ij}}{x_{ij}^3} \right] \left( \hat{p}_i \cdot \hat{p}_j \right) \nonumber \\
    &+\frac{3\gamma}{4} \left[ \frac{\cos x_{ij}}{k_0r_{ij}} -3 \frac{\sin x_{ij}}{x_{ij}^2} -3 \frac{\cos x_{ij}}{x_{ij}^3} \right] \left( \hat{p}_i \cdot \hat{r}_{ij} \right)\left( \hat{p}_j \cdot \hat{r}_{ij} \right) \, .
\end{eqnarray}

\subsection{Final Expression and Single-excitation approximation}

The final expression of the master equation is
\begin{eqnarray}
  \label{final}
  \fl \qquad
  \frac{d\hat{\rho}_S}{dt} = &-i \left[ \hat{H}_0 + \hat{\Delta}, \hat{\rho}_S \right] +\sum_{i,j} \Gamma_{ij} \left( 1 + f_Sn_S(\omega_0) \right) \left[ \hat{\sigma}_j^- \hat{\rho}_S \hat{\sigma}_i^+  -\frac{1}{2} \left\{ \hat{\sigma}_i^+ \hat{\sigma}_j^-, \hat{\rho}_S \right\} \right] \nonumber \\
  &+\sum_{i,j} \Gamma_{ij} f_Sn_S(\omega_0)  \left[ \hat{\sigma}_j^+ \hat{\rho}_S \hat{\sigma}_i^-  -\frac{1}{2} \left\{ \hat{\sigma}_i^- \hat{\sigma}_j^+, \hat{\rho}_S \right\} \right]
\end{eqnarray}
and, defining the parameters $x_{ij}=\omega_0r_{ij}/c$ and $\gamma=\frac{4}{3}\mu^2\frac{\omega_0^3}{c^3}$ we have
\begin{eqnarray}
    \Gamma_{ij} = &\frac{3\gamma}{2} \left[ \frac{\sin x_{ij}}{x_{ij}} + \frac{\cos x_{ij}}{x_{ij}^2} - \frac{\sin x_{ij}}{x_{ij}^3} \right] \left( \hat{p}_i \cdot \hat{p}_j \right) + \nonumber \\
    &+ \left[ -\frac{\sin x_{ij}}{x_{ij}} -3 \frac{\cos x_{ij}}{x_{ij}^2} + 3\frac{\sin x_{ij}}{x_{ij}^3} \right] \left( \hat{p}_i \cdot \hat{r}_{ij} \right) \left( \hat{p}_j \cdot \hat{r}_{ij} \right) \, .
\end{eqnarray}
As regards the Hamiltonian term, neglecting the renormalization of the site energies, we have
\begin{eqnarray}
  \hat{H}_0 + \hat{\Delta} = \frac{{\omega_0}}{2} \sum_i \hat{\sigma}_i^z + \sum_{\substack{i,j \\ i \ne j}} \Delta_{ij} \hat{\sigma}_i^+ \hat{\sigma}_j^-
\end{eqnarray}
where the coupling terms are given by~\eqref{deltaij}.

The master equation \eqref{final} acts on the full Hilbert space spanned by the $N$ sites, having dimension $2^N$ and including all the possible numbers of excitations (from none to $N$ excitations). However, in this manuscript we focus on the weak fluence regime, where the photon absorption rate is much smaller than the excitation decay rate, so that no more than one excitation at a time is present in the system. In this regime we perform the \emph{single-excitation approximation}, \textit{i.e.}
we neglect all the states with more than one excitation and we consider only: the state $\ket{0}$, where all the sites are in their ground state, and the $N$ single-excitation states of the form $\ket{j} = \hat{\sigma}_j^+\ket{0}$, where only the $j$th site is excited while all the other ones are in their ground state. In this $(N+1)$-dimensional subspace, each $\sigma_j^+$ operator acts only on $\ket{0}$ resulting in $\sigma_j^+\ket{0}=\ket{j}$, while each $\sigma_j^-$ operator acts only on $\ket{j}$ giving $\sigma_j^-\ket{j}=\ket{0}$.
Therefore, we write the master equation in the single-excitation approximation by replacing $\hat{\sigma}_j^+ \to \ket{j}\bra{0}$ and $\hat{\sigma}_j^- \to \ket{0}\bra{j}$ into~\eqref{final}. For readability, we also drop the subscript ``S'' from $\hat{\rho}_S$, and we have the single-excitation master equation
\begin{eqnarray}
  \label{single}
  \fl \qquad \qquad
  \frac{d\hat{\rho}}{dt} = &-i \left[ \hat{H}_0 + \hat{\Delta}, \hat{\rho} \right] +\sum_{i,j} \Gamma_{ij} \left( 1 + f_Sn_S(\omega_0) \right) \left[ \rho_{ji} \ket{0}\bra{0}   -\frac{1}{2} \left\{ \ket{i}\bra{j}, \hat{\rho} \right\} \right] \nonumber \\
  &+\sum_{i,j} \Gamma_{ij} f_Sn_S(\omega_0)  \left[ \rho_{00} \ket{j}\bra{i}  -\frac{1}{2} \delta_{ij} \left\{ \ket{0}\bra{0}, \hat{\rho} \right\} \right]
\end{eqnarray}
with the single-excitation Hamiltonian
\begin{eqnarray}
  \hat{H}_0 + \hat{\Delta} = {\omega_0} \sum_i \ket{i}\bra{i} + \sum_{\substack{i,j \\ i \ne j}} \Delta_{ij} \ket{i}\bra{j} \, .
\end{eqnarray}

Finally, let us consider the particular case where there exist a common eigenbasis $\ket{\alpha}$ for both $\left( \hat{H}_0 + \hat{\Delta} \right)$ and $\sum_{i,j}\Gamma_{ij}\ket{i}\bra{j}$ such that
\begin{eqnarray}
  \bra{\alpha} \hat{H}_0 + \hat{\Delta} \ket{\beta} &= \omega_\alpha \delta_{\alpha \beta} \\
  \bra{\alpha} \left( \sum_{i,j} \Gamma_{ij}\ket{i}\bra{j} \right) \ket{\beta} &= \gamma_\alpha \delta_{\alpha \beta} \, .
\end{eqnarray}
We can then write~\eqref{single} in that basis:
\begin{eqnarray}
  \frac{d\hat{\rho}}{dt} = &-i \sum_\alpha \omega_\alpha \left[ \ket{\alpha}\bra{\alpha}, \hat{\rho} \right] \nonumber \\
  &+\sum_{\alpha} \gamma_{\alpha} \left( 1 + f_Sn_S(\omega_0) \right) \left[ \rho_{\alpha\alpha} \ket{0}\bra{0}   -\frac{1}{2} \left\{ \ket{\alpha}\bra{\alpha}, \hat{\rho} \right\} \right] \nonumber \\
  &+\sum_{\alpha} \gamma_{\alpha} f_Sn_S(\omega_0)  \left[ \rho_{00} \ket{\alpha}\bra{\alpha}  -\frac{1}{2} \left\{ \ket{0}\bra{0}, \hat{\rho} \right\} \right] \, .
\end{eqnarray}
If we consider the diagonal elements, which describe the dynamics of the populations of $\ket{0}$ and of the eigenstates $\ket{\alpha}$, that part of the master equation can be mapped into a Pauli master equation, which reads
\begin{eqnarray}
  \frac{d\rho_{00}}{dt} &= \sum_\alpha B_\alpha\left( \rho_{\alpha\alpha} - \rho_{00} \right) + \sum_\alpha \gamma_\alpha \rho_{\alpha\alpha} \\
  \frac{d\rho_{\alpha\alpha}}{dt} &= B_\alpha\left( \rho_{00} - \rho_{\alpha\alpha} \right) - \gamma_\alpha \rho_{\alpha\alpha}
\end{eqnarray}
where we have defined the absorption and stimulated emission rates $B_\alpha = f_Sn_S(\omega_0)\gamma_\alpha$.

\section{Effective three-level model}
\label{sec:3lev}

Here we provide a detailed derivation of the three-level model introduced in the main text.

Let us consider a system made of $N$ sites (molecules) and let us perform our analysis in the single-excitation approximation, meaning that we consider just the ground state of the whole system $\ket{0}$ and the states $\ket{j}$ where only the $j$-th site is excited. Let us now consider a basis where the single-excitation subspace is diagonal, and let us call $\ket{\alpha}$ the eigenstates of the single-excitation subspace.

The system is assumed to be coupled to an incoming radiation field, which induces absorption and stimulated emission to each eigenstate $\ket{\alpha}$ with a pumping rate $B_\alpha$. Moreover, we consider the spontaneous emission of excitation by radiation from each eigenstate $\ket{\alpha}$, with a fluorescence rate $\gamma_\alpha$. Finally, we add one level (labelled ``RC'') which is coupled to each $\ket{\alpha}$ state with a transfer rate $T_\alpha^{RC}$ and where excitation can be collected to a trapping environment (sink), modeled by a trapping rate $\kappa$.
The RC can also absorb excitation from the radiation field with a rate $B_{RC}$, and it also has stimulated and spontaneous emission rates, $B_{RC}$ and $\gamma$.

\paragraph{Neglect coherences}
As a first approximation, we assume that the coherences between the eigenstates do not play a role in the transport process. We then write a rate equation for the population of the ground state and for the RC state:
\begin{subequations}
  \label{rate}
  \begin{eqnarray}
    \frac{dP_0(t)}{dt} = &- \sum_\alpha B_\alpha P_0(t) -B_{RC} P_0(t) + \sum_\alpha B_\alpha P_\alpha(t) \nonumber \\
                 &+ B_{RC} P_{rc}(t) + \sum_\alpha \gamma_\alpha P_\alpha(t) + (\kappa+\gamma) P_{rc}(t) \\
    \frac{dP_{rc}(t)}{dt} = &- \sum_\alpha T_\alpha^{RC} P_{rc}(t) + \sum_\alpha T_\alpha^{RC} P_\alpha(t) \nonumber \\
                 &+B_{RC}P_0(t)-B_{RC}P_{rc}(t)- (\kappa+\gamma) P_{rc}(t) \, ,  
  \end{eqnarray}
\end{subequations}
where $P_0$ is the population of the ground state, $P_{rc}$ is the population of the RC and $P_\alpha$ is the population of the $\alpha$-th excitonic eigenstate.

\paragraph{Thermal equilibrium}
Secondly, let us assume that the excitonic subspace is at thermal equilibrium with a temperature $T$. Formally, we define the population in the whole excited subspace as
\begin{eqnarray}
  P_e(t) = \sum_\alpha P_\alpha(t) \, ,
\end{eqnarray}
so that the trace preservation condition is
\begin{eqnarray}
  \label{trace}
  P_0(t) + P_e(t) + P_{rc}(t) = 1~.
\end{eqnarray}
We impose thermal equilibrium in the aggregate as
\begin{eqnarray}
  \label{therm}
  P_\alpha(t) = P_e(t) p_\alpha \qquad \text{with} \qquad p_\alpha = \frac{e^{-E_\alpha/(k_BT)}}{Z}
\end{eqnarray}
where $E_\alpha$ is the energy of the $\alpha$-th excitonic eigenstate and
\begin{eqnarray}
\label{partZ}
    Z=\sum_\alpha e^{-E_\alpha/(k_BT)}
\end{eqnarray}
is the partition function.
By substituting Eq.~\eqref{therm} into Eq.~\eqref{rate} we have
\begin{subequations}  \label{rate2}
  \begin{eqnarray}
  \frac{dP_0(t)}{dt} = &- \left(\sum_\alpha B_\alpha\right) P_0(t) + \left(\sum_\alpha B_\alpha p_\alpha \right) P_e(t) + \left(\sum_\alpha \gamma_\alpha p_\alpha \right) P_e(t) \nonumber \\
  &+ (\kappa+\gamma) P_{rc}(t) -B_{RC}P_0(t)+B_{RC}P_{rc}(t) \\
  \frac{dP_{rc}(t)}{dt} = &- \left(\sum_\alpha T_\alpha^{RC}\right) P_{rc}(t) + \left(\sum_\alpha T_\alpha^{RC} p_\alpha \right) P_e(t) \nonumber \\
                 &+B_{RC}P_0(t)-B_{RC}P_{rc}(t)- (\kappa+\gamma) P_{rc}(t)
\end{eqnarray}
\end{subequations}

\paragraph{Steady-state solution and current}
We now express Eq.~\eqref{rate2} in terms of thermal averages of the rates ($\braket{X}=\sum_\alpha X_\alpha p_\alpha$) and by defining the total absorption rate, $B_{TOT} = \sum_\alpha B_\alpha$, and the total transfer rate to the RC, $T_{TOT}^{RC} = \sum_\alpha T_\alpha^{RC}$, so that
\begin{subequations}
\label{3levrate}
  \begin{eqnarray}
  \frac{dP_0(t)}{dt} = &- B_{TOT} P_0(t) + \left\langle B \right\rangle P_e(t) - B_{RC}P_0(t) \nonumber \\
                 &+B_{RC}P_{rc}(t)+ \left\langle \gamma \right\rangle P_e(t) + (\kappa+\gamma) P_{rc}(t) \\
  \frac{dP_{rc}(t)}{dt} = &- T_{TOT}^{RC} P_{rc}(t) + \left\langle T^{RC} \right\rangle P_e(t) +B_{RC}P_0(t) \nonumber \\
                 &-B_{RC}P_{rc}(t) - (\kappa+\gamma) P_{rc}(t) \\
  1 = &P_0(t) + P_e(t) + P_{rc}(t)~,
\end{eqnarray}
\end{subequations}
where the last equation is Eq.~\eqref{trace}. The steady-state solution is obtained by setting the time derivatives to zero, and it is reached at long time (``$t=\infty$''), so that we have
\begin{eqnarray}
  \label{pe}
  \fl
  P_{rc}(\infty) = &\cfrac{ B_{TOT} + B_{RC} }{B_{TOT} + \gamma + \kappa + 2B_{RC} + \left( B_{TOT} + B_{RC} + \left\langle B \right\rangle + \left\langle \gamma \right\rangle \right) \frac{ T_{TOT}^{RC} + B_{TOT}\frac{B_{RC} + \gamma + \kappa }{B_{TOT} + B_{RC}} }{\left\langle T^{RC} \right\rangle  + B_{RC} \frac{ \left\langle B \right\rangle + \left\langle \gamma \right\rangle}{B_{TOT} + B_{RC}}} } \\
  \fl
  P_e(\infty) = &\cfrac{ T_{TOT}^{RC} + B_{TOT}\cfrac{B_{RC} + \gamma + \kappa }{B_{TOT} + B_{RC}} }{\left\langle T^{RC} \right\rangle  + B_{RC} \cfrac{ \left\langle B \right\rangle + \left\langle \gamma \right\rangle}{B_{TOT} + B_{RC}}} P_{rc}(\infty)
\end{eqnarray}
Finally, we define the stationary current trapped into the sink as
\begin{eqnarray}
  \label{cur}
  I_3 = \kappa P_{rc}(\infty)~.
\end{eqnarray}
By substituting Eq.~\eqref{pe} into Eq.~\eqref{cur} we have
\begin{eqnarray}
  \label{curr}
  \fl
  I_3 = \cfrac{ \kappa \left( B_{TOT} + B_{RC} \right) }{B_{TOT} + \gamma + \kappa + 2B_{RC} + \left( B_{TOT} + B_{RC} + \left\langle B \right\rangle + \left\langle \gamma \right\rangle \right) \frac{ T_{TOT}^{RC} + B_{TOT}\frac{B_{RC} + \gamma + \kappa }{B_{TOT} + B_{RC}} }{\left\langle T^{RC} \right\rangle  + B_{RC} \frac{ \left\langle B \right\rangle + \left\langle \gamma \right\rangle}{B_{TOT} + B_{RC}}} }~.
\end{eqnarray}

\subsection{Equivalence with the multi-chromophoric F\"orster resonance energy transfer (MC-FRET)}

Here we show that the transfer rates $\braket{T^{RC}}$ and $T^{RC}_{TOT}$ between ring and RC are exactly the multi-chromophoric F\"orster resonance energy transfer (MC-FRET) rates~\cite{silbeyPRL2004,caoNJP2013}, also known as generalized F\"orster theory~\cite{scholesARPC2003}. In the MC-FRET framework, the transfer rate from a donor aggregate ``D'' to an acceptor ``A'' is expressed in terms of their emission and absorption spectra, $E(\omega)$ and $I(\omega)$ respectively. An aggregate absorption spectrum is~\cite{caoNJP2013} $I(\omega) \propto \sum_\alpha |\vec{p}_\alpha|^2 I_\alpha(\omega)$, where $|\vec{p}_\alpha|^2$ and $I_\alpha(\omega)$ are, respectively, the dipole strength and the normalized lineshape for each $\alpha$ eigenstate. The emission spectrum on the other hand is $E(\omega) \propto \sum_\alpha |\vec{p}_\alpha|^2 E_\alpha(\omega)$, where the emission lineshapes are multiplied by the thermal populations $p_\alpha$, see Eqs.~\eqref{therm} and~\eqref{partZ}, namely $E_\alpha(\omega)=p_\alpha I_\alpha(\omega)$. The MC-FRET rate is usually expressed as~\cite{silbeyPRL2004,caoNJP2013,scholesARPC2003}
\begin{eqnarray}
  \label{mcfret}
    K_{D,A} = \sum_{\alpha \in D} \sum_{\beta \in A} \frac{|\braket{\alpha|H_S|\beta}|^2}{2\pi\hbar^2} \int_{-\infty}^\infty E_\alpha(\omega) I_\beta(\omega) d\omega~,
\end{eqnarray}
where $\braket{\alpha|H_S|\beta}$ is the Hamiltonian matrix element between the $\alpha$ donor eigenstate and the $\beta$ acceptor eigenstate and the normalization condition is $\int_{-\infty}^\infty I_\alpha(\omega)d\omega=2\pi$.

For high temperature and short bath correlation time~\cite{caoNJP2013}, we can neglect the phonon-induced Stokes and anti-Stokes shifts and approximate all the absorption lines as Lorentzians
\begin{eqnarray}
  I_\alpha(\omega) = \frac{2\Gamma_\phi}{\Gamma_\phi^2 + (\omega-\omega_\alpha)^2}
\end{eqnarray}
peaked on the eigenstate frequency $\omega_\alpha$ and with a dephasing-induced linewidth $\Gamma_\phi$. Under this assumption, the overlap integral in Eq.~\eqref{mcfret} is analytically computed as
\begin{eqnarray}
  \int_{-\infty}^\infty E_\alpha(\omega) I_\beta(\omega) d\omega = p_\alpha \frac{4\pi\Gamma_\phi}{4\Gamma_\phi^2 + (\omega_\alpha-\omega_\beta)^2}~.
\end{eqnarray}
Therefore, we can express the MC-FRET rate in Eq.~\eqref{mcfret} as
\begin{eqnarray}
\label{mcfret2}
  K_{D,A} = \sum_{\alpha \in D} \sum_{\beta \in A} p_\alpha \bar{K}_{\alpha,\beta}~,
\end{eqnarray}
where the transfer rates between an $\alpha$ donor eigenstate and a $\beta$ acceptor eigenstate are
\begin{eqnarray}
\label{mcfret3}
  \bar{K}_{\alpha,\beta} = \frac{|\braket{\alpha|H_S|\beta}|^2}{\hbar^2} \frac{4\Gamma_\phi}{4\Gamma_\phi^2 + (\omega_\alpha-\omega_\beta)^2}~.
\end{eqnarray}
Note that these rates are symmetric, $\bar{K}_{\beta,\alpha}=\bar{K}_{\alpha,\beta}$, while the $K_{D,A}$ rate usually are non-symmetric, due to the different thermal populations $p_\alpha$ between the donor and acceptor.

The transfer rates between ring and RC from our rate equations, Eqs.~\eqref{3levrate}, are exactly equivalent to the MC-FRET rates in Eq.~\eqref{mcfret2}, in fact we have two cases:
\begin{enumerate}
    \item for the transfer from the ring to the RC, the ring is the donor and the RC the acceptor, so the sum over $\beta\in A$ runs over the single RC state and we have
    \begin{eqnarray}
      K_{ring,RC} = \sum_{\alpha\in ring} p_\alpha \bar{K}_{\alpha,rc}~,
    \end{eqnarray}
    which is exactly $\braket{T^{RC}}$;
    \item for the transfer from the RC to the ring we have the opposite situation, so the sum over the the donor states $\alpha \in D$ runs over the single RC state, whose normalized population is trivially $p_\alpha=1$, and so we have
    \begin{eqnarray}
      K_{RC,ring} = \sum_{\beta\in ring}  \bar{K}_{rc,\beta}~,
    \end{eqnarray}
    which is exactly $T_{TOT}^{RC}$.
\end{enumerate}

\subsection{Transfer between ring and RC}

Since the coupling between the ring and the RC is weak, we compute the incoherent transfer rates between the ring eigenstates and the RC using the MC-FRET rates in Eq.~\eqref{mcfret3}, $T_\alpha^{RC}\propto|\braket{\alpha | \Delta | rc}|^2$, proportional to the squared coupling between the two states. As we show in~\ref{app:Oc}, the coupling is non-vanishing only for the eigenstate $\ket{E_2}$, that is resonant with the RC, where the coupling scales as $\Omega_C\propto (\cos\theta)/N^{5/2}$, see Eq.~\eqref{OC}. Therefore, the transfer rate in Eq.~\eqref{mcfret3} can be written in the form
\begin{eqnarray}
    \label{TRC}
    T_2^{RC} = \tau_{RC}^{-1} \left( \frac{32}{N}\right)^5 \cos^2\theta
\end{eqnarray}
where $\tau_{RC}$ represents the transfer time in the reference case $N=32$ and $\theta=0$ (representing the natural purple bacteria LHI complex, see~\ref{sec:lh}). We have verified numerically that the scaling in Eq.~\eqref{TRC} holds for $N=16,32,64$ and for different values of $\theta$. By comparing Eq.~\eqref{mcfret3} and Eq.~\eqref{TRC} we can find a proportionality between the dephasing $\Gamma_\phi$ in the MC-FRET approach and the reference ring-RC transfer time $\tau_{RC}$ in our approach. We recall that $\ket{E_2}$ and $\ket{rc}$ are at resonance and the matrix element $\Omega_C=\braket{E_2|\Delta|rc}$ is given by Eq.~\eqref{OC}, so that the MC-FRET Eq.~\eqref{mcfret3} is written explicitly as
\begin{eqnarray}
\label{mcfretring}
  \bar{K}_{E_2,rc} = \frac{\Omega_C^2}{\hbar^2\Gamma_\phi} = \frac{32\pi^6\mu^4 d^6}{\hbar^2\Gamma_\phi} \frac{\cos^2\theta}{N^5}~.
\end{eqnarray}
By setting Eq.~\eqref{mcfretring} equal to Eq.~\eqref{TRC} we have that $\cos^2\theta/N^5$ cancels since it is present in both equations, and we obtain the relation
\begin{eqnarray}
  \Gamma_\phi = \frac{\pi^6\mu^4d^6}{32^4\hbar^2}\tau_{RC}~.
\end{eqnarray}
Substituting the parameter values used in this manuscript ($\mu^2=519310$~cm$^{-1}$\AA$^3$ and $d=0.1$~\AA$^{-1}$) we have
\begin{eqnarray}
\label{gphitaurc}
  \Gamma_\phi = (8.8~\text{ps}^{-2}) \tau_{RC}~.
\end{eqnarray}

In order to determine the value of $\tau_{RC}$, now we consider the ring+RC system in absence of absorption, emission and trapping, where the excitation can only be exchanged between the ring and the RC. In such situation, Eq.~\eqref{3levrate} simplifies to
\begin{subequations}
  \begin{eqnarray}
    \frac{dP_e(t)}{dt} = &- \left\langle T^{RC} \right\rangle P_e(t) + T_{TOT}^{RC} P_{rc}(t) \\
    \frac{dP_{rc}(t)}{dt} = &- T_{TOT}^{RC} P_{rc}(t) + \left\langle T^{RC} \right\rangle P_e(t)
  \end{eqnarray}
\end{subequations}
where the two rates are related by $\braket{T^{RC}}=T_{TOT}^{RC}p_*$ (here $p_*$ is the Boltzmann population of the eigenstate $\ket{E_2}$). If one excitation is initially present on the ring, the time evolution of the probability to be on the RC is
\begin{eqnarray}
\label{Prctau}
    P_{rc}(t)=\frac{p_*}{1+p_*} \left[ 1 - e^{-T_{TOT}^{RC}(1+p_*)t} \right]~,
\end{eqnarray}
and it allows to determine the value of $\tau_{RC}$, as explained below.

\begin{figure}[!htb]
    \centering
    \includegraphics[width=0.8\columnwidth]{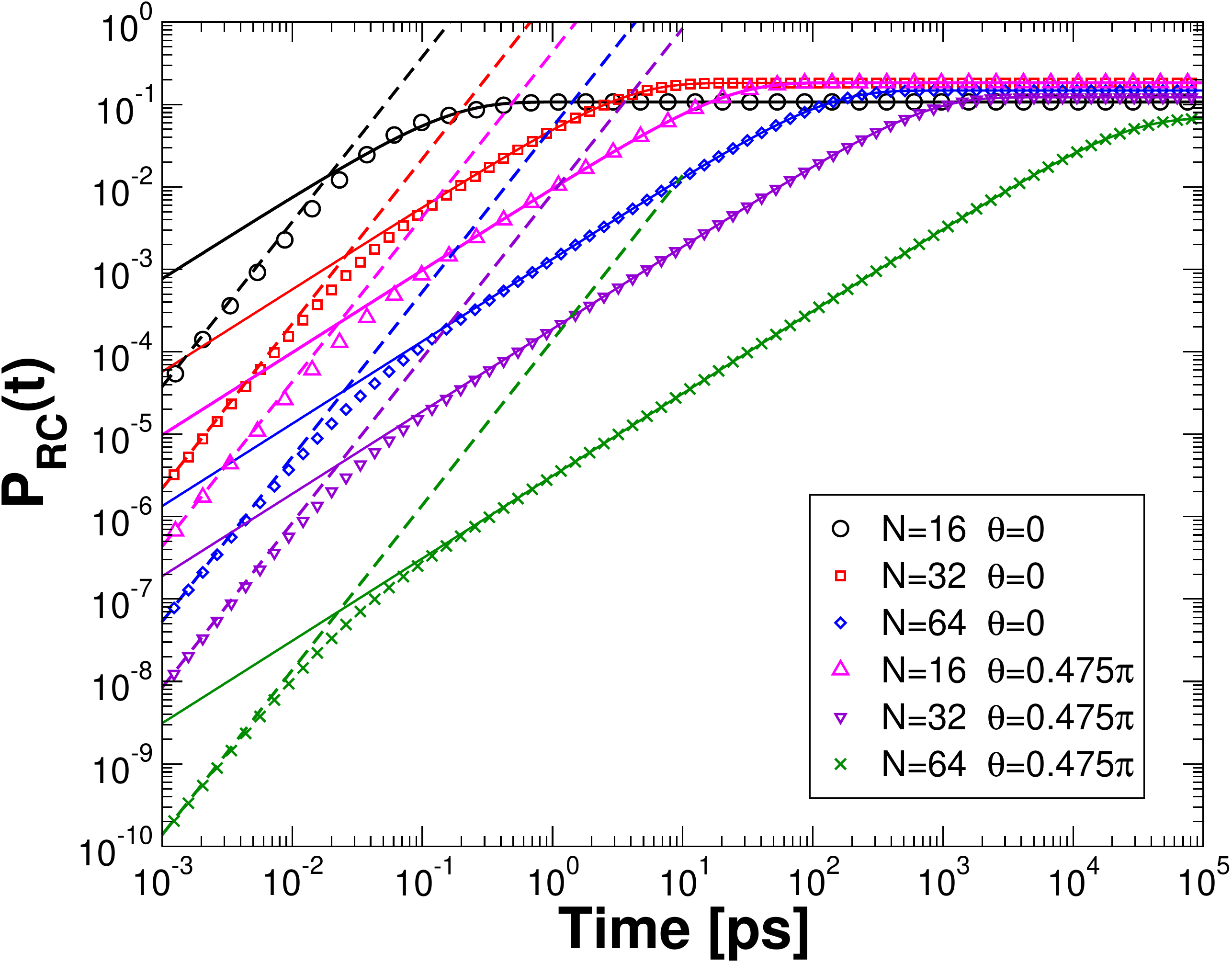}
    \caption{Population of the RC as computed with the master equation, Eq.~\eqref{eq:mas} (symbols) in absence of absorption ($\Omega_R=0$), emission ($\Gamma_{mn}=0$) and trapping ($\kappa=0$). Best fits with the three-level solution, Eq.~\eqref{Prctau}, are shown as continuous lines, with fit parameters in Table~\ref{tab:fit}. The initial quadratic growths $P_{rc}(t)\approx p_*\Omega_C^2t^2/\hbar^2$ [with $\Omega_C$ given by Eq.~\eqref{OC}] are shown as dashed lines. Parameters for the master equation: $E_R=200$~cm$^{-1}$, $\omega_c=333$~cm$^{-1}$, $T=300$~K.
    }
    \label{fig:tauRC}
\end{figure}

\begin{table}[!hbt]
    \centering
    \begin{tabular}{cccc}
        \toprule
        $N$ & $\theta$ & $\tau_{RC}$ [ps] & $p_*$ \\
        \midrule
        16 & 0 & 5.0 & 0.120 \\
        32 & 0 & 3.9 & 0.223 \\
        64 & 0 & 4.0 & 0.173 \\
        16 & $0.475\pi$ & 4.5 & 0.224 \\
        32 & $0.475\pi$ & 4.5 & 0.139 \\
        64 & $0.475\pi$ & 4.5 & 0.073 \\
        \bottomrule
    \end{tabular}
    \caption{Fit parameters for Fig.~\ref{fig:tauRC}, using Eq.~\eqref{Prctau}.}
    \label{tab:fit}
\end{table}

In Fig.~\ref{fig:tauRC} we compute the time evolution of the ring+RC system using the master equation (see~\ref{sec:mas}) in absence of absorption, emission and trapping, and in presence of a thermal bath with the standard parameters used in this work (see caption).
We initialize the system with one excitation at thermal equilibrium on the ring, i.e.
\begin{eqnarray}
    \rho(0)=\sum_\alpha \frac{e^{-E_\alpha/(k_BT)}}{Z} \ket{E_\alpha}\bra{E_\alpha}
\end{eqnarray}
and we compute the time evolution of the population of the RC, $P_{rc}(t)=\braket{rc|\rho(t)|rc}$ (symbols in Fig.~\ref{fig:tauRC}). We fit the results obtained this way with the three-level solution Eq.~\eqref{Prctau}, leaving $\tau_{RC}$ and $p_*$ as fitting parameters. As one can see from the figure, $P_{rc}(t)$ obtained from the master equation (symbols) grows initially quadratically in time (dashed lines) as
\begin{eqnarray}
\label{quad}
  P_{rc}(t) \approx p_*\frac{\Omega_C^2t^2}{\hbar^2}~.
\end{eqnarray}
This quadratic growth is the pure quantum-mechanical time evolution given by the ring $\ket{E_2}$ eigenstate (with initial occupation probability $p_*$) resonant with $\ket{rc}$. For times larger than $\approx 0.01-0.1$~ps, the time evolution is instead well captured by Eq.~\eqref{Prctau} (continuous lines), with the fitting parameters reported in Table~\ref{tab:fit}. We observe that, in the reference case $N=32$, $\theta=0$ (representing the natural LHI system), there is a perfect fit with $\tau_{RC}=3.9$~ps, and therefore we choose this value for all our simulations with the three-level model across the manuscript. The corresponding dephasing value for the MC-FRET rates is obtained from Eq.~\eqref{gphitaurc}, and it has the value $\Gamma_\phi=34$~ps$^{-1}$.
We also note that $\tau_{RC}$ varies by about $20\%$ when $\theta$ is changed (see Table~\ref{tab:fit}). Such $\pm 20\%$ variation in $\tau_{RC}$ produces variations in the steady-state current, and we show those variations as shaded areas in Figs.~\ref{fig:tilt}, \ref{fig:theta} and \ref{fig:kappasmall}.
Finally, we determine the critical time $\tau_\phi$ when the dynamics changes from quadratic quantum-mechanical growth to the linear growth predicted by the incoherent three-level model. This transition happens at the crossing point between Eq.~\eqref{quad} and the initial, linear growth of Eq.~\eqref{Prctau}, that for short times is approximated as $P_{rc}(t)\approx p_*T_{RC}^{TOT}t$. Matching the linear and quadratic expressions for $P_{rc}(t)$ and using Eq.~\eqref{mcfretring} for $T_{RC}^{TOT}$, we have that $\Omega_C^2$ cancels, so that the crossing point happens at $\tau_\phi=\Gamma_\phi^{-1}$. With the choice of $\tau_{RC}$ and the corresponding $\Gamma_\phi$ used in this manuscript, we have $\tau_\phi=0.03$~ps, independent of $N$ or $\theta$, as one can see in Fig.~\ref{fig:tauRC}.

\subsection{Parametrization for D-configuration, LH-configuration and Sunlight configuration}

Here we express the relevant rates present in Eqs.~\eqref{3levrate} as a function of the system parameters, in absence of disorder. Due to the symmetry, simple analytical expressions can be derived, as it is shown below. Note however that, in presence of disorder, the following expressions are not valid and one must use instead the general definitions of the rates [see Eqs.~\eqref{rate2}, the paragraph below those equations and Eqs.~\eqref{3levrate}].

Let us start with the D- and LH- configurations, where absorption takes place from a polarized CW laser source.

\subsubsection{D-configuration}

For the D-configuration the laser source is polarized along the $z$ axis and resonant with the ring eigenstate having the highest energy. Thus we compute the quantities in Eq.~\eqref{curr} as
\begin{subequations}
  \label{parD}
  \begin{eqnarray}
    B_{TOT} &= T_L = \frac{N\Omega_R^2\Gamma_T \sin^2\theta}{2[\Gamma_T^2+(\omega-\omega_N)^2]} \\
    \left\langle B \right\rangle &=T_L p_N \approx 0\\
    \left\langle \gamma \right\rangle &= \gamma_N p_N + \gamma_2 p_2 + \gamma_3 p_3 \approx N\gamma \cos^2\theta \, p_*\\
    T_{TOT}^{RC} &= T_2^{RC} \approx \tau_{RC}^{-1} \left( \frac{32}{N}\right)^5 \cos^2\theta \\
    \left\langle T^{RC} \right\rangle &= T_2^{RC} p_2 \approx \tau_{RC}^{-1} \left( \frac{32}{N}\right)^5 \cos^2\theta \, p_* \\
    B_{RC} &= \frac{(\hat{p}_{rc}\cdot\hat{z})^2\Omega_R^2\Gamma_{T,RC}}{2[\Gamma_{T,RC}^2+(\omega-\omega_{\rm rc})^2]} = 0
      \end{eqnarray}
\end{subequations}
where $p_N$ and $p_2 = p_3 =: p_*$ are, respectively, the Boltzmann populations for the eigenstate with the highest energy, and for the first and second excited eigenstates, which are degenerate and, thus, they have the same Boltzmann population $p_*$.
Here, $\hat{p}_{rc}=\hat{y}$ is the unit vector indicating the dipole direction of the RC, while $\Gamma_{T,RC}$ is the dephasing rate between $\ket{0}$ and $\ket{rc}$, and it is given by Eq.~\eqref{GTsite}. Since $\hat{p}_{rc}$ is orthogonal to the polarization $\hat{z}$ of the laser, we have $B_{RC}=0$. The phenomenological parameter $\tau_{RC}$ represents the transfer time between the ring and the RC for $N=32$ and $\theta=0$, as explained above. In the range of parameters that we have studied in our simulations, at $T \approx 300$~K we are justified to neglect the factor $p_N$.

\subsubsection{LH-configuration}

For the LH-configuration the laser source is polarized along the $y$ axis and resonant with the first and second (degenerate) excited states of the excitonic subspace. The difference between the D-configuration and the LH-configuration is in the polarization and frequency of the laser source. The only parameters which have different values between the two configurations are $B_{TOT}$, $\left\langle B \right\rangle$ and $B_{RC}$. Specifically, $B_{RC}$ here is different from zero, because the dipole of the RC is parallel to the laser polarization, $(\hat{p}_{rc}\cdot\hat{y})^2=1$, and therefore for the LH-configuration we have
\begin{subequations}
  \label{parLH}
  \begin{eqnarray}
    B_{TOT} &= T_L = \frac{N\Omega_R^2\Gamma_T \cos^2\theta}{4[\Gamma_T^2+(\omega-\omega_2)^2]} \\
    \left\langle B \right\rangle &= T_L p_2 = \frac{N\Omega_R^2\Gamma_T \cos^2\theta}{4[\Gamma_T^2+(\omega-\omega_2)^2]} p_*\\
    \left\langle \gamma \right\rangle & \approx N\gamma \cos^2\theta \, p_*\\
    T_{TOT}^{RC} &\approx \tau_{RC}^{-1} \left( \frac{32}{N}\right)^5 \cos^2\theta \\
    \left\langle T^{RC} \right\rangle &\approx \tau_{RC}^{-1} \left( \frac{32}{N}\right)^5 \cos^2\theta \, p_* \\
    B_{RC} &= \frac{\Omega_R^2\Gamma_{T,RC}}{2[\Gamma_{T,RC}^2+(\omega-\omega_{\rm rc})^2]}~.
      \end{eqnarray}
\end{subequations}

\subsubsection{Sunlight}

For the case of absorption from natural sunlight, the incoming light is isotropic, unpolarized, incoherent, and it covers a broad black-body spectrum at $T_S\approx 6000$~K. Each eigenstate has a corresponding absorption and stimulated emission rate $B_\alpha=f_Sn_S(\omega_0)\gamma_\alpha$, see~\ref{sec:sun}. The only parameters which have different values with respect to the laser case are $B_{TOT}$, $\left\langle B \right\rangle$ and $B_{RC}$. Specifically, for the natural sunlight we have
\begin{subequations}
  \label{parLH2}
  \begin{eqnarray}
    B_{TOT} &= N\gamma f_Sn_S(\omega_0) \\
    \left\langle B \right\rangle &\approx N\gamma f_Sn_S(\omega_0) \cos^2\theta \, p_*\\
    \left\langle \gamma \right\rangle & \approx N\gamma \cos^2\theta \, p_*\\
    T_{TOT}^{RC} &\approx \tau_{RC}^{-1} \left( \frac{32}{N}\right)^5 \cos^2\theta \\
    \left\langle T^{RC} \right\rangle &\approx \tau_{RC}^{-1} \left( \frac{32}{N}\right)^5 \cos^2\theta \, p_* \\
    B_{RC} &= \gamma f_Sn_S(\omega_0)~.
      \end{eqnarray}
\end{subequations}

\section{Validity of the three-level model for weak and strong trapping rate}
\label{sec:klarge}

\begin{figure*}[!htb]
  \centering
  \includegraphics[width=\linewidth]{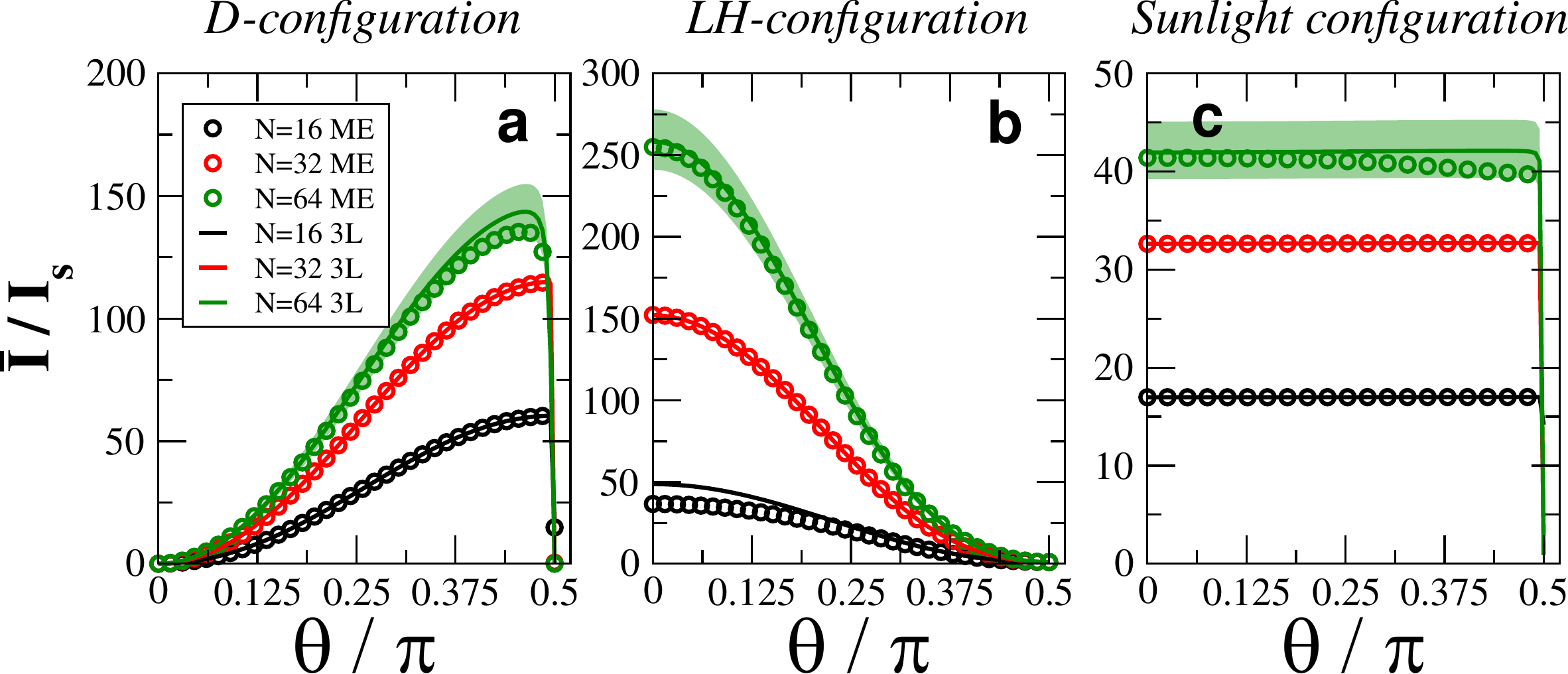}
  \caption{{\it Validity of the         three-level model for large trapping, $\kappa=10^4\gamma$}.
  Transmitted normalized current at the peak laser frequency, $\bar{I}/I_s$, \emph{\it vs.} $\theta$ for D-configuration (a), LH-configuration (b), and Sunlight configuration (c). All simulations are done at room temperature ($T=300$~K). Different values of $N$ have
  been considered, varying the radius of the ring to keep the
  density fixed.
  Symbols represent the current $\bar{I}$, see Eq.~\eqref{eq:cur}, obtained from the master equation (ME), see Eq.~\eqref{eq:mas}, while the solid curve is the three-level approximation $\bar{I}_3$ (3L), see Eq.~\eqref{jj}.
  Shaded areas represent variations in $\bar{I}_3$ produced by $\pm 20\%$ variations in $\tau_{RC}$.
  Parameters for all panels: $\kappa=10^4\gamma$, $\Omega_R=4.68\gamma$ 
  (laser intensity: $1365$~W/m$^2$, same as natural sunlight), 
  $\tau_{RC}=3.9$~ps (for three-level model).
  }
  \label{fig:theta}
\end{figure*}

\begin{figure*}[!htb]
  \centering
  \includegraphics[width=\linewidth]{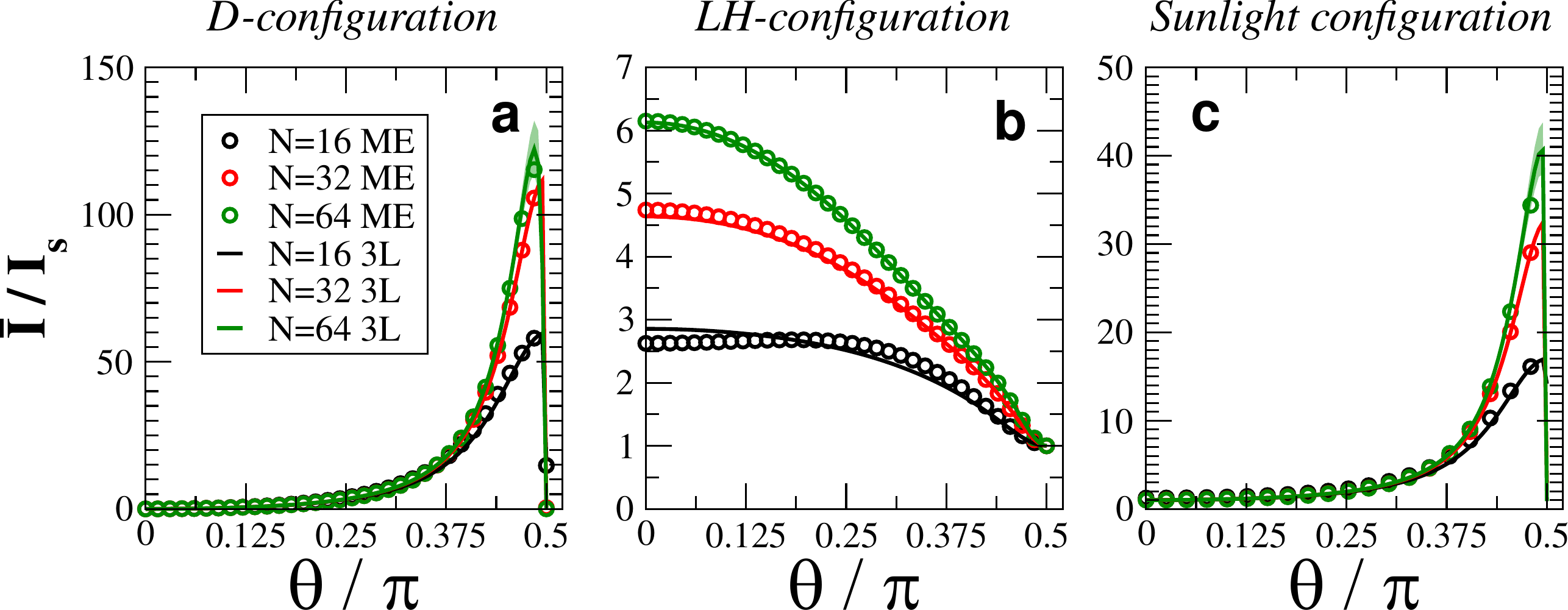}
  \caption{{\it Validity of the three-level model for small trapping, $\kappa=10^{-4}\gamma$}.
    Transmitted normalized current at the peak laser frequency, $\bar{I}/I_s$, \emph{\it vs.} $\theta$ for D-configuration (a), LH-configuration (b), and Sunlight configuration (c). All simulations are done at room temperature ($T=300$~K). Different values of $N$ have
    been considered, varying the radius of the ring to keep the
    density fixed.
    Symbols correspond to the the current $\bar{I}$, see Eq.~\eqref{eq:cur}, obtained solving the master equation (ME), see Eq.~\eqref{eq:mas}, while the solid curve is the three-level approximation $\bar{I}_3$ (3L), see Eq.~\eqref{jj}.
    Shaded areas represent variations in $\bar{I}_3$ produced by $\pm 20\%$ variations in $\tau_{RC}$.
    Parameters for all panels: $\kappa=10^{-4}\gamma$, $\Omega_R=4.68\gamma$ 
  (laser intensity: $1365$~W/m$^2$, same as natural sunlight), 
$\tau_{RC}=3.9$~ps (for three-level model).
  }
  \label{fig:kappasmall}
\end{figure*}

\begin{figure}[!htb]
  \centering
  \includegraphics[width=0.59\columnwidth]{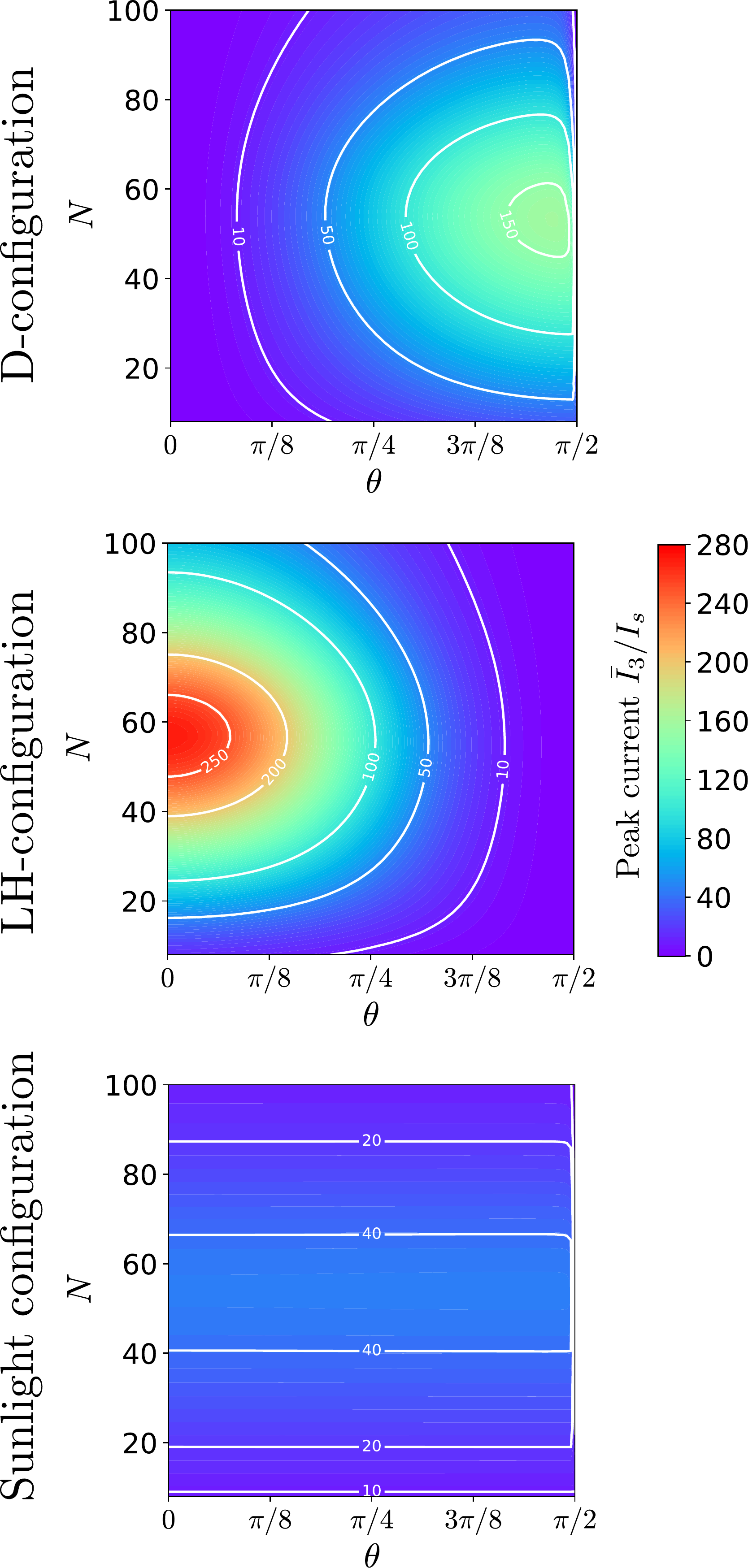}
  \caption{\textit{Peak current vs. $\theta$ and $N$ for large trapping, $\kappa=10^4\gamma$}. Normalized current at the peak laser frequency $\bar{I}_3/I_s$ obtained from the three-level model, see Eq.~\eqref{jj}. Different illumination conditions 
  are used (see figure). For the sunlight configuration (lowest panel), $\bar{I}_3$ coincides with $I_3$, since there is no laser frequency. Parameters: $\kappa=10^4\gamma$, $\Omega_R=4.68\gamma$ 
  (laser intensity: $1365$~W/m$^2$, same as natural sunlight), 
$\tau_{RC}=3.9$~ps, $T=300$~K.
}
  \label{fig:coop2}
\end{figure}

In the main text we have shown that the three-level model reproduces the peak current very well for the trapping rate $\kappa=10\gamma$. Here we show that the three-level model works very well also for weaker and stronger trapping rates.

Specifically, in Fig.~\ref{fig:theta} we plot the peak current as a function of $\theta$ for the D-, LH- and Sunlight- configurations for large trapping rate, $\kappa=10^4\gamma$. As one can see, the results obtained with the master equation (symbols) are very close to the three-level model results (lines).
Some discrepancies are present in the LH-configuration (Fig.~\ref{fig:theta}b) for $N=16$ and $\theta \approx 0$ because the three-level model in that regime does not capture the ring-RC transfer process. In fact for those parameters, as one can see from Fig.~\ref{fig:tauRC}, the ring-RC population transfer follows purely quantum-mechanical quadratic dynamics, and therefore the three-level model is not expected to work.
Deviations are also seen for large $N$ in the D- and Sunlight configurations. As discussed in the main text and in~\ref{sec:3lev}, those deviations are due to our choice of a constant, $\theta$-independent $\tau_{RC}$ parameter in our simulations, while Table~\ref{tab:fit} shows that $\tau_{RC}$ can vary by up to $20\%$ with $\theta$. Once we account for $\pm 20\%$ variations in $\tau_{RC}$, we obtain a confidence interval for the three-level current (see shaded areas in figure) that include the master equations results.
For the Sunlight configuration (Fig.~\ref{fig:theta}c), for strong trapping, $\kappa=10^4\gamma$, the current is independent of $\theta$, because in this limit the current it is ultimately determined just by the absorption rate, which is $Nf_Sn_S\gamma$, independent of $\theta$.

On the other hand, in Fig.~\ref{fig:kappasmall} we show the peak current as a function of $\theta$ for the D-, LH- and Sunlight- configurations for small trapping rate, $\kappa=10^{-4}\gamma$. Also here, the results obtained with the master equation (symbols) are nearly identical to the three-level model results (lines), thus justifying our use of the three-level model in the main text.

Moreover, in Fig.~\ref{fig:coop2} we show the peak current obtained from the three-level model as a function of $\theta$ and $N$ for large trapping, $\kappa=10^4\gamma$. A similar figure is present in the main text, see Fig.~\ref{fig:coop}, showing the effectiveness of the D-configuration for intermediate trapping values, $\kappa=10\gamma$. Here, see Fig.~\ref{fig:coop2}, we show that for large trapping, $\kappa=10^4\gamma$, the D-configuration is actually performing worse than the LH-configuration.

\bibliographystyle{iopart-num}
\bibliography{biblio}

\end{document}